\documentclass[useAMS,fleqn]{aa}
\usepackage{amsmath}
\usepackage{amsfonts}
\usepackage{amssymb}
\usepackage{bm}
\usepackage{caption}
\usepackage{float}
\usepackage{graphicx}
\usepackage{longtable}
\usepackage{marginnote}
\usepackage{mathptmx}
\usepackage{multirow}
\usepackage{pdflscape}
\usepackage{placeins}
\usepackage{rotating}
\usepackage{subcaption}
\usepackage{siunitx}
\usepackage{tabularx}
\usepackage{threeparttable}

\DeclareMathAlphabet{\mathcal}{OMS}{cmsy}{m}{n}

\usepackage{natbib,twoopt}
\usepackage[breaklinks=true]{hyperref} 
\bibpunct{(}{)}{;}{a}{}{,} 
\makeatletter
\newcommandtwoopt{\citeads}[3][][]{\href{http://adsabs.harvard.edu/abs/#3}%
{\def\hyper@linkstart##1##2{}%
\let\hyper@linkend\@empty\citealp[#1][#2]{#3}}}
\newcommandtwoopt{\citepads}[3][][]{\href{http://adsabs.harvard.edu/abs/#3}%
{\def\hyper@linkstart##1##2{}%
\let\hyper@linkend\@empty\citep[#1][#2]{#3}}}
\newcommandtwoopt{\citealtads}[3][][]{\href{http://adsabs.harvard.edu/abs/#3}%
{\def\hyper@linkstart##1##2{}%
\let\hyper@linkend\@empty\citealt[#1][#2]{#3}}}
\newcommandtwoopt{\citetads}[3][][]{\href{http://adsabs.harvard.edu/abs/#3}%
{\def\hyper@linkstart##1##2{}%
\let\hyper@linkend\@empty\citet[#1][#2]{#3}}}
\newcommandtwoopt{\citeyearads}[3][][]%
{\href{http://adsabs.harvard.edu/abs/#3}
{\def\hyper@linkstart##1##2{}%
\let\hyper@linkend\@empty\citeyear[#1][#2]{#3}}}
\makeatother

\hypersetup{colorlinks=true,linkcolor=blue,citecolor=blue,filecolor=blue,urlcolor=blue}

\def\reff@jnl#1{{\rm#1\/}}

\def\aj{\reff@jnl{AJ}}                  
\def\araa{\reff@jnl{ARA\&A}}            
\def\apj{\reff@jnl{ApJ}}                        
\def\apjl{\reff@jnl{ApJ}}               
\def\apjs{\reff@jnl{ApJS}}              
\def\ao{\reff@jnl{Appl.Optics}}         
\def\apss{\reff@jnl{Ap\&SS}}            
\def\aap{\reff@jnl{A\&A}}                       
\def\apjl{\reff@jnl{ApJ}}               
\def\aapr{\reff@jnl{A\&A~Rev.}}         
\def\aaps{\reff@jnl{A\&AS}}             
\def\azh{\reff@jnl{AZh}}                        
\def\baas{\reff@jnl{BAAS}}              
\def\jrasc{\reff@jnl{JRASC}}            
\def\memras{\reff@jnl{MmRAS}}           
\def\mnras{\reff@jnl{MNRAS}}            
\def\pra{\reff@jnl{Phys. Rev. A}}         
\def\prb{\reff@jnl{Phys. Rev. B}}         
\def\prc{\reff@jnl{Phys. Rev. C}}         
\def\prd{\reff@jnl{Phys. Rev. D}}         
\def\prl{\reff@jnl{Phys. Rev. Lett}}      
\def\pasp{\reff@jnl{PASP}}              
\def\pasj{\reff@jnl{PASJ}}              
\def\qjras{\reff@jnl{QJRAS}}            
\def\skytel{\reff@jnl{S\&T}}            
\def\solphys{\reff@jnl{Solar~Phys.}}    
\def\sovast{\reff@jnl{Soviet~Ast.}}     
\def\ssr{\reff@jnl{Space~Sci.Rev.}}     
\def\zap{\reff@jnl{ZAp}}                        
\def\nat{\reff@jnl{Nature}}             

\def\p#1by#2{{\partial{#1} \over \partial{#2}}}
\def\pp#1by#2#3{{\partial^2{#1} \over \partial{#2}\partial{#3}}}
\def\d#1by#2{{{\rm d}{#1} \over {\rm d}{#2}}}
\def\dd#1by#2#3{{{\rm d}^2{#1} \over {\rm d}{#2}{\rm d}{#3}}}

\begin{document}
\title{Diffuse radio emission in MACS~J0025.4$-$1222: the effect of a major merger on bulk separation of ICM components}
\titlerunning{A new double-relic system in MACS~J0025.4$-$1222}


\author{
 C.~J.~Riseley\inst{\ref{i0},\ref{i1},\ref{i2}} \and
 A.~M.~M.~Scaife\inst{\ref{i1}} \and M.~W.~Wise\inst{\ref{i3},\ref{i4}} \and A.~O.~Clarke\inst{\ref{i1}}
 }
\institute{
CSIRO Astronomy \& Space Science, 26 Dick Perry Avenue, Kensington, WA 6151, Australia. \email{chris.riseley@csiro.au}\label{i0}
\and
Jodrell Bank Centre for Astrophysics, Alan Turing Building, School of Physics and Astronomy, The University of Manchester, Oxford Road, Manchester, M13 9PL, U.K. \label{i1}
\and
School of Physics \& Astronomy, University of Southampton, Highfield, Southampton, SO17 1BJ, U.K. \label{i2}
\and
Netherlands Institute for Radio Astronomy (ASTRON), Postbus 2, 7990 AA Dwingeloo, The Netherlands. \label{i3}
\and
Astronomical Institute Anton Pannekoek, University of Amsterdam, Postbus 94249, 1090 GE Amsterdam, The Netherlands. \label{i4}
}
\authorrunning{C.~J.~Riseley et al.}

\date{Accepted ---; received ---; in original form \today}


\abstract
	{Mergers of galaxy clusters are among the most energetic events in the Universe. These events have significant impact on the intra-cluster medium, depositing vast amounts of energy -- often in the form of shocks -- as well as heavily influencing the properties of the constituent galaxy population. Many clusters have been shown to host large-scale diffuse radio emission, known variously as radio haloes and relics. These sources arise as a result of electron (re-)acceleration in cluster-scale magnetic fields, although the processes by which this occurs are still poorly understood.}  
	{We present new, deep radio observations of the high-redshift galaxy cluster MACS~J0025.4$-$1222, taken with the GMRT at 325 MHz, as well as new analysis of all archival \emph{Chandra} X-ray observations. We aim to investigate the potential of diffuse radio emission and categorise the radio population of this cluster, which has only been covered previously by shallow radio surveys.} 
	{We produce low-resolution maps of MACS~J0025.4$-$1222 through a combination of \emph{uv-}tapering and subtracting the compact source population. Radial surface brightness and mass profiles are derived from the \emph{Chandra} data. We also derive a 2D map of the ICM temperature.} 
	{For the first time, two sources of diffuse radio emission are detected in MACS~J0025.4$-$1222, on linear scales of several hundred kpc. Given the redshift of the cluster and the assumed cosmology, these sources appear to be consistent with established trends in power scaling relations for radio relics. The X-ray temperature map presents evidence of an asymmetric temperature profile and tentative identification of a temperature jump associated with one relic.} 
	{We classify the pair of diffuse radio sources in this cluster as a pair of radio relics, given their consistency with scaling relations, location toward the cluster outskirts, and the available X-ray data.}


\keywords{radio continuum: general -- galaxies: clusters: individual (MACS~J0025.4$-$1222) -- X-rays: galaxies: clusters }

\maketitle

\section{Introduction}
Clusters and super-clusters of galaxies comprise the largest gravitationally-bound structures in the Universe, following the profile of dark matter distribution left over from high redshift. In the hierarchical model of structure formation, these objects grow through a number of processes, from constant accretion of matter from the local environment, to periodic consumption of small groups of galaxies, to violent merger events involving collisions with other galaxy clusters. During merger events, vast amounts of gravitational energy ($\sim10^{64}$ erg; e.g. \citealt{2008SSRv..134...93F}) are deposited into the intracluster medium (ICM) during the timescale necessary for clusters to cross ($\sim\,1$ Gyr; \citealt{bj2014}).

The most massive clusters of galaxies have masses of the order of $10^{15} \, \rm{M}_{\odot}$, of which typically $\sim70-80$ per cent is dark matter, and the remaining baryonic matter is split between the hot ICM gas ($\sim15-20$ per cent) and galaxies (a few per cent). The ICM gas is hot ($T \sim10^8$ K) and sparse ($n_{\rm{gas}}\sim10^{-3} \, \rm{cm}^{-3}$) and known to be permeated by a large-scale magnetic field. Major merger events strongly disrupt the structure and properties of galaxy clusters. They are believed to both trigger and quench periods of star formation in cluster member galaxies (e.g. \citealt{ma2010}) as well as causing both large-scale disruption and heating of the ICM. 

In a small number of galaxy clusters, merger events have been shown to cause physical separation in the components of the ICM -- for example, in the Bullet Cluster (1E~0657-56) there is clear separation between the peaks of the dark matter and baryonic matter distribution (e.g. \citealt{2006ApJ...648L.109C,2006ApJ...652..937B}). In terms of observational indicators, lensing effects strongly trace the total mass distribution (which is dominated by dark matter), X-ray emission from galaxy clusters traces the thermal gas component, and radio emission traces the non-thermal components -- magnetic fields and the cosmic ray population. 

Energy is primarily dissipated into the ICM in the form of shocks, as well as bulk motion of the ICM (e.g. \citealt{bj2014}). Historically, shocks have mainly been detected in galaxy clusters through observations at X-ray wavelengths, typically presenting as sharp jumps in the surface brightness, temperature, and density of the ICM (see \citealt{bj2014} for a recent review). However, toward the cluster outskirts, where stronger shocks are expected, the X-ray flux density is significantly lower than in the cluster centre, and shocks are far more difficult to detect. 

Radio relics are a class of large-scale diffuse radio source associated with merging clusters. Observational data is broadly consistent with an interpretation of diffusive shock acceleration (DSA; e.g. \citealt{feretti2012,bj2014}) mechanism: radio relics are usually aligned perpendicular to the merger axis, toward the outskirts of the cluster, and they generally exhibit a high polarization fraction (indicative of highly-ordered magnetic fields) with suggestions that the magnetic field is aligned with the outward-propagating shock front.

The association between relics and shocks has been confirmed in only a handful of clusters to-date -- for example Abell 754 \citep{2011ApJ...728...82M}, Abell 3376 \citep{2012PASJ...64...67A}, and Abell 3667 (e.g. \citealt{2010ApJ...715.1143F,2012PASJ...64...49A}). However, questions are emerging regarding the nature of the connection between shocks and radio relics as some observations have shown that radio relics are not always co-located with X-ray shocks (e.g. \citealt{2013MNRAS.433..812O}). As such, clarifying the picture of the relationship between X-ray and radio properties of galaxy clusters is crucial to better understanding the processes by which these Mpc-scale sources of diffuse synchrotron emission come into being.

Many merging clusters have also been shown to radio haloes., amorphous, large-scale ($\sim$Mpc size) radio sources associated with the ICM. Like relics, haloes are very diffuse (typically $\sim0.1 - 1\,\mu$Jy arcsec$^{-2}$ at 1.4 GHz). Unlike relics, however, these objects generally exhibit polarization of less than a few per cent. Two exceptions to this are Abell 2255 (e.g. \citealt{2011A&A...525A.104P}) and MACS J0717.5+3745 (e.g. \citealt{2009A&A...503..707B}) which both host haloes that exhibit a high degree of polarization. However, this may be due to relics viewed in projection (in the case of MACS J0717.5+3745; see \citealt{2009A&A...505..991V}) or highly-polarized filamentary structures within the halo (in the case of Abell 2255; see \citealt{2005A&A...430L...5G}). 

Two principal models of halo generation exist: turbulent acceleration models and secondary (or hadronic) models. Within the turbulent acceleration framework (e.g. \citealt{2001MNRAS.320..365B,2001ApJ...557..560P}) low-energy ($\sim1-10$ GeV) electrons from the ICM are accelerated up to the radio emitting regime ($\gtrsim10$ GeV) by merger turbulence. Conversely, hadronic models (e.g. \citealt{1980ApJ...239L..93D,1999APh....12..169B}) suggest the electrons responsible for the observed synchrotron emission are injected into the ICM following inelastic collisions between relativistic protons and thermal ions. 

The amorphous nature and low polarization fraction of haloes are naturally explained within the context of the turbulent acceleration models, as the electrons responsible for the emission are located throughout the cluster ICM (explaining the typical similarity in extent between halo size and the X-ray emission) and the turbulence disrupts any large-scale ordering of magnetic fields (required for a high polarization fraction). The large size of both relics and haloes requires efficient \emph{in-situ} acceleration, as the lifetime of synchrotron-emitting electrons is significantly less than the typical timescale required for electrons to diffuse over the linear size of these objects \citep{1977ApJ...212....1J}.

MACS~J0025.4$-$1222 is a reasonably familiar massive cluster at high redshift. It is perhaps most well-known for being the second cluster where significant offset was found between the peaks in the dark matter and baryonic matter distributions \citep{bradac2008}. This offset is interpreted as strong evidence for the low interaction cross-section of dark matter, as in the case of the Bullet cluster (e.g. \citealt{2006ApJ...648L.109C,2006ApJ...652..937B}).

\subsection{MACS~J0025.4$-$1222}
MACS~J0025.4$-$1222 (hereafter MACS0025; J2000 right ascension and declination 00$^{\rm{h}}$25$^{\rm{m}}$29.38$^{\rm{s}}$ $\SI{-12}{\degree}22^{\prime}37.0^{\prime\prime}$) was identified in the Massive Cluster Survey (MACS; \citealt{2001ApJ...553..668E}) sample of high-luminosity, high-redshift clusters \citep{2007ApJ...661L..33E}. MACS0025 is a double-cluster system believed to have undergone a recent major merger event. Based on a sample of 108 galaxies within 1.5 Mpc of the cluster centre, \cite{bradac2008} find the galaxy redshift distribution is consistent with a single Gaussian centred at a redshift of $z = 0.5857$; the velocity dispersion is $\sigma_{\rm{cl}} = 835^{+58}_{-59}$ km s$^{-1}$. The redshift difference between the two sub-clusters is $\Delta z = 0.0005 \pm 0.0004$; this indicates that the merger event is occurring close to the plane of the sky -- within $\SI{5}{\degree}$ \citep{bradac2008}.

The sub-clusters of MACS0025 have very similar total masses: for the North-West sub-cluster, the total mass is $M = 2.6^{+0.5}_{-1.4} \times 10^{14} {\rm{M}_{\odot}}$; for the South-East sub-cluster the total mass is $M = 2.5^{+1.0}_{-1.7} \times 10^{14} {\rm{M}_{\odot}}$, measured within 300 kpc of the brightest cluster galaxy (BCG) of each sub-cluster \citep{bradac2008}. With such similar masses, theory suggests that a major merger should generate a pair of similar shocks propagating out from the cluster centre. 

However, X-ray observations to-date have been unable to detect any significant shocks -- \cite{bradac2008} present 1-dimensional radial temperature measurements only. Additionally, MACS0025 is relatively unexplored at radio wavelengths: \cite{2014arXiv1412.0285P} report the only published radio observations to-date. Their observations were conducted with the Giant Metrewave Radio Telescope (GMRT) at 610 and 235 MHz; no diffuse emission was detected, however the integration time of 5 hours will have been a limiting factor when observing this high-redshift cluster.

\cite{ma2010} perform a deep study of the morphological, spectroscopic and photometric properties of galaxies within the MACS0025 region using the \emph{Hubble Space Telescope} (\emph{HST}) and \emph{Keck}. Based on 436 galaxy spectra, they identify 212 cluster members within 4 Mpc of the cluster centre. They find the global fraction of spiral and lenticular galaxies to be among the highest observed in high-redshift clusters. Additionally, they find six of the 212 cluster members to be post-starburst galaxies, all concentrated in the cluster centre, between the dark matter peaks. They propose that the starburst phase of these galaxies was both triggered and extinguished by the cluster merger, and that first core passage occurred 0.5--1 Gyr ago.

In this work we present a multi-wavelength study of MACS0025, based on new, deep radio observations with the GMRT and re-processed archival \emph{Chandra} data. Throughout this work, we assume a concordance cosmology of H$_{0}=73$ km s$^{-1}$ Mpc$^{-1}$, $\Omega_{\rm{m}} = 0.27$, $\Omega_{\rm{vac}} = 0.73$. All errors are quoted to 1-$\sigma$. We adopt the spectral index convention that $S \propto \nu^{\alpha}$ and we take the redshift of MACS0025 to be 0.5857 \citep{bradac2008}. At this redshift, an angular distance of 1$^{\prime\prime}$ corresponds to 6.416 kpc \citep{2006PASP..118.1711W}.

\begin{table*}
\centering
\caption{Summary of the GMRT observations of MACS0025. \label{tab:obs_summary}}
\scalebox{0.9}{
\begin{tabular}{cccccp{6.5cm}}
\hline
 & & \\ 
Date 	& Start time & Hours observed & \# Antennas & Antennas missing & Comments     \\
 		& (IST) & & & & \\
\hline 
2014 Jan 12 & 13:00 & 8.25 & 29 & C10 & S02 down for maintenance 14:50 -- 16:26. C10 started observing at 18:09. C00--C03 no fringes between 16:53 and 18:13. Reset at 18:18 due to correlator problem.\\
2014 Jan 13 & 18:09 & 4.00 & 28 & C10, W02 & C10 (W02) started observing at 18:35 (18:57). \\
2014 Feb 01 & 15:00 & 6.00 & 28 & C14, W02 & Issues with power level settings forced reset at 16:18, data taken beforehand were rejected. \\
\hline
\end{tabular}
}
\end{table*}

\section{Observations \& Data Reduction}
\subsection{GMRT Data}
\subsubsection{Observations}
MACS0025 was observed with the GMRT for a total of 3 nights during 2014 January and February, for a total of approximately 15 hours. Data were taken with the GMRT Software Backend (GSB; \citealt{2010ExA....28...25R}) in 325 MHz mode, with the default settings. Data were recorded every 8.05 seconds, with an acquisition bandwidth of 33 MHz, and 256 channels. Two polarization products were collected. 

Throughout all observations, a minimum of 28 antennas were functional; any missing antennas were taken for painting, or down due to problems with the hardware. Two major failures occurred during the observing runs -- antenna fringe failures on Jan 12 resulted in a system reset, whereas significant power issues occurred during the first hour of observing on Feb 01. Compromised data were discarded, for a total integration time on target of around 12 hours. A summary of each night's observations, including integration time, number of working antennas, and any general comments are listed in Table \ref{tab:obs_summary}. 

\begin{figure}[!htb]
	\begin{center}
	\begin{tabular}{r}
		\includegraphics[width=0.45\textwidth]{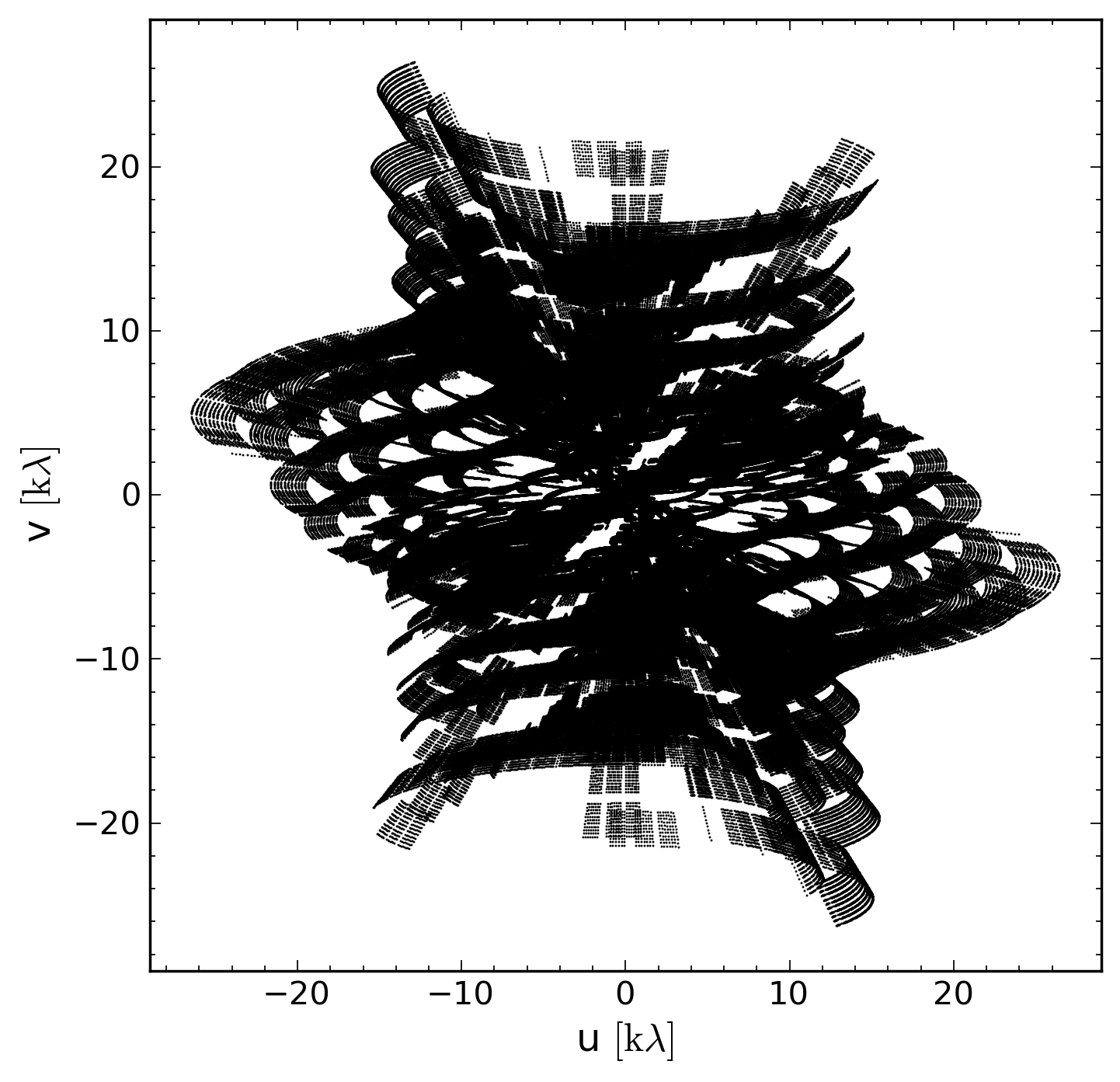} \\
		\includegraphics[width=0.443\textwidth]{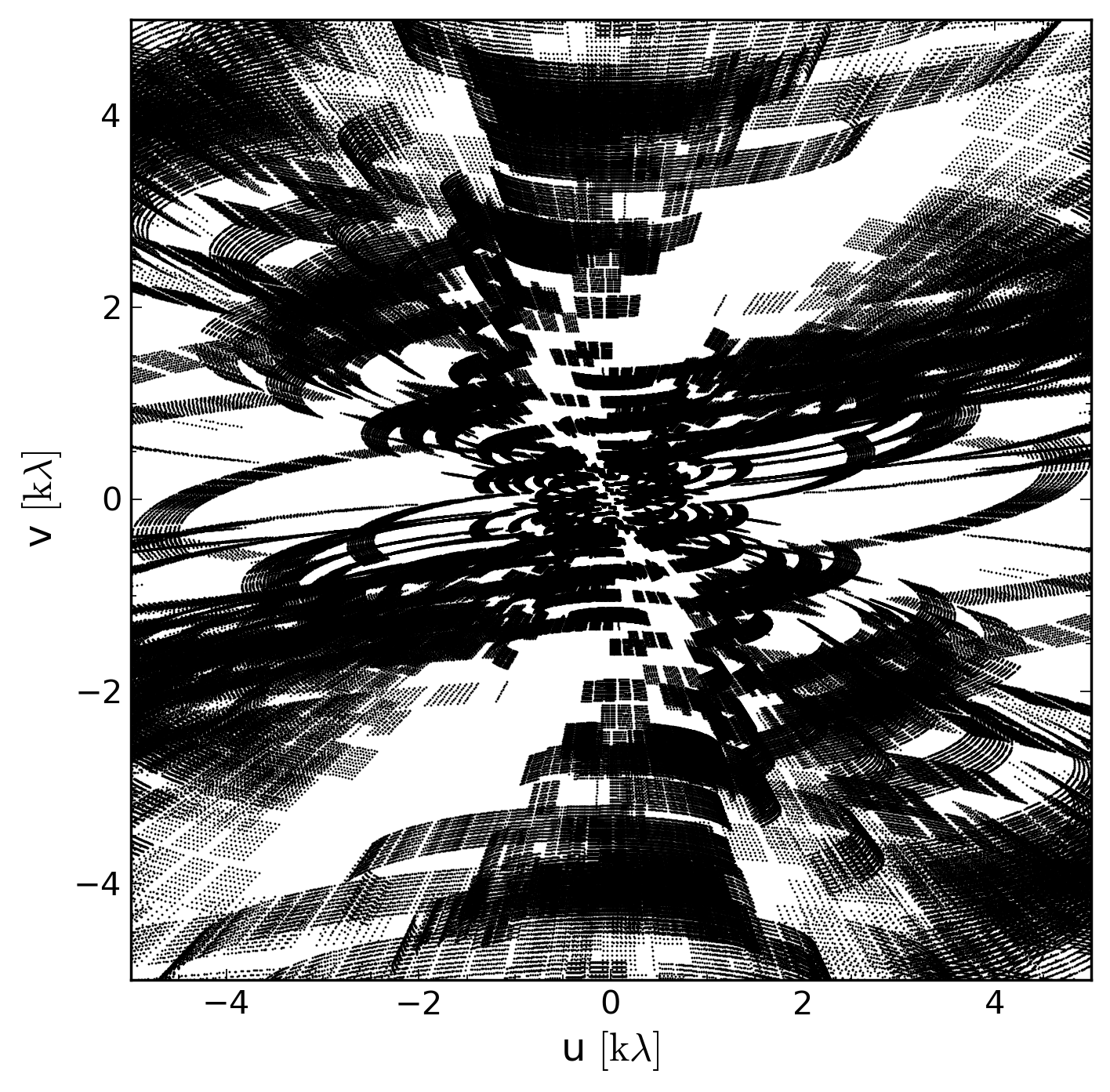} \\
	\end{tabular}
	\end{center}
\caption{\emph{uv}-coverage of GMRT observations of MACS0025. \emph{Top:} Full \emph{uv-}range. \emph{Bottom:} Zoom on the central portion of the \emph{uv-}plane, illustrating the coverage on short baselines. Only one out of every ten data points are plotted.}
\label{fig:uvcovg}
\end{figure}

On each night, the observations of the interleaved phase calibrator (0018-127; selected from the VLA calibrator manual\footnote{\url{http://www.aoc.nrao.edu/~gtaylor/csource.html}}) and target (MACS0025) were bracketed at the beginning and end of the run by a 15-minute scan on a primary calibrator source: either 3C~147, 3C~48 or 3C~468.1, depending on availability. The final \emph{uv-}coverage of the GMRT observations of MACS0025 is shown in Figure \ref{fig:uvcovg}, where both the full \emph{uv-}plane and a zoom on the short ($<5\rm{k}\lambda$) baselines is presented.

\subsubsection{Data Reduction}
\defcitealias{2012MNRAS.423L..30S}{SH12} 

The data were reduced using the Source Peeling and Atmospheric Modelling (\texttt{SPAM}) software \citep{2009A&A...501.1185I} which employs NRAO Astronomical Image Processing Software (\texttt{AIPS}) tasks through the \texttt{ParselTongue} interface \citep{2006ASPC..351..497K}. \texttt{SPAM} employs the \citet{2012MNRAS.423L..30S} flux density scale (hereafter \citetalias{2012MNRAS.423L..30S}) which provides models for a set of six calibrator sources designed to be used at low frequencies (particularly with LOFAR; \citealt{2013A&A...556A...2V}). These sources all have flux densities that are stable over long periods, with well-understood spectral energy distributions (SED) and are compact compared to the resolution of LOFAR, to enable use of simple calibration models. In \citetalias{2012MNRAS.423L..30S}, calibrators are selected from the 3C \citep{1959MmRAS..68...37E} and revised 3C (3CR; \citealt{1962MmRAS..68..163B}) catalogues based on three criteria. Firstly, they must be at declination North of $\delta = \SI{20}{\degree}$. Secondly, they must have an integrated flux density greater than 20 Jy at 178 MHz, and thirdly, they must have an angular diameter less than 20 arcsec. 

The radio source source 3C~468.1 is available as a bandpass calibrator for the GMRT, although it is less well documented than 3C~48, 3C~147 and 3C~286. 3C~468.1 is also not on the \citetalias{2012MNRAS.423L..30S} flux scale. However, flux density measurements between 38 MHz and 10.7 GHz are available from the literature, and in this work we use the measurements to extend the \citetalias{2012MNRAS.423L..30S} to include 3C~468.1 -- see Appendix \ref{sec:3c468}. 

\cite{2009A&A...501.1185I} describes data reduction with \texttt{SPAM} in detail; however here we will summarise the process. Following removal of edge channels, the data were averaged by a factor 4 in frequency (yielding 64 channels of width 502.8 kHz) and 2 in time, as a compromise between improving data processing speed and mitigating bandwidth-/time-smearing effects. Prior to calibration, the data were visually inspected for strong RFI and bad antennas/baselines. Calibration solutions were derived for the primary calibrators 3C~147, 3C~48 and 3C~468.1 and applied to the field, following standard techniques in \texttt{SPAM}. Properties of these calibrator sources are presented in Table \ref{tab:calibrators}.

The interleaved calibrator was not used during the reduction process\footnote{The interleaved calibrator was used to track data quality variation and atmospheric effects during the observing run itself.}. Instead, \texttt{SPAM} performs an initial phase calibration (and astrometry correction) using a sky model derived from the NRAO VLA Sky Survey (NVSS; \citealt{1998AJ....115.1693C}) before proceeding with three rounds of direction-independent phase-only self-calibration and imaging.

\begin{table}
\centering
\caption{Properties of calibrator sources used in this work, with flux density measurements and spectral index values on the \cite{2012MNRAS.423L..30S} flux density scale. \label{tab:calibrators}}
\scalebox{0.93}{
\begin{tabular}{lcccc}
\hline
Name & RA & DEC & $S_{325 \, \rm{MHz}}$ & $\alpha$ \\
& (J2000) & (J2000) & [Jy] & \\
\hline
3C~147 & 05$^{\rm{h}}$42$^{\rm{m}}$36.26$^{\rm{s}}$ & $\SI{+49}{\degree}$51$^{\prime}$07.08$^{\prime\prime}$ & 52.925 & $-0.516$\\
3C~48 & $01^{\rm{h}}37^{\rm{m}}41.30^{\rm{s}}$ & $\SI{+33}{\degree}09^{\prime}35.13^{\prime\prime}$ & 43.742 & $-0.608$ \\
3C~468.1 & 23$^{\rm{h}}$50$^{\rm{m}}$54.85$^{\rm{s}}$ & $\SI{+64}{\degree}$40$^{\prime}$19.50$^{\prime\prime}$ & 24.105 & $-0.859$ \\
\hline
\end{tabular}
}
\end{table}

Following the self-calibration, \texttt{SPAM} identifies strong sources within the primary beam FWHM that are suitable for direction-dependent calibration \& ionospheric correction. Only sources with well-defined astrometry are selected, in this case yielding a catalogue of approximately 20 sources. Subsequently, \texttt{SPAM} performs direction-dependent calibration on a per-facet basis, using the solutions to fit a global ionospheric model, as described by \cite{2009A&A...501.1185I}. Throughout the reduction process, multiple automated flagging routines are used between cycles of imaging and self-calibration in order to reduce residual RFI and clip outliers. Overall, 55 per cent of the data were flagged; whilst this is high, this level of flagging is not uncommon for GMRT data at this frequency.

\subsubsection{Imaging}
During the self-calibration and imaging cycles, images were made using an \texttt{AIPS} \texttt{ROBUST} of $-1.0$ as a trade-off between sensitivity and resolution\footnote{An \texttt{AIPS} \texttt{ROBUST} = $-5.0$ indicates pure uniform weighting, for maximum resolution; \texttt{ROBUST} = $+5.0$ indicates pure natural weighting, for maximum sensitivity.}. All imaging was performed with facet-based wide-field imaging as implemented in \texttt{SPAM}. Three different images were produced following calibration: a full-resolution image including the full \emph{uv-}range, with an \texttt{AIPS} \texttt{ROBUST} $-1.0$; a high-resolution image made using baselines longer than $2\rm{k}\lambda$, made to filter out emission on $\sim$Mpc scales; and a low-resolution image made after subtracting the clean-component model of the high-resolution image, with a \emph{uv-}taper of $5\rm{k}\lambda$ to improve sensitivity to diffuse emission. 

\begin{figure*}
	\centering
	\includegraphics[width=0.95\textwidth]{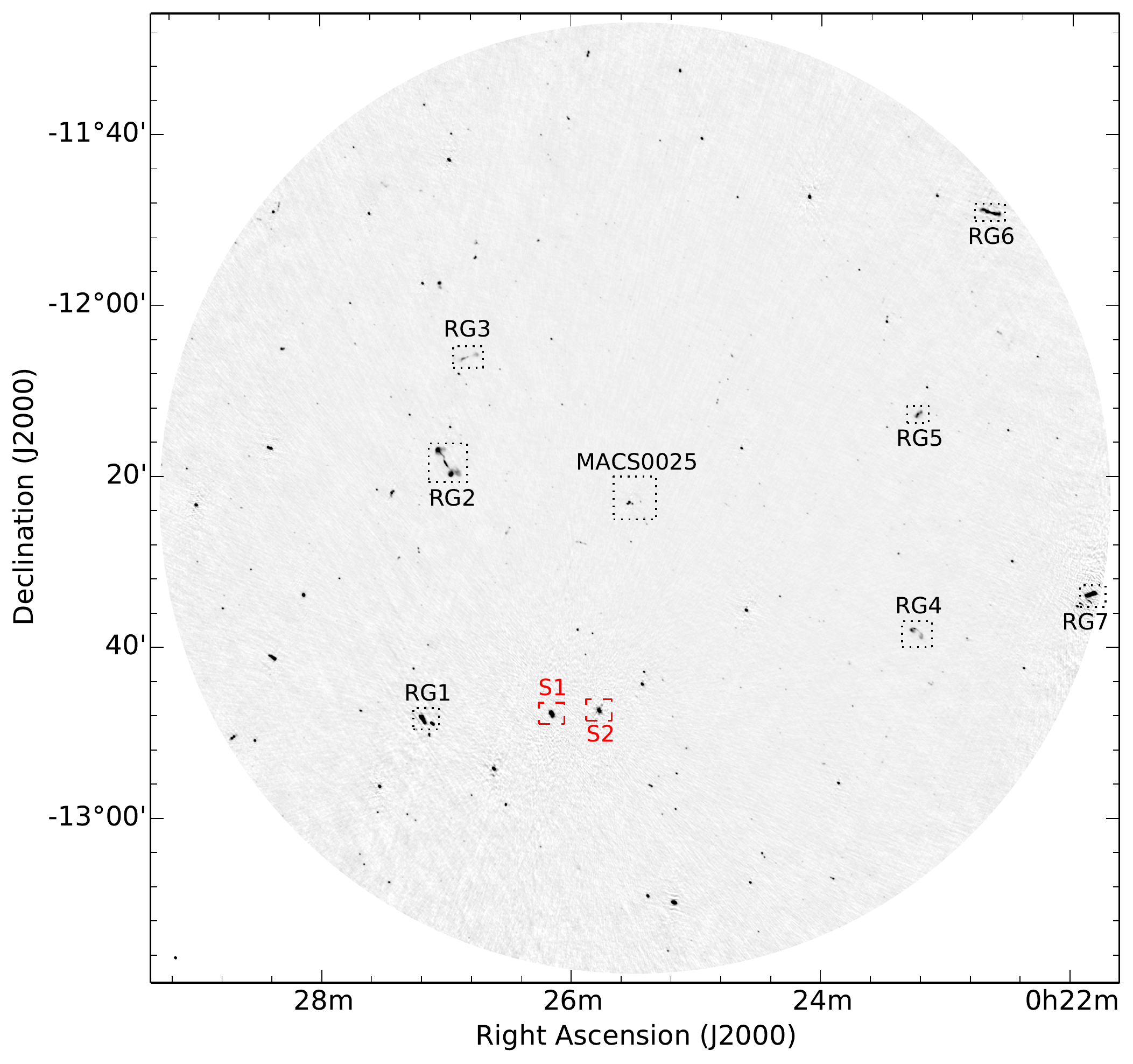}
	\caption{Full field of view of the GMRT at 325 MHz, at a resolution of $10.1\times8.3$ arcsec. Image noise toward the field centre is $81\,\mu$Jy beam$^{-1}$. Colour scale saturates at 10 mJy beam$^{-1}$ in order to emphasise emission from radio galaxies in the field. A number of the more extended radio galaxies are identified: these are further discussed in Appendix \ref{sec:app2}. MACS0025 is also identified, as are the two brightest compact sources in the field, S1 and S2, which have integrated flux densities of $2.70\pm0.27$ and $0.94\pm0.09$ Jy respectively at 325 MHz.}
	\label{fig:macs_full}
\end{figure*}

In Figure \ref{fig:macs_full}, we present the entire field-of-view of the GMRT observations of MACS0025, imaged with an \texttt{AIPS} \texttt{ROBUST} of $-1.0$. The MACS0025 field is rich, with many complex radio sources, including a number of radio galaxies (RG) hosting emission on a wide variety of scales. A number of these are highlighted in Figure \ref{fig:macs_full}, and we discuss these sources further in Appendix \ref{sec:app2}, where we present postage stamps of these sources (Figure \ref{fig:macs_rgpostage}) and mark any potential hosts identified in the literature. Given the locations of these sources and the redshifts of identified hosts, none of these are associated with MACS0025.

Additionally, there are two bright radio sources located reasonably far to the South of the field centre. These are identified as S1 and S2 in Figure \ref{fig:macs_full}; in the literature they are respectively known as PKS~0023$-$13 and MRC~0023$-$12B. With integrated flux densities of $2.70\pm0.27$ and $0.94\pm0.09$ Jy for S1 and S2 respectively\footnote{These were measured using \texttt{imfit} in \textsc{CASA} 4.1.}, these sources caused severe artefacts in the field. While calibration with \texttt{SPAM} has significantly reduced these artefacts, our dynamic range is still limited. We measure an image noise of $81 \, \mu$Jy beam$^{-1}$ toward the field centre, whereas the predicted thermal noise for this observation is around $20 \, \mu$Jy beam$^{-1}$.

\subsection{Chandra Data}

\subsubsection{Observations}
MACS0025 has been observed by {\em Chandra} with the Advanced CCD Imaging Spectrometer \citep[ACIS;][]{Garmire03} a total of five times. All exposures were taken with the ACIS-I imaging configuration and therefore have uniform spectral properties. We have extracted and reprocessed all of this existing data for the subsequent analysis. Table~\ref{tab:obsids} lists the relevant characteristics of the individual exposures.

\begin{table}                                                                                               
\begin{minipage}[t]{0.5\textwidth}
\caption{Observation log of $Chandra$ data for MACS0025}
\label{tab:obsids}                                                                                           
\centering
\renewcommand{\footnoterule}{}  
\begin{tabular}{lccr}                                                                                       
\hline                                                                                                 
OBSID   &  Instrument &   Date  &  Exposure (ksec)\footnote{Exposure times reflect the final exposure after screening for flares.} \\
\hline 
3251  &  ACIS-I  &  2002-11-11  &  18.9 \\
5010  &  ACIS-I  &  2004-08-09  &  24.8 \\
10413 &  ACIS-I  &  2008-10-16  &  75.1 \\
10786 &  ACIS-I  &  2008-10-18  &  13.9 \\
10797 &  ACIS-I  &  2008-10-21  &  23.9 \\
\hline
Total &          &              & 156.6  \\
\hline 
\end{tabular}
\end{minipage}
\end{table}

\subsubsection{Data Reduction}
The individual datasets were reprocessed using CIAO 4.6 and CALDB 4.6.1.1 to apply the latest gain and calibration corrections. Standard filtering was applied to the event files to remove bad grades and pixels. Each of the individual ObsIDs were also examined for the presence of strong background flares. None of the exposures showed evidence for appreciable flare events, however, so no additional filtering was required. These reprocessed event files were used in all subsequent imaging and spectral analyses. The final, combined exposure time after all filtering was 156.6 ksec.

\subsubsection{Imaging}
\label{sec:imaging}
A surface brightness mosaic of the field around MACS0025 was constructed by reprojecting the individual event files to a common tangent point on the sky and then combining them. To correct for sensitivity variations across the field, instrument and exposure maps were made for each ObsID individually and combined after re-projection to flat-field the resulting mosaic.

For the instrument maps, the spectral weighting was determined by fitting a single temperature, MEKAL thermal model \citep{Mewe85,Liedahl95} plus foreground Galactic absorption to the total integrated spectrum from the central 90 arcsec region around MACS0025. The Galactic absorption was modelled as neutral gas with solar abundances and a column density fixed at N$_H = 2.38 \times 10^{20}$ cm$^{-2}$ as determined by the LAB Survey of Galactic H I \citep{Kalberla05}. This model gives a reasonable fit to the data with a reduced $\chi^2 = 1.04$ and best-fit temperature and metallicity of $kT = 8.5$ keV and $Z = 0.23$, respectively.

\begin{figure*}
\centering
\includegraphics[width=\textwidth]{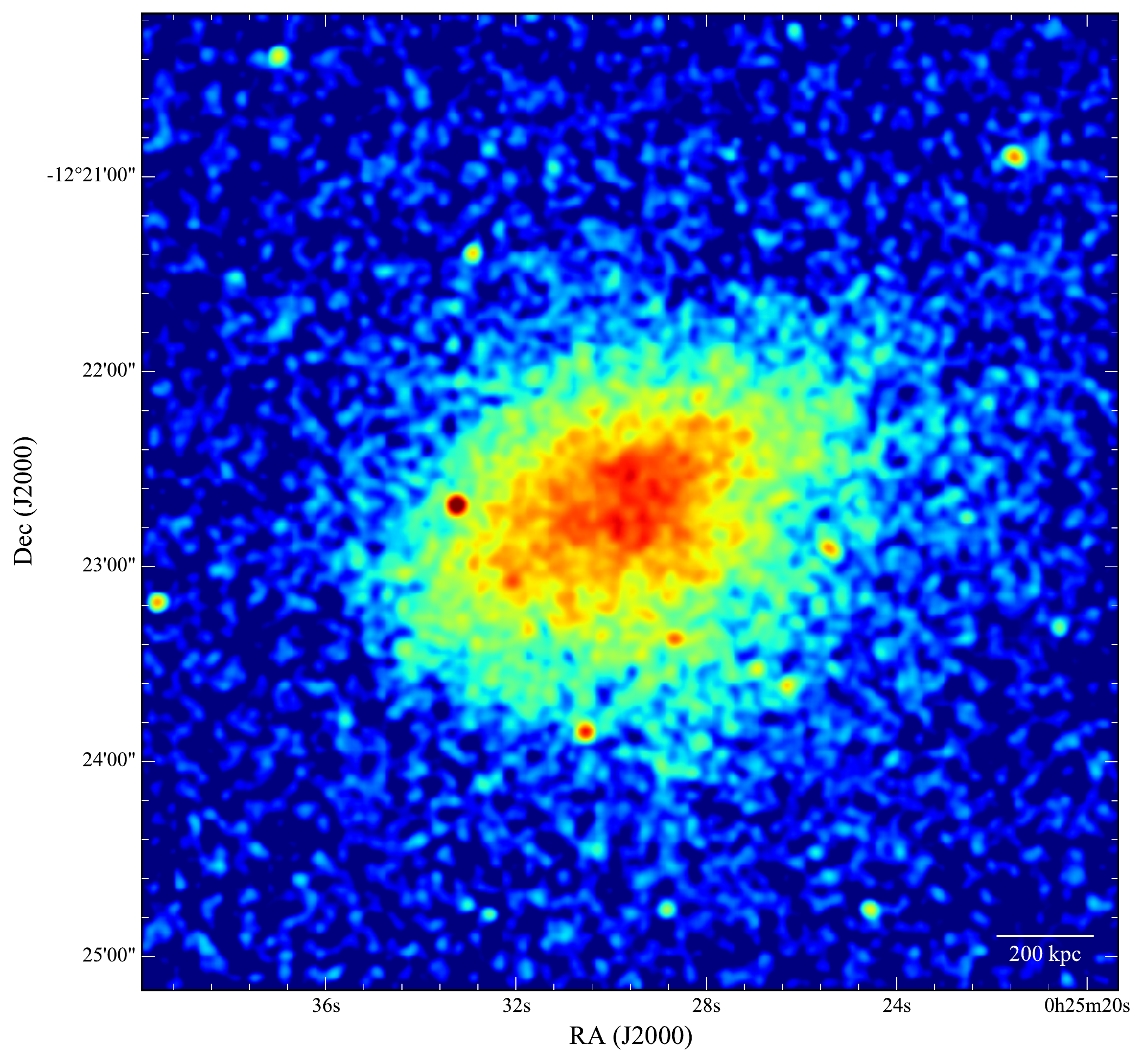}
\caption{\small Exposure-corrected, background-subtracted 0.5-7.0 keV surface brightness mosaic for the central region around MACS0025 based on 156 ksec of existing {\it Chandra} archival data. The field measures $5 \times 5$ arcmin$^2$ and has been smoothed with a $\sigma \sim 3\arcsec$ Gaussian.}
\label{fig:fov}
\end{figure*}

Individual background event files were created for each dataset from the standard ACIS blank-sky event files following the procedure described in \cite{Vikhlinin05} and combined after re-projection to form a background mosaic. The resulting background-subtracted, exposure-corrected mosaic for the energy range 0.5-7.0 keV is shown in Figure~\ref{fig:fov}.

\subsubsection{Spectral Analysis}
We have performed spatially-resolved spectral analyses of the X-ray emission in MACS0025 considering both 1-D and 2-D adaptive binning. For the 2-D analysis, the contour binning algorithm of \cite{Contour_Sanders_2006} was used, while for the 1-D analysis a series of radially symmetric annuli were defined, centred on the peak of the X-ray flux density ($00^{\rm{h}}25^{\rm{m}}29.380^{\rm{s}}~\SI{-12}{\degree}22^{\prime}37.06^{\prime\prime}$). We note that due to the combination of the adaptive binning process and the signal-to-noise requirement on the extracted spectra, the definition of the resulting extraction regions are relatively insensitive to the exact choice of X-ray centroid. In both cases, all spectral extraction regions were defined adaptively so as to contain a specified signal-to-noise ratio after background subtraction. Within each region, source and background spectra, effective area, and response matrix files were created for each OBSID individually and then combined into a single, summed set of files for analysis using the CIAO tool {\tt combine\_spectra}. All subsequent spectral analysis was done using the {\tt Sherpa} \citep{Sherpa01} fitting package in CIAO.

\section{Results}
\begin{figure}
	\centering
	\begin{subfigure}[t]{0.49\textwidth}
		\includegraphics[width=\textwidth]{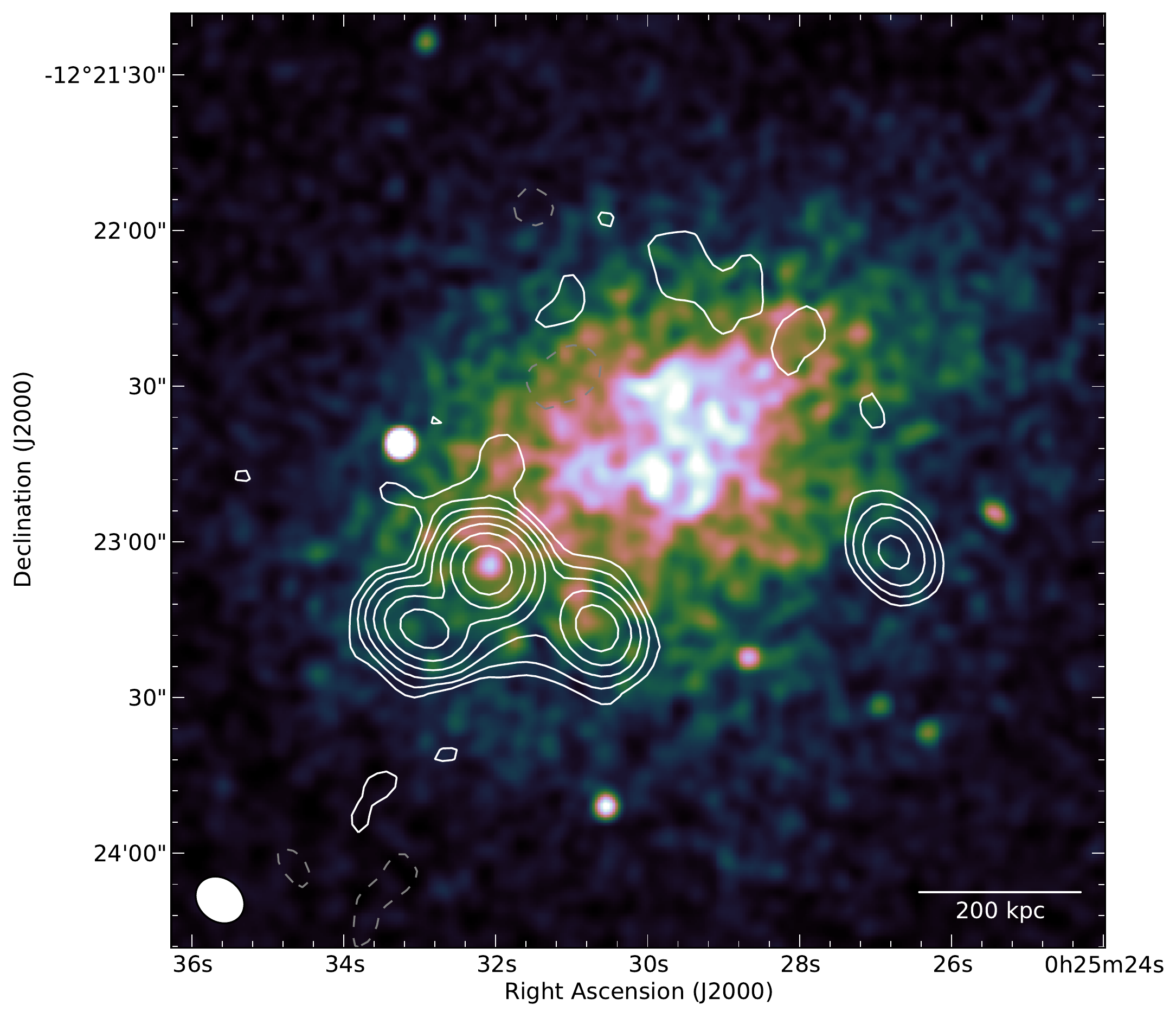}
	\end{subfigure}
	\begin{subfigure}[b]{0.49\textwidth}
		\includegraphics[width=\textwidth]{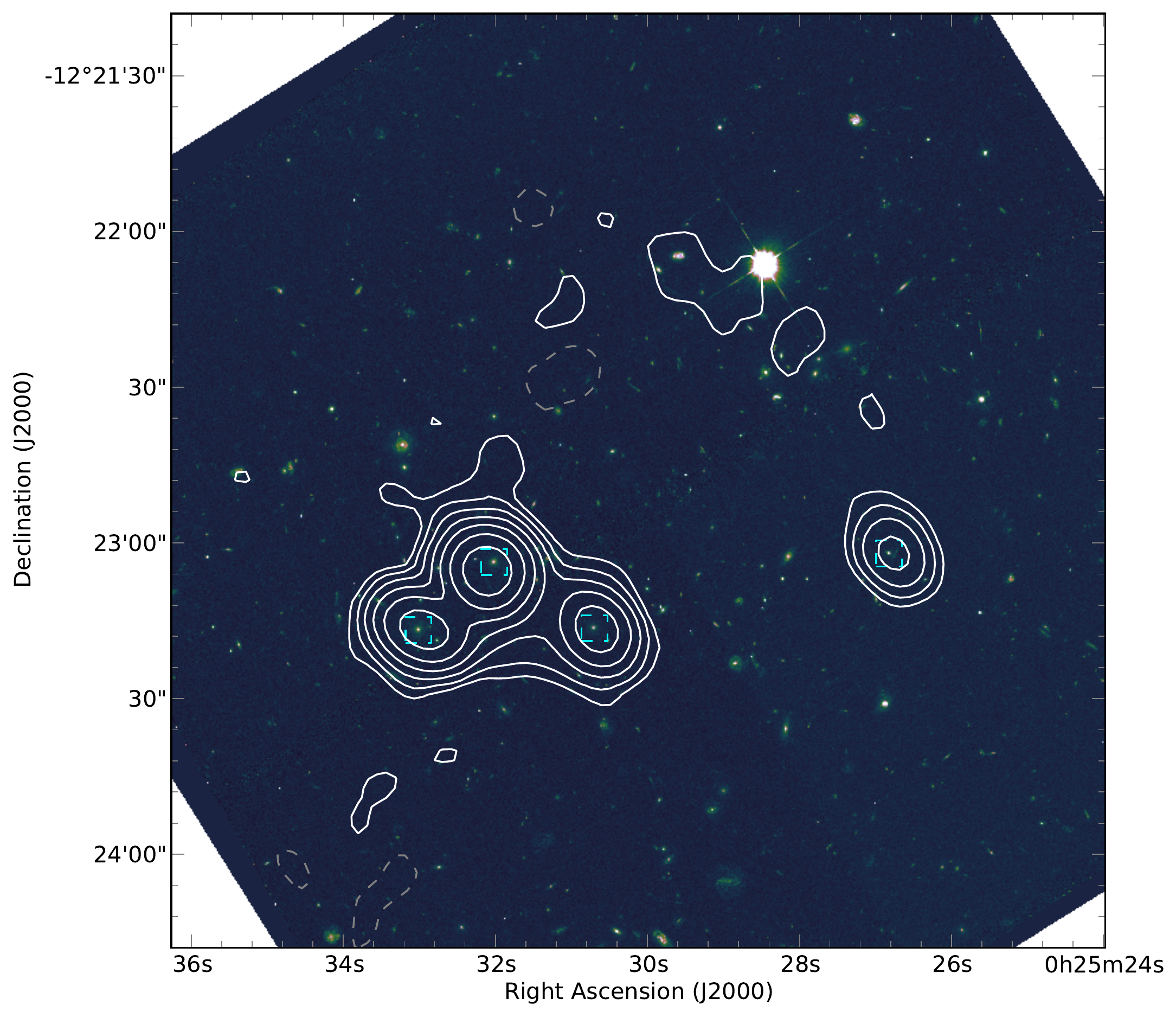}
	\end{subfigure}
\caption{Composite images of MACS0025. \emph{Top panel:} Colour image is X-ray surface brightness as per Figure \ref{fig:fov}. \emph{Bottom panel:} Colour image is \emph{HST} image of the cluster centre (proposal 10703, P.I. Ebeling H.). Cyan boxes identify optical hosts for the compact radio sources. In both panels, contours are 325 MHz GMRT data, imaged with an \texttt{AIPS} \texttt{ROBUST} $-1.0$, at levels [1, 2, 4, 8, 16, 32, 64]$\times5\sigma_{\rm{FR}}$ where $\sigma_{\rm{FR}} = 81 \,\, \mu$Jy beam$^{-1}$. The $-5\sigma$ contour is presented in gray. The synthesised beam is $10.1\times 8.3$ arcsec at PA $\SI{51.1}{\degree}$, indicated by the unfilled ellipse in the lower-left corner. }
\label{fig:macs_fullres}
\end{figure}

We present multi-wavelength images of MACS0025 in Figure \ref{fig:macs_fullres}. Contours correspond to the full-resolution (\texttt{AIPS} \texttt{ROBUST} $-1.0$) radio data at 325 MHz from the GMRT. In the top panel, the colour image is the X-ray surface brightness from archival \emph{Chandra} data as per Figure \ref{fig:fov} (details of which are shown in Table \ref{tab:obsids}) displayed using the `cubehelix' colour scheme \citep{2011BASI...39..289G}. The bottom panel presents the optical image of MACS0025 from \emph{HST} data (proposal 10703, P.I. Ebeling H.) reduced using the default \emph{HST} pipeline. The synthesised beam of the full-resolution GMRT image is $10.1\times 8.3$ arcsec; the measured off-source noise is $81 \, \mu$Jy beam$^{-1}$.

\subsection{Radio-Optical Cross Identification}
\begin{figure}
	\centering
	\begin{subfigure}[t]{0.45\textwidth}
	\includegraphics[width=\textwidth]{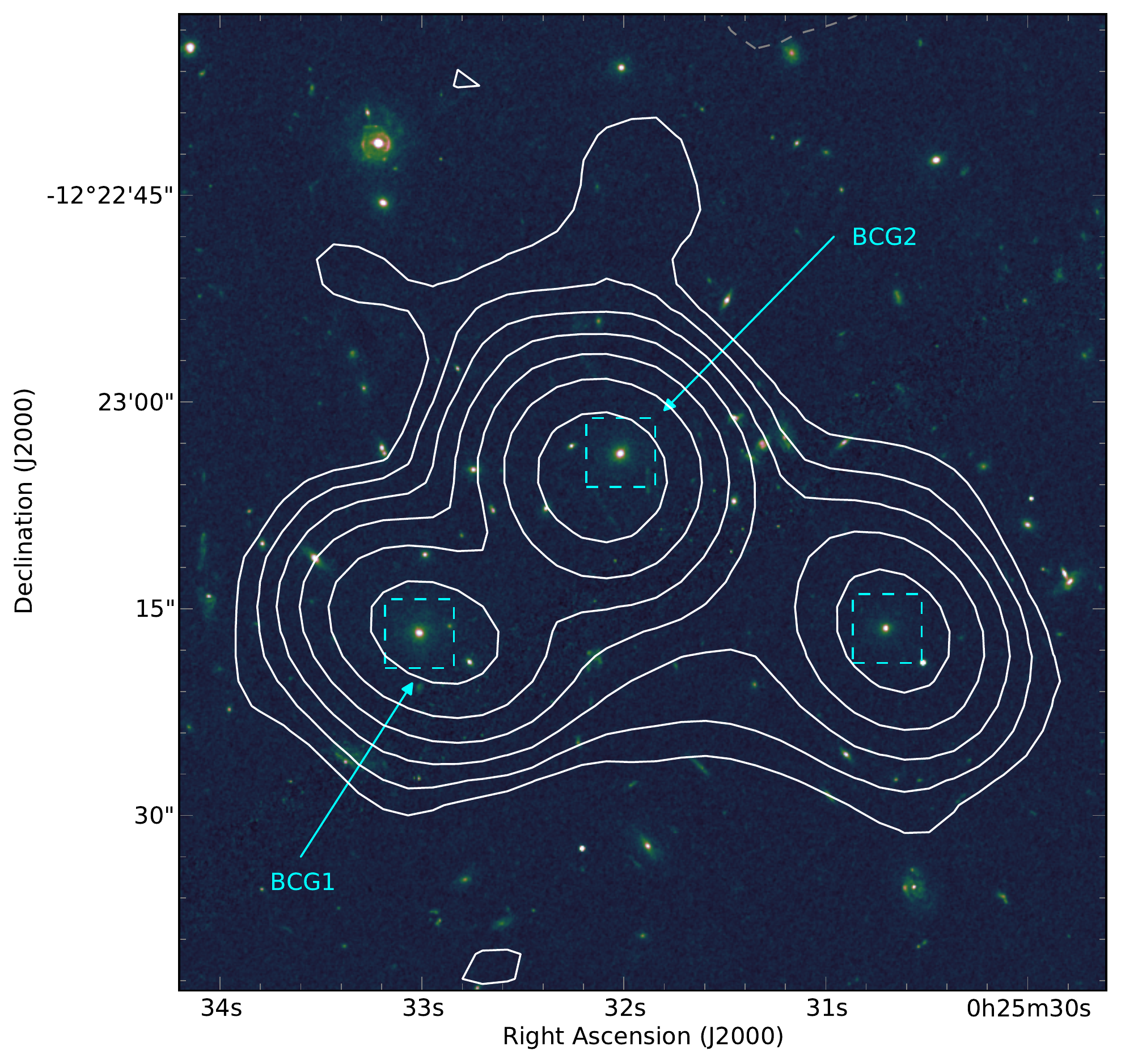}
	\caption{ \label{fig:triplet} }
	\end{subfigure}
	\begin{subfigure}[t]{0.45\textwidth}
	\includegraphics[width=\textwidth]{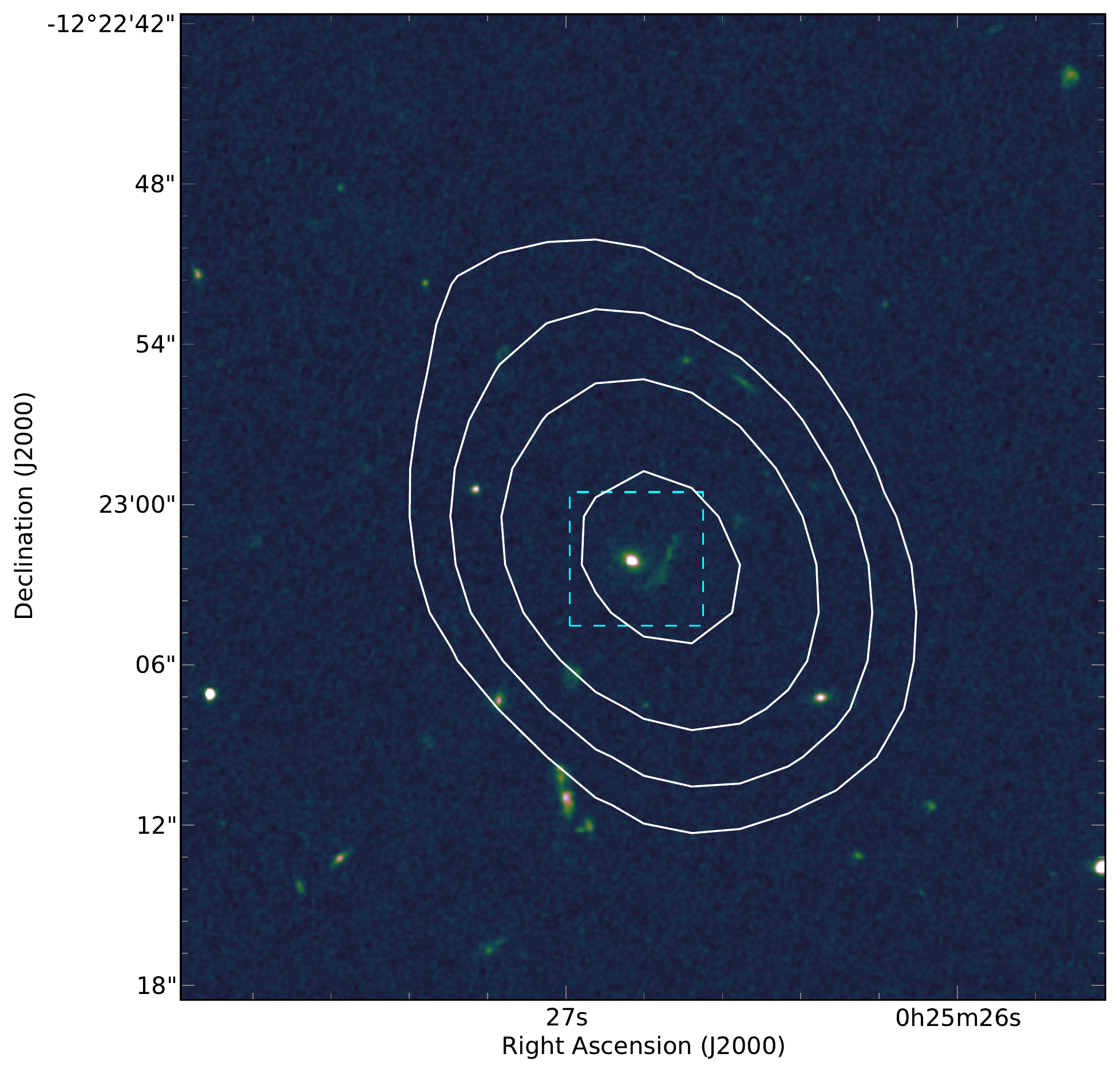}
	\caption{ \label{fig:singlet} }
	\end{subfigure}
\caption{Postage stamps of radio/optical associations for compact radio sources in MACS0025. Panel (a) presents the triplet of compact radio sources in the SE sub-cluster of MACS0025, panel (b) presents the single compact radio source in the NW sub-cluster. Contours are full-resolution GMRT data as per Figure \ref{fig:macs_fullres}. The colorscale is \emph{HST} data as per Figure \ref{fig:macs_fullres}. All selected optical hosts (cyan boxes) are cluster member galaxies, identified as absorption-line ellipticals by \protect\cite{ma2010}.}
\label{fig:optical_postage}
\end{figure}

From Figure \ref{fig:macs_fullres}, it can be seen that we recover emission from multiple compact sources in the region of MACS0025, and that there is little indication of any diffuse emission at this resolution. The triplet of compact radio sources South East of the cluster centre all appear to be associated with cluster-member galaxies, as identified by \cite{bradac2008} and \cite{ma2010}. Figure \ref{fig:optical_postage} presents postage stamp images of the compact radio sources in MACS0025. 

\begin{figure*}
\centering
\includegraphics[width=0.9\textwidth]{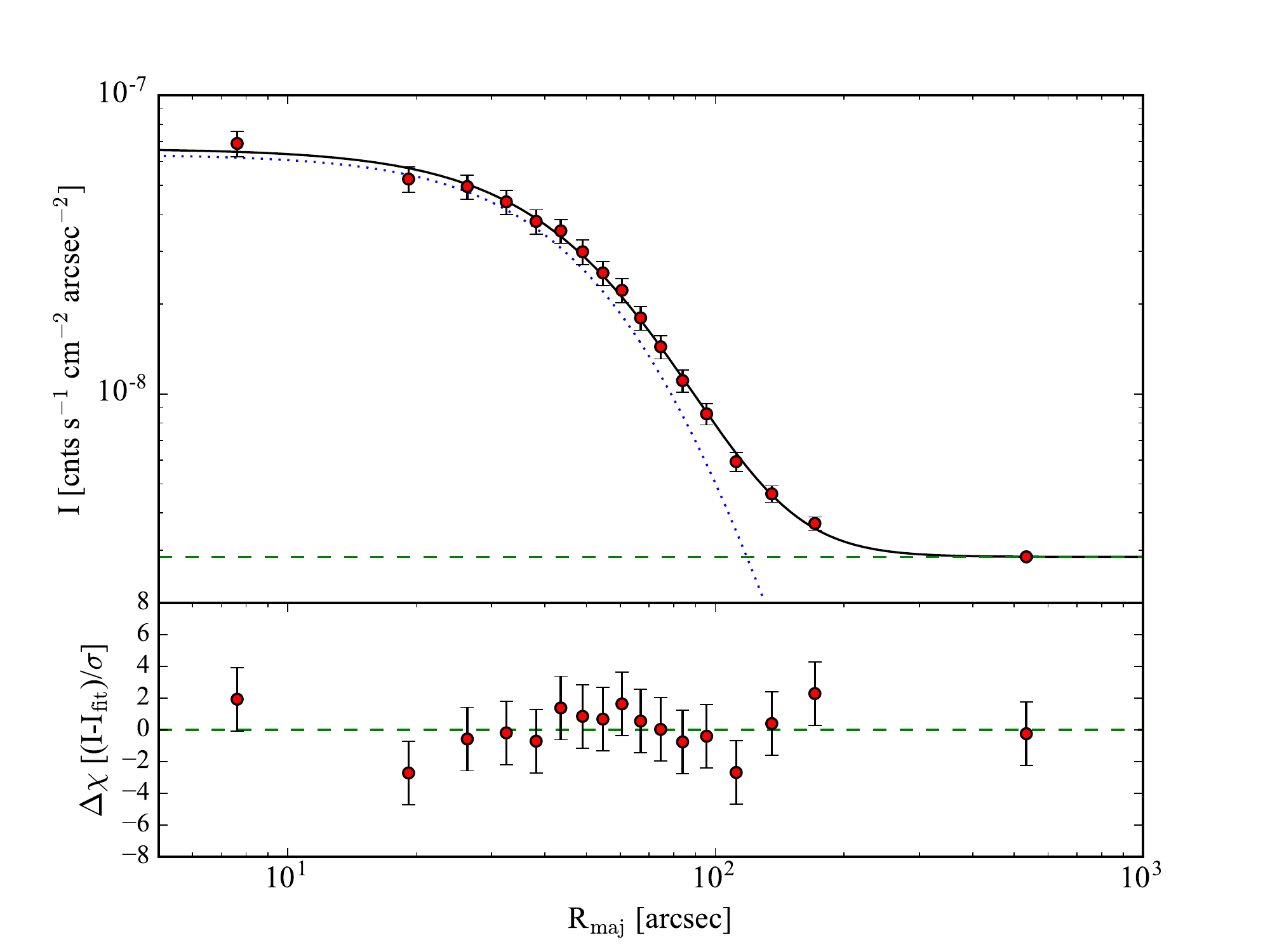}
\caption{\small
Extracted surface brightness profile for MACS0025. In the upper panel, the solid line shows the best-fit model to the overall surface brightness. The background level is denoted by the horizontal, dashed line while the dotted curve indicates the underlying, background-subtracted model. The lower panel indicates the difference between data and model relative to the error at each radius. For both panels, the radius, $R_{maj}$, corresponds to the radial distance along the semi-major axis of the assumed elliptical distribution. The model is discussed in the text.}
\label{fig:sb}
\end{figure*}

The host galaxies of two of the compact radio sources to the SE of MACS0025 are the two BCGs in this sub-cluster (as identified by \citealt{bradac2008}). These BCGs are identified in Figure \ref{fig:triplet}. Additionally, from Figure \ref{fig:macs_fullres} the central radio source of the triplet appears to be associated with a faint X-ray point source, which may suggest a powerful AGN. From the spectroscopic analysis of \cite{ma2010} the hosts of these compact radio sources are all absorption-line elliptical galaxies.


\subsection{X-ray Surface Brightness Profile}\label{sec:sbprofile}
The X-ray surface brightness distribution in MACS0025 shows a clear elliptical structure with the major axis of the ellipse oriented along what is presumed to be the merger axis in the system (see Figure~\ref{fig:fov}). A radial surface brightness profile was computed by defining a set of elliptical annuli centered on the peak of the X-ray emission. As discussed above, these annuli have been defined adaptively to include a minimum number of counts after background subtraction, thereby assuring a uniform signal-to-noise raio (SNR) in each bin. A fixed axial ratio of $b/a = 2$ was assumed, where $b$ and $a$ are taken to be the major and minor axes of the elliptical distribution, respectively. The position angle (PA) of the X-ray emission was taken to be $\SI{120}{\degree}$ as measured counter-clockwise from north.

The resulting elliptical surface brightness profile is shown in Figure~\ref{fig:sb}. An isothermal $\beta$ profile \citep{Cavaliere76, Sarazin77} of the form:
\begin{equation}
I(r) = I_{0} \left( 1 + {r^2 \over r_{c}^2} \right)^{-3\beta + \frac{1}{2}} + I_{B}
\label{eqn:beta}
\end{equation}
provides a reasonable representation of the azimuthally averaged surface brightness profile where $I_{0}$, $r_c$, and $\beta$ are the peak central sufrace brightness, core radius, and exponential falloff at large radii, respectively. $I_B$ is a constant representing the contribution of the background.
Given this model, we find best-fit values of $I_{0} = 6.42 \pm 0.14 \times 10^{-8}$ cnts s$^{-1}$ cm$^{-2}$, $r_{c} = 72\arcsec \pm 3 \arcsec$, and $\beta = 0.94 \pm 0.04$ for these parameters. The corresponding best-fit value for the background surface brightness is $I_{B} = 2.85 \pm 0.01 \times 10^{-9}$ cnts s$^{-1}$ cm$^{-2}$.

\subsection{Mass Profiles}
\label{sec:mass}

As we discuss in Section 4.1.2, \cite{gasperin2014} find evidence for a correlation between the observed 1.4 GHz radio power for large-scale radio relics observed in clusters and the mass of the host cluster as inferred from SZ measurements with Planck \citep{planck_clusters}. For an assumed spectral index, we can in principle use the GMRT observations presented here to assess whether the observed power of the radio relics in the MACS0025 cluster is consistent with this correlation. Unlike the sample presented in \cite{gasperin2014}, however, the MACS0025 cluster was not detected by Planck, so no SZ-derived mass estimates are available. Instead we have here utilized the observed X-ray properties, under the assumption of hydrostatic equilibrium, to derive a mass for the cluster.

\begin{figure*}
\centering
\includegraphics[width=\textwidth]{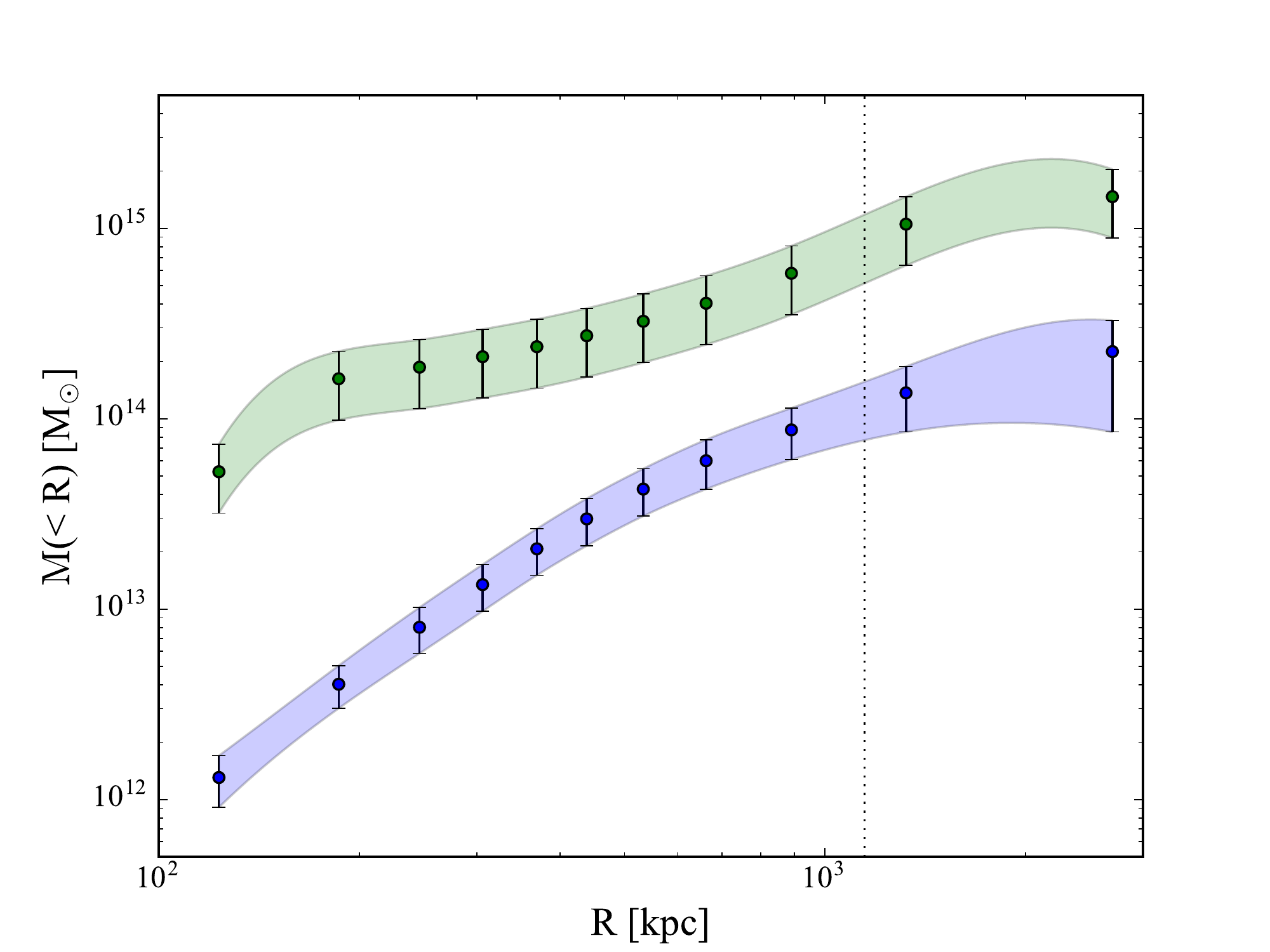}
\caption{\small Integrated mass profiles, $M(< r)$, for MACS~J0025.4$-$1222 assuming hydrostatic equilibrium and a prolate, ellipsoidal geometry. The upper curve represents the total gravitating mass within radius, $r$, and the lower curve give the total gas mass. The vertical dotted line indicates the $R_{500}$ radius as discussed in the text.}
\label{fig:mass}
\end{figure*}

Given radial profiles for the gas properties in MACS0025, we can derive both the gas mass and total gravitating mass in the MACS0025 system. Assuming hydrostatic equilibrium and spherical symmetry, the total cluster mass, $M$, contained with a radius $r$ is given by
\begin{equation}
	M(<r) = - \frac{k \,T(r) \,r^2}{\mu m_p \,G} ~\biggl( \frac{1}{n_e} \frac{dn_e}{dr} ~+ ~\frac{1}{T} \frac{dT}{dr} \biggr)
\label{eqn:totmass}
\end{equation}
where $T(r)$ and $n_{e}(r)$ are the gas temperature and electron number density at radius $r$, respectively \citep{Fabricant80}.
Similarly, the gas mass can be calculated directly using
\begin{equation}
	M_{gas}(<r) = \mu m_p ~\int n_e(r) ~dV
\label{eqn:gasmass}
\end{equation}
where $dV$ is the volume of the shell containing gas at density $n_{e}(r)$, and $\mu m_p$ is the mean mass per particle $1.0 \times 10^{-24}$\,g. 

In order to determine $kT$ and $n_e$ as a function of position in the cluster, we have extracted spectra for a series of elliptical annuli centered on the peak of the X-ray emission for the ellipsoidal geometry described in \S\ref{sec:sbprofile}. We have used a SNR criteria of 40 which results in $\sim$1600 net counts per bin after background subtraction in each region. Within each extraction region, a single temperature, MEKAL thermal model \citep{Mewe85,Liedahl95} was fit to the combined spectrum from that region. The absorption was set to the nominal Galactic value of N$_H = 2.38 \times 10^{20}$ cm$^{-2}$ and the abundance was fixed to a value of $Z = 0.23$ based on the fit to the mean spectrum as discussed in \S\ref{sec:imaging}.

The gas temperature $kT$ at each radius is determined from the spectral fit directly. The electron density, $n_e$, can be calculated directly from the normalization of the fitted MEKAL model using
\begin{equation}
N_{MEKAL} = 1.0 \times 10^{-14} \frac{\mu}{4 \pi D^2_A (1+z)^2} ~\int n^2_e ~dV
\label{eqn:mekal}
\end{equation}
where $D_A$ is the angular diameter distance at the cluster redshift $z$. For the assumed ellipsoidal geometry, the projected volume associated with each annuli can be estimated as $(2/3)$\,$S L$ where $S$ is the area of a given elliptical annulus with inner and outer semimajor axes of $r_1$ and $r_2$, respectively. The longest line-of-sight distance through the volume is denoted by $L$ and, assuming a prolate ellipsoidal geometry with major to minor axis ratio of 2, can be expressed as $L=2 (r^2_2 - r^1_1)^{1/2}$ \citep{Henry2004,Mahdavi2005}. This approximate method of deprojection assumes that the emission along the line-of-sight is dominated by material in the volume and that the contribution from material in front and behind the volume is negligible.

With these assumptions, the resulting total gas and gravitating mass as a function of radius along the cluster semimajor is shown in Figure~\ref{fig:mass}. Based on these profiles, we estimate a total gravitating mass $M_{500} = 8.44 \pm 3.16 \times 10^{14}~{\rm{M}}_{\odot}$, where $R_{500} = 1.15$~Mpc. This value agrees with the total mass estimate (lensing mass plus gas mass) $M_{\rm{tot}} = 5.65 \pm 2.2 \times 10^{14}~{\rm{M}}_{\odot}$ from \cite{bradac2008}. We note that \cite{bradac2008} measure the lensing masses within radii of 300 kpc of the sub-cluster BCGs, while the gas mass was measured within 500 kpc of the ICM X-ray peak.

\subsection{Low-Resolution Radio Imaging}
To emphasise any diffuse emission associated with the cluster merger in MACS0025, we have subtracted the clean-component model of the compact radio sources associated with the cluster, made using the high-resolution image (not presented in this paper). Subsequently, the data were re-imaged, applying a \emph{uv-}taper of $5\,{\rm{k}}\lambda$ to improve our sensitivity to large-scale radio emission. The cluster-member radio galaxies remain unresolved in both the full- and high-resolution images; consequently, flux density measurements are consistent between the full- and high-resolution images. As such, we are confident that any remaining flux density associated with these cluster member RG should be negligible.

\begin{figure}
\includegraphics[width=0.49\textwidth]{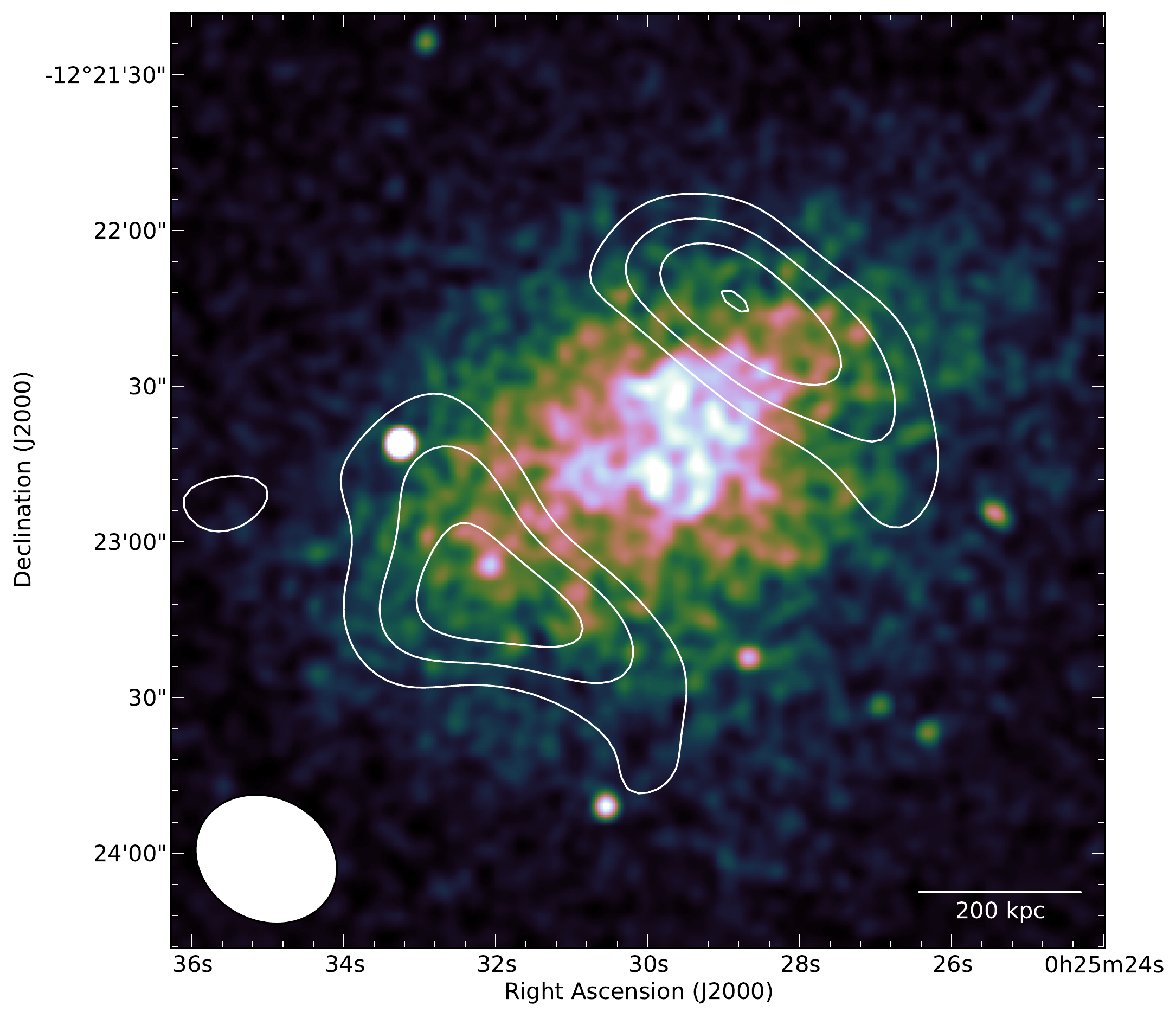}
\caption{Composite image of MACS0025. Colour plot is X-ray surface brightness as per Figure \ref{fig:fov}. Contours are source-subtracted low-resolution GMRT data at 325 MHz, at levels [5,7,9,11]$\times \sigma_{\rm{LR}}$, where $\sigma_{\rm{LR}} = 270 \,\, \mu$Jy beam$^{-1}$. The synthesised beam in the low-resolution image is $28.2 \times 23.7$ arcsec at PA $\SI{61.2}{\degree}$, indicated by the unfilled ellipse in the lower-left corner.}
\label{fig:macs_lowres}
\end{figure}

The low resolution, source subtracted image is presented in Figure \ref{fig:macs_lowres}, where the synthesised beam size is $28.2\times 23.7$ arcsec; the image rms is $270 \, \mu$Jy beam$^{-1}$. As can be seen in Figure \ref{fig:macs_lowres}, we recover significant radio emission associated with the ICM on large angular scales. This represents the first detection of significant diffuse radio emission from this cluster. 

Two sources of diffuse emission are present in Figure \ref{fig:macs_lowres}, to the North-West (NW) and South-East (SE) of the cluster centre. The NW (SE) source has largest angular size of approximately 100 (90) arcsec; at the redshift of MACS0025, this corresponds to a largest linear size (LLS) of 640 (577) kpc. This is consistent with the typical size of many radio relics (e.g. \citealt{bonafede2012,gasperin2014}) and haloes (e.g. \citealt{venturi2007,2008A&A...484..327V,2013A&A...557A..99K,kale2015}). 

These sources both exhibit an approximately arc-like morphology, are located toward the cluster periphery, and appear perpendicular to the proposed merger axis. As such, we identify this as a candidate double-relic system, and refer henceforth to these sources as the NW and SE relic. We use the fitting routine \texttt{fitflux} \citep{2007BASI...35...77G} to measure the flux density of the diffuse radio sources in Figure \ref{fig:macs_lowres}. We recover an integrated flux density of $6.54\pm0.72$ $(8.88\pm1.01)$ mJy for the NW (SE) relic\footnote{The uncertainty here comes from the sum of five per cent of the integrated flux density plus the standard deviation of seven fits, plus the off-source image noise, added in quadrature.}.


\section{Analysis}
Given that this is the first detection of large-scale, diffuse radio emission from MACS0025, we cannot compare directly with previous work. As of 2016 January, approximately 47 clusters are known to host radio relics, of which around 16 are known to host \emph{a pair} of radio relics. In this section, we will examine how this new candidate double-relic system fits in with the known population. We include relics from both larger samples and surveys (e.g. \citealt{feretti2012,gasperin2014,kale2015}) as well as individual cluster studies (e.g. \citealt{2012A&A...546A.124V,2015MNRAS.453.3483D,shimwell2015,tmc16}).

\subsection{Proposed Scenario: Two Radio Relics}
From both observations of clusters as well as simulations, it is well established that major merger events with an approximately even mass ratio should generate pairs of outward-propagating shocks, which should consequently generate a pair of radio relics (e.g. \citealt{vw2011}, although the merger they consider has a mass ratio around 2:1). Previous observations of MACS0025 have suggested that the masses of the merging clusters are approximately equal: $M \simeq 2.5\times10^{14} \, {\rm{M}}_{\odot}$ \citep{bradac2008}. 

The morphology of the radio emission in Figure \ref{fig:macs_lowres} appears to be consistent with this interpretation, with both diffuse sources exhibiting approximately arc-like morphology. Additionally, both relics are detected at the periphery of the X-ray emission from the ICM, also consistent with both previous detections of relics and the suggestion that the merger event is occurring close to the plane of the sky \citep{bradac2008,ma2010}.

These relics have not been detected at other frequencies, so we cannot directly measure a spectral index. In the NVSS, emission is detected associated with the triplet of compact radio sources, but no emission is recovered at the $3\sigma$ level from the location of the NW relic. Likewise, we found no positive flux below the $3\sigma$ level in the NVSS image using \texttt{fitflux}; however, we can derive an upper limit from the surface brightness.

At 325 MHz, the emission from the NW relic peaks at 2.98 mJy beam$^{-1}$, which corresponds to a surface brightness of around $15.7\,\mu$Jy arcsec$^{-2}$ given the beam size. Taking the $3\sigma$ limit from the NVSS (1.35 mJy beam$^{-1}$, or $\sim2.4\,\mu$Jy arcsec$^{-2}$) implies an upper limit of $\alpha = -1.3$ for the spectral index of the NW relic. This is consistent with the known population -- for example, the `September 2011 relic collection' of \cite{feretti2012} has a mean spectral index $\langle \alpha \rangle = -1.3$. 

We use this spectral index to scale the flux density of our relics to 1.4 GHz and derive the radio power. Given the redshift of MACS0025 ($z = 0.5857$; \citealt{bradac2008}) we find $P_{\rm{1.4~GHz}} = 1.29 \pm 0.14 \times10^{24}$ W Hz$^{-1}$ for the NW relic and $P_{\rm{1.4~GHz}} = 1.76 \pm 0.20 \times10^{24}$ W Hz$^{-1}$ for the SE relic, with our cosmology.

\subsubsection{Power Scaling: Radio Power vs. X-ray Luminosity}
In their review, \cite{feretti2012} categorise cluster radio relics as `elongated' and `roundish', with the forming bearing the morphological hallmarks typical of the well-known giant radio relics, and the latter exhibiting a more rounded, regular morphology. Like `elongated' relics, `roundish' relics also exhibit steep spectra, an apparent lack of an optical counterpart, as well as a high polarization fraction. \cite{feretti2012} include radio phoenixes under the umbrella of `roundish' relics. Some examples of roundish relics are those in A584b \citep{2006MNRAS.368..544F}, AS753 \citep{2003AJ....125.1095S} and A1664 \citep{govoni2001}. It is worth noting that no roundish relics have been detected beyond around $z\simeq0.2$, whereas MACS0025 is at redshift $z=0.5857$ \citep{bradac2008}. We do not include `roundish' relics in our analysis, as they may include a variety of different classes of source.

\begin{figure}
	\includegraphics[width=0.49\textwidth]{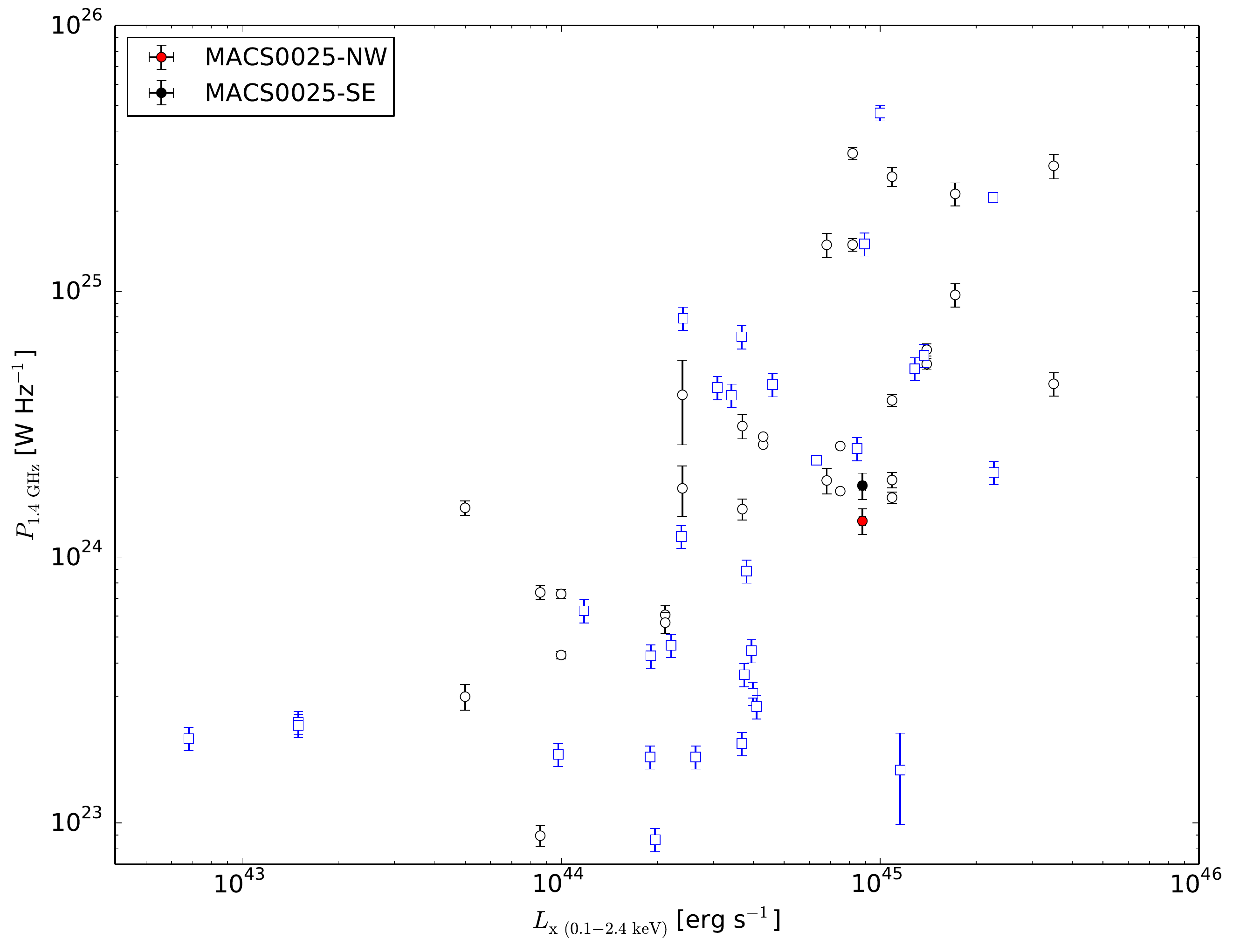}
	\caption{Power scaling relation in the $L_{\rm{x}}$--$P_{\rm{1.4 \, GHz}}$ plane for galaxy cluster radio relics. Open circles indicate clusters hosting \emph{a pair} of radio relics from the literature. Blue squares denote clusters hosting a single radio relic. Filled black (red) circles indicate the radio power and X-ray luminosity for the SE (NW) diffuse radio sources hosted by MACS0025.}
	\label{fig:relic_power}
\end{figure}

The X-ray luminosity of MACS0025 is $L_{\rm{x}} = 8.8\pm0.2\times10^{44}$ erg s$^{-1}$ \citep{2007ApJ...661L..33E}. In Figure \ref{fig:relic_power} we present the $L_{\rm{x}}/P_{\rm{1.4 \, GHz}}$ plane for galaxy clusters hosting radio relics in the literature. Clusters hosting single radio relics are marked in blue, those hosting a pair are marked in black. From Figure \ref{fig:relic_power}, the relics hosted by MACS0025 are consistent with the known population. 

\begin{figure}
	\begin{subfigure}[t]{0.478\textwidth}
	\includegraphics[width=\textwidth]{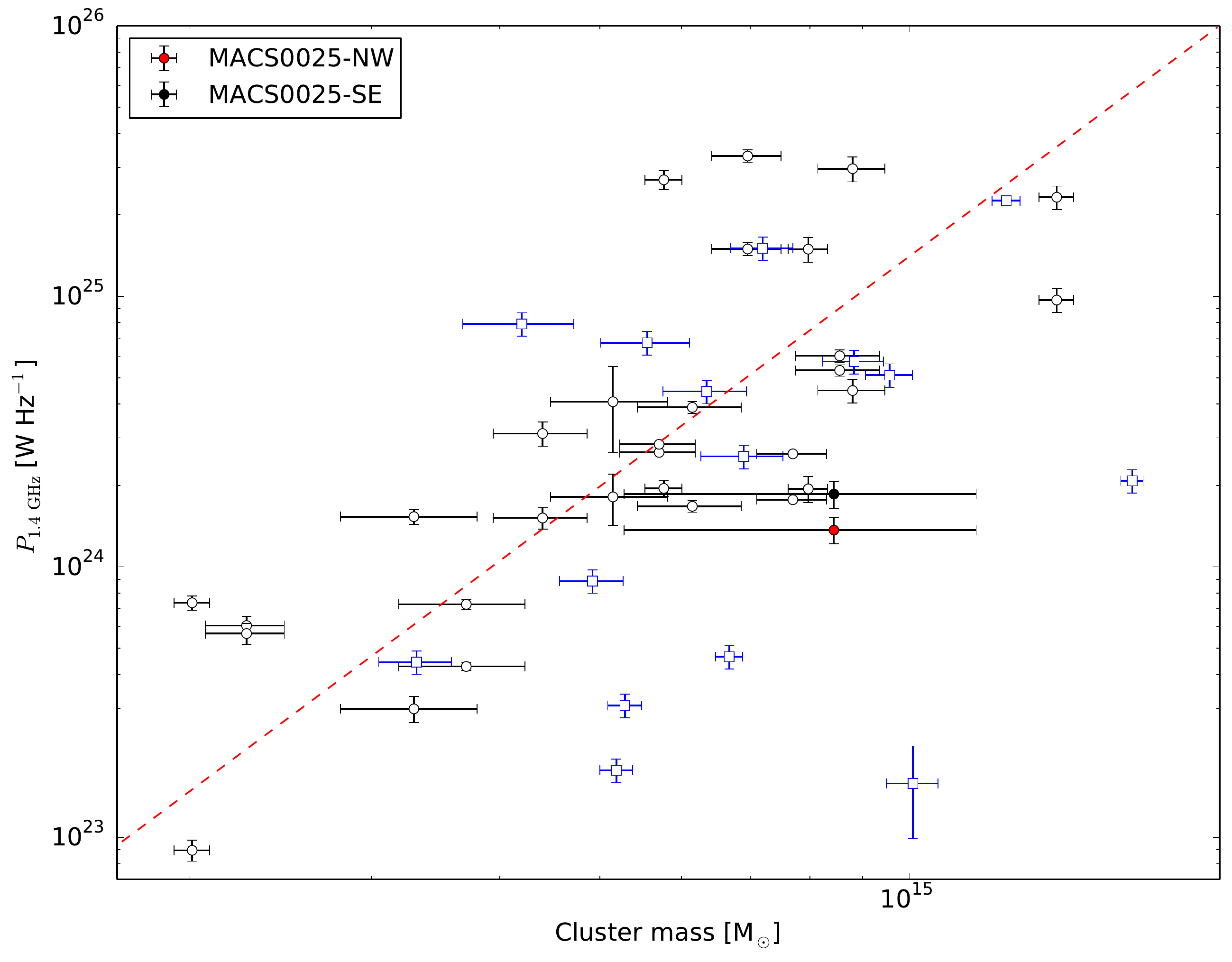}
	\caption{ \label{fig:radio_mass} }
	\end{subfigure}
	\begin{subfigure}[b]{0.49\textwidth}
	\includegraphics[width=\textwidth]{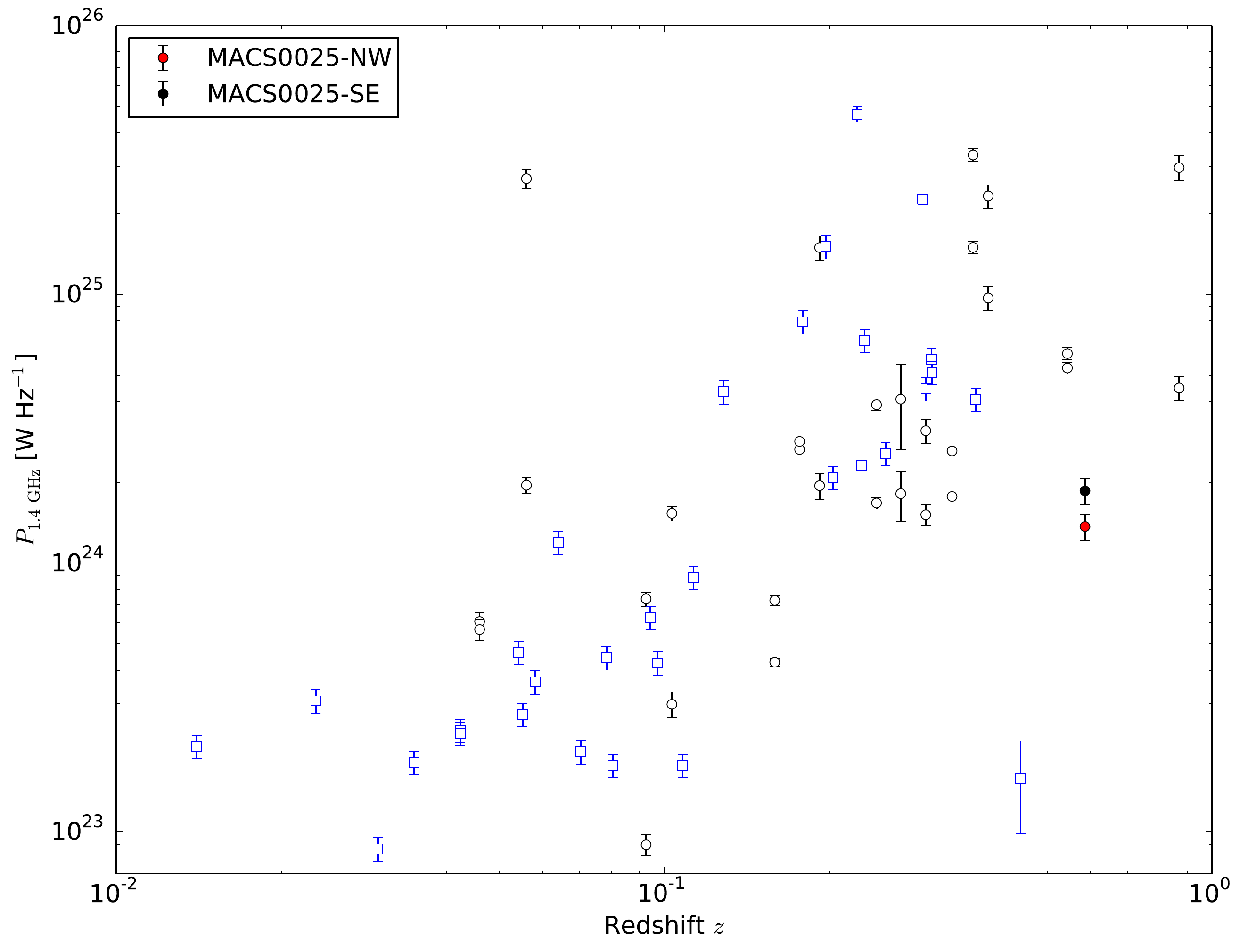}
	\caption{ \label{fig:radio_z} }
	\end{subfigure}
\caption{The dependence of cluster relic radio power on (a) cluster mass and (b) redshift. Open circles indicate clusters hosting a pair of radio relics from the literature. Blue squares denote clusters hosting a single radio relic. Filled black (red) circles indicate the position of the SE (NW) diffuse radio sources hosted by MACS0025. The dashed red line in panel (a) denotes the $M_{500}/P_{1.4~\rm{GHz}}$ relation from \protect\cite{gasperin2014}. Mass estimates are via the SZ-effect from the \emph{Planck} cluster catalogue \protect\citep{planck_clusters}, or via a scaling relation based on the X-ray luminosity ($L_{{\rm{x}} \,[0.1-2.4\,\rm{keV}]}$; \protect\citealt{gasperin2014}) for all clusters except MACS0025, where the mass is derived in \S\ref{sec:mass}.}
\label{fig:relic_relations}
\end{figure}

\subsubsection{Power Scaling: Radio Power vs. Cluster Mass}
\cite{gasperin2014} investigate the possibility of a relation between cluster mass\footnote{Derived via the Sunyaev-Zel'dovich effect from the \emph{Planck} cluster catalogue \citep{planck_clusters}} $(M_{500})$ and radio power at 1.4 GHz, finding find a strong correlation. This relationship takes the form $P_{\rm{1.4 \, GHz}} \propto M_{500}^{2.83}$ for clusters hosting double-relics, and still holds (albeit with greater scatter) when clusters hosting a single relic are included. 

In Figure \ref{fig:radio_mass} we reproduce the data from \cite{gasperin2014} with the addition of a MACS0025 as well as some other clusters hosting relics that have been detected following \cite{gasperin2014}. MACS0025 was not detected by \emph{Planck}; instead we use the total gravitating mass derived from the X-ray data, $M_{500} = 8.44 \pm 3.16 \times10^{14}~{\rm{M}}_{\odot}$. 

We note that \cite{2015ApJ...806...18C} find a mass estimate $M_{2500} = 2.38^{+0.66}_{-0.50}\times10^{14} \, {\rm{M}_{\odot}}$ for MACS0025 from the Bolocam X-ray SZ (BOXSZ; \citealt{2013ApJ...768..177S}) sample. However, this is measured over a smaller aperture than the mass measurements used by \cite{gasperin2014}. From Figure \ref{fig:relic_relations}, the relics hosted by MACS0025 are consistent with the scatter in the $M_{500} / P_{\rm{1.4\,GHz}}$ plane for known relics, although they lie far from the relation derived by \cite{gasperin2014}.

\subsubsection{Power Scaling: Radio Power vs. Redshift}
In Figure \ref{fig:radio_z} we present the radio power at 1.4 GHz as a function of redshift for the known population of radio relics. The relics in MACS0025 appear to be consistent with the general trend between radio power and redshift. To-date, only two other clusters beyond redshift $z = 0.5$ are known to host Mpc-scale radio emission: MACS~J1149.5$+$2223 at $z = 0.544$ \citep{bonafede2012} and ACT-CLJ0102$-$4915 at $z = 0.855$ (`El Gordo'; \citealt{lindner2014}). For high-redshift clusters, MACS0025 appears to be slightly low in the $P_{\rm{1.4\,GHz}}/z$ plane, although this is likely due to selection effects and/or small number statistics. 

In general, the apparent trend exhibited in the $P_{\rm{1.4\,GHz}}/z$ plane is severely impacted by selection effects -- the surface brightness sensitivity of historic cluster studies has not been sufficient to detect faint relics at high redshift. The fact that the relics in MACS0025 lie slightly below the apparent correlation may suggest we are probing deeper than the majority of previous high-redshift cluster studies.

Overall, from Figures \ref{fig:relic_power} and \ref{fig:relic_relations} it is appears that the relics hosted by MACS0025 are consistent with established trends in power-scaling relations for the known population of relics. Note that the relic in MACS~J2243.3$-$0935 \citep{tmc16} appears slightly low in all power scaling planes; this may be due to the suggested nature as an \emph{infall} relic rather than a relic generated by a merger shock, or a result of improved data reduction techniques.

\begin{figure*}
\centering
\includegraphics[width=0.8\textwidth]{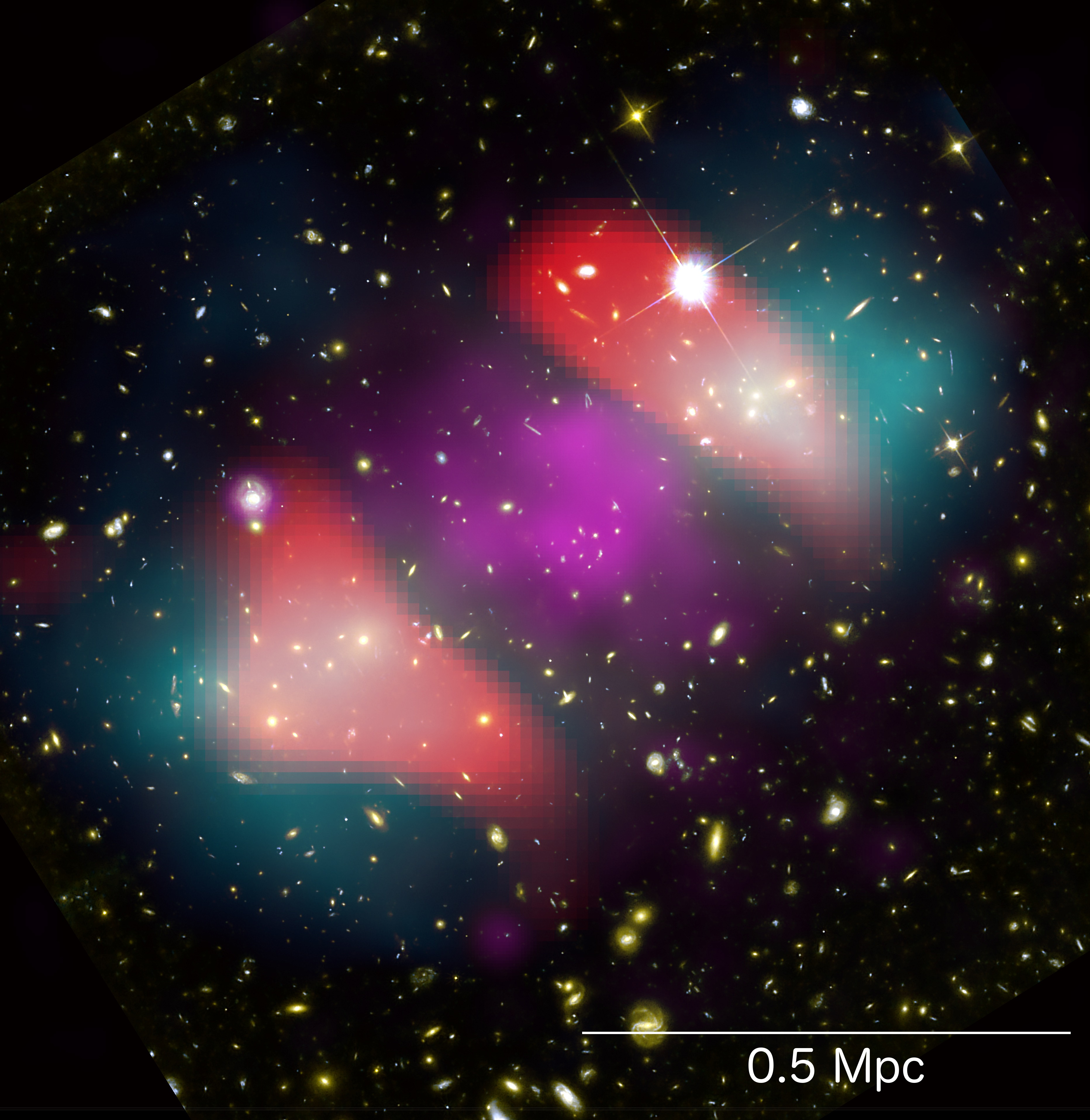}
\caption{False-colour image of MACS0025. Background image is an RGB optical image from \emph{HST}. Cyan denotes the \emph{total} mass distribution traced by lensing (dominated by the dark matter). Purple denotes the \emph{baryonic} mass distribution (dominated by the ICM plasma). Red denotes the diffuse radio emission at 325 MHz. Image is in the J2000 coordinate system, north is up and east is to the left. X-ray and radio data are from this work, optical and lensing data are from \protect\cite{bradac2008}. The scale bar denotes the angular extent corresponding to a physical distance of 0.5 Mpc at the redshift of MACS0025.}
\label{fig:macs_multi}
\end{figure*}

\section{Discussion}
\subsection{Bulk components of the ICM}
In Figure \ref{fig:macs_multi}, we present a RGB optical image of MACS0025 taken with \emph{HST} (from \citealt{bradac2008}) overlain with false-colour multi-wavelength data. Figure \ref{fig:macs_multi} shows the relative distributions of the total mass (as traced by gravitational lensing, which is dominated by dark matter, from \citealt{bradac2008}) and the baryonic mass (as traced by the X-ray surface brightness, based on the new analysis presented in this work). The lensing mass is shown in cyan, the baryonic mass in purple. Also overlain is the radio emission recovered by the GMRT at 325 MHz, in red. 

MACS0025 was the second galaxy cluster where significant separation was detected between the total mass and baryonic mass distributions. MACS0025 joins the Bullet cluster as part of a growing population of clusters that both host diffuse radio emission and exhibit clear separation between baryonic matter and dark matter components. \cite{shan2010} derive the offset between the lensing and X-ray centroids for a sample of 38 galaxy clusters; in their sample are a small number of other galaxy clusters with significant offset ($\gtrsim150$ kpc) that are also known to host both a radio relic and halo: A2163 \citep{feretti2001} and A2744 \citep{govoni2001,orru2007}. 

From \cite{bradac2008} the offsets between the lensing peaks and the baryonic matter peak (defined as the X-ray centroid) are 0.5 and 0.82 arcmin; at the redshift of MACS0025, these correspond to a physical distance of 192 and 316 kpc, respectively, for the NW (SE) sub-cluster. These offsets are comparable to the offsets for the Bullet cluster (201 kpc), A2163 (141 kpc) and A2744 (238 kpc) from \cite{shan2010}\footnote{Note that the measured \emph{angular} offsets from \protect\cite{shan2010} were converted to a \emph{physical} size using our cosmology.}.

\subsection{Large-Scale Temperature Structure}
\label{sec:tmap}
In order to search for signatures of the merger in the underlying temperature structure in MACS0025, we have performed a two-dimensional spectral analysis. As mentioned previously, the spectral extraction regions were defined using the contour binning algorithm of \cite{Contour_Sanders_2006} based on the X-ray surface brightness image and such that each region had a fixed SNR. This technique trades off spectral accuracy against extraction region size and by extension the ability to resolve smaller-scale temperature structures. We have used a criteria of SNR$\sim$30 which results in $\sim$900 net counts per bin after background subtraction in each region. This choice represents a compromise between obtaining reasonable accuracy in the fitted temperature values while simultaneously preserving modest angular resolution.

In each of the resulting extraction regions, the standard set of source spectrum, background spectrum, effective area, and response matrix files were created for each OBSID contributing to that region. These individual files were then combined into a single, set of summed files for each region with the CIAO tool {\tt combine\_spectra} and used for all subsequent fitting. At each point in the resulting map, we have fit a single temperature, MEKAL thermal model \citep{Mewe85,Liedahl95} plus foreground Galactic absorption to the combined spectrum from that region. The Galactic absorption was fixed to a value of N$_H = 2.38 \times 10^{20}$ cm$^{-2}$ as discussed previously in \S\ref{sec:imaging}. 

\begin{figure*}
\centering
\includegraphics[width=\textwidth]{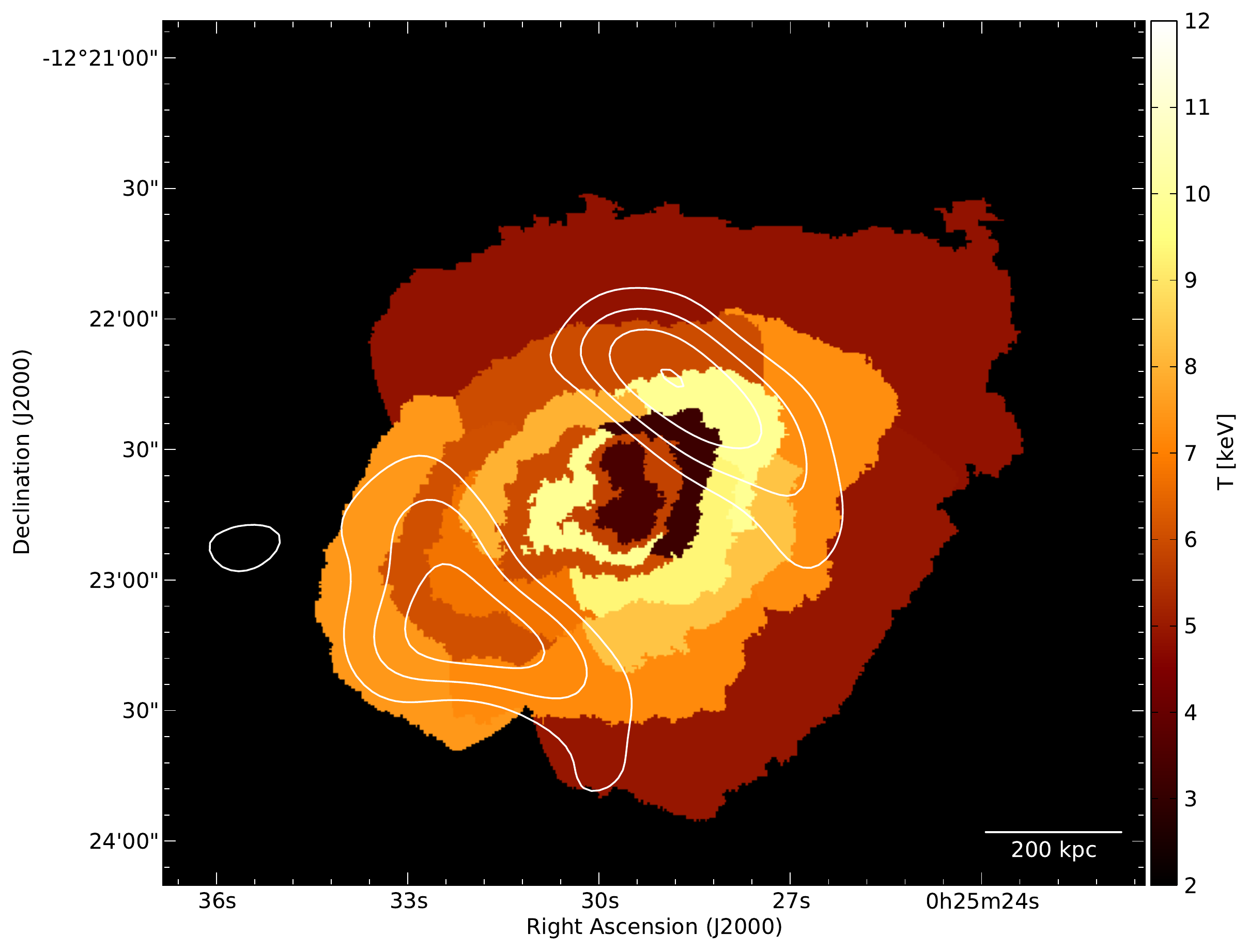}
\caption{X-ray temperature map of MACS0025, with low-resolution radio contours from the GMRT, as per Figure \ref{fig:macs_lowres}. }
\label{fig:macs_kt}
\end{figure*}

For the elemental abundance, we have derived temperature maps where the abundance was both fixed and allowed to vary spatially. The maps are qualitatively similar in both cases; however, the fitted abundances are not well-constrained in the spatially varying case and the errors on the derived temperature values are also higher. Given the higher temperature range ($\sim$6--10 keV) observed in MACS0025, the fits are relatively insensitive to the abundance. Consequently, to determine the best constraints on the temperature values, we have fixed the abundance to a value of $Z = 0.23$ based on the fit to the mean spectrum discussed above in \S\ref{sec:imaging}. The final temperature map for MACS0025 is shown in Figure~\ref{fig:macs_kt}.

The overall temperature structure shown in Figure~\ref{fig:macs_kt} seems generally consistent with the presence of a strong merger in MACS0025. We find regions of higher temperature roughly aligned with the merger axis. We have been cautious in interpreting the maps as the contour binning algorithm combined with the lower signal-to-noise of the data can tend to enhance apparent "shell-like" substructures in the temperature. With this caveat in mind, we note that the map shown in Figure~\ref{fig:macs_kt} does seem to exhibit preferentially hotter gas along the merger axis presumably due to shock heating associated with the merger. However, given the limitations of the data, and the fact that the typical error in Figure~\ref{fig:macs_kt} is of the order of 30 to 50 per cent, this correspondence must be seen as more suggestive than definitive.

\subsection{Shocks in the ICM?}
As can be seen in Figure \ref{fig:macs_kt}, we have a tentative detection of regions of enhanced temperature in the ICM of MACS0025. However, as discussed, the errors are sufficiently large that Mach number constraints from the X-ray data will not provide tight constraints on the viability of this being a merger shock. 

Additionally, the temperature structure exhibited in Figure \ref{fig:macs_kt} is reminiscent of a cold front, with the cooler (unshocked) region of gas downstream of the hotter (shocked) region. However, from earlier, there is no evidence of surface brightness  discontinuity that is typically associated with cold fronts (see \citealt{2007PhR...443....1M} for a thorough review of cold fronts in galaxy clusters). 

We can estimate the Mach number based on the radio spectral index, following the formalism of \cite{be1987}. For simple shocks,

\begin{equation} \label{eq:mach}
	\mathcal{M} = \sqrt{  \frac{ 2 \alpha_{\rm{inj}} + 3 }{ 2 \alpha_{\rm{inj}} - 1}  }
\end{equation}

Where $\alpha_{\rm{inj}}$ is the power-law spectral index of cosmic ray electrons (CRe) at the shock front. Earlier, we estimated the upper limit spectral index to be $\alpha < -1.3$; hence, Equation \ref{eq:mach} suggests a Mach number $\mathcal{M} < 1.87$. This is typical of the Mach numbers both predicted from simulations of cluster merger shocks and from observations at radio wavelengths, which suggest Mach numbers $1 \leq \mathcal{M} \leq 3$ (e.g. \citealt{feretti2012,bj2014}).

\subsection{Should MACS0025 Host a Radio Halo?}
Whilst radio haloes and relics are strongly associated with merging clusters, the exact conditions necessary to generate these objects are far from clear. Cluster mergers deposit vast quantities of energy into the ICM and cause significant turbulence, yet not all merging clusters host a RH. Likewise, the link between merger shocks and relics is generally supported by observational evidence, but many merging clusters with prominent ICM shocks do not host relics.

\cite{bj2014} estimate the timescale required to generate a RH following the turbulent acceleration framework, deriving $\tau_{\rm{acc}} \sim 100$ Myr. Likewise, \cite{bj2014} estimate the lifetime of radio-emitting electrons for the Coma cluster using viable magnetic field estimates, deriving $\tau_{\rm{CRe}} \sim {\rm{few}} \times100$ Myr. For comparison, \cite{brunetti2009} estimate the lifetime of RH based on a statistical sample of around 19 galaxy clusters, deriving $\tau_{\rm{RH}} \sim 1.3$ Gyr. Based on the optical properties of the cluster, \cite{ma2010} conclude that core passage in MACS0025 occurred $0.5-1$ Gyr ago, suggesting that sufficient time should have elapsed to generate a RH.

From examination of double-relic systems in the literature, it appears that time since core passage ($\tau_{\rm{CP}}$) and mass ratio cannot be used alone to suggest whether a cluster should host a RH. For example, the system CIZA J2242.8+5301 (hereafter CIZAJ2242) hosts a pair of relics but no RH (e.g. \citealt{2010Sci...330..347V}). Simulations suggest that this system involves a pair of merging clusters with a mass ratio approximately 2:1, with core passage occurring around 1 Gyr ago \citep{vw2011}. 

In contrast, `El Gordo' hosts a pair of relics and a halo \citep{lindner2014}. Weak lensing observations of El Gordo suggest that the sub-clusters also have a mass ratio approximately 2:1 \citep{jee2014}. A dynamical analysis of what is known about El Gordo is presented by \cite{ng2015}, who indicate two potential scenarios for the merger in El Gordo: either the system is observed shortly after first core passage ($\tau_{\rm{CP}}\sim0.46$ Gyr) or the clusters are again on approach, having turned around ($\tau_{\rm{CP}}\sim0.91$ Gyr). \cite{ng2015} suggest that the latter scenario is favoured if the time-averaged sub-cluster velocities exceed the shock velocity of the corresponding relic.

Both clusters also appear to have similar total masses: \cite{gasperin2014} find total masses $M_{\rm{tot}} = 7.97\pm0.35\times10^{14} \, (8.80\pm0.66\times10^{14}) \, {\rm{M}}_{\odot}$ for CIZAJ2242 (El Gordo). It should be noted that these mass measurements were derived through different means -- via the SZ effect for El Gordo (\citealt{planck_clusters}) and via the scaling relation between X-ray luminosity and total mass (\citealt{pratt2009}) for CIZAJ2242. The velocity dispersions of the sub-clusters involved in the merger events also appear to be consistent: from \cite{jee2014} the spectroscopically-measured velocity dispersions of the NW (SE) sub-cluster is $\sigma_{\rm{v}} = 1290\pm134 \, (1089\pm200)$ km s$^{-1}$ for El Gordo, compared to $\sigma_{\rm{v}} = 1160^{+100}_{-90} \, (1080^{+100}_{-700})$ km s$^{-1}$ \citep{dawson2015}.

There is no evidence of a RH in MACS0025 at the $2\sigma$ level or higher, although an upper limit to the radio power can be derived. The sensitivity of the GMRT image at 325 MHz (Figure \ref{fig:macs_lowres}) is $270\,\mu$Jy beam$^{-1}$. Given the restoring beam FWHM $\theta = 28.2\times23.7$ arcsec, the corresponding $2\sigma$  surface brightness limit ${\rm{SB}}_{\rm{lim}} = 1.03\, \mu$Jy arcsec$^{-2}$. Assuming a spherical geometry, a 1 Mpc RH would appear to be around 155 arcsec in diameter at the redshift of MACS0025, given the adopted cosmology. With a typical RH spectral index $\alpha = -1.3$, this suggests a limiting radio power $P_{\rm{1.4\,GHz,\,lim}} \simeq 4\times10^{24}$ W Hz$^{-1}$. This does not place a tight constraint on the power of any potential RH in MACS0025.

The apparent lack of a RH in MACS0025 is not unexpected. For example, the statistical study of RH occurrence rates in clusters performed by \cite{2015A&A...580A..97C} suggests that only $\sim20-30$ per cent of clusters with similar mass to MACS0025 host RH. Furthermore, previous surveys (e.g. \citealt{kale2015}) indicate that there are many clusters in the luminosity range of MACS0025 which do not host a radio halo.

\section{Conclusions}\label{sec:CONC}
In this paper, we have presented the first deep radio observations of the high-redshift merging cluster MACS~J0025.4$-$1222, performed using the GMRT at 325 MHz. The large FOV allows us to detect a population of radio galaxies that exhibit a wide range of morphologies. We also present a new analysis of all available \emph{Chandra} X-ray data on this cluster.

Following subtraction of the compact radio source population, we recover two sources of diffuse emission on scales of several hundred kpc. Based on their location toward the cluster outskirts and their approximately arc-like morphology, we believe that this constitutes a new double-relic system, discovered for the first time in this cluster. We have shown that these relics are consistent with established power-scaling trends for the known relic population. 

We derive an upper-limit spectral index $\alpha<-1.3$ for the NW relic; this is consistent with the spectral index of known relics. This implies a Mach number that is also consistent with what is expected from weak shocks. The X-ray data exhibit an asymmetric surface brightness profile, extended along the merger axis; we also derive a 2D temperature map of MACS~J0025.4$-$1222, which suggests the presence of regions of enhanced temperature along the merger axis. 

We tentatively identify a region co-located with the NW relic which appears to exhibit a sharp temperature discontinuity. However, its structure is perhaps more reminiscent of a cold front than a merger shock, as the cooler region of gas lies downstream of the hotter region. The uncertainties on the X-ray temperature are too significant to provide tight constraints on the Mach number.

\subsection{Acknowledgements}
We thank our anonymous referee for their careful review, which has improved the quality of our manuscript. We also thank the operators and engineers of the GMRT who made these observations possible. The GMRT is operated by the National Centre for Radio Astrophysics (NCRA) of the Tata Institute of Fundamental Research (TIFR), India. CJR wishes to thank H.~Intema for many helpful conversations during the data reduction process. CJR gratefully acknowledges funding support from the United Kingdom Science \& Technology Facilities Council (STFC). AMS gratefully acknowledges support from the European Research Council under grant ERC-2012-StG-307215 LODESTONE.

This work has made use of the NASA/IPAC Extragalactic Database (NED) and the NASA Astrophysics Data System (ADS). Many Figures in this work have employed the `cubehelix' colour scheme \citep{2011BASI...39..289G}.

\bibliographystyle{aa}
\bibliography{cjriseley_macsj0025}

\begin{thebibliography}{108}
\expandafter\ifx\csname natexlab\endcsname\relax\def\natexlab#1{#1}\fi

\bibitem[{{Akamatsu} {et~al.}(2012{\natexlab{a}}){Akamatsu}, {de Plaa},
  {Kaastra}, {Ishisaki}, {Ohashi}, {Kawaharada}, \&
  {Nakazawa}}]{2012PASJ...64...49A}
{Akamatsu}, H., {de Plaa}, J., {Kaastra}, J., {et~al.} 2012{\natexlab{a}},
  \pasj, 64, 49

\bibitem[{{Akamatsu} {et~al.}(2012{\natexlab{b}}){Akamatsu}, {Takizawa},
  {Nakazawa}, {Fukazawa}, {Ishisaki}, \& {Ohashi}}]{2012PASJ...64...67A}
{Akamatsu}, H., {Takizawa}, M., {Nakazawa}, K., {et~al.} 2012{\natexlab{b}},
  \pasj, 64, 67

\bibitem[{{Alam} {et~al.}(2015){Alam}, {Albareti}, {Allende Prieto}, {Anders},
  {Anderson}, {Anderton}, {Andrews}, {Armengaud}, {Aubourg}, {Bailey}, \&
  et~al.}]{2015ApJS..219...12A}
{Alam}, S., {Albareti}, F.~D., {Allende Prieto}, C., {et~al.} 2015, \apjs, 219,
  12

\bibitem[{{Baars} {et~al.}(1977){Baars}, {Genzel}, {Pauliny-Toth}, \&
  {Witzel}}]{1977A&A....61...99B}
{Baars}, J.~W.~M., {Genzel}, R., {Pauliny-Toth}, I.~I.~K., \& {Witzel}, A.
  1977, \aap, 61, 99

\bibitem[{{Bennett}(1962)}]{1962MmRAS..68..163B}
{Bennett}, A.~S. 1962, \memras, 68, 163

\bibitem[{{Blandford} \& {Eichler}(1987)}]{be1987}
{Blandford}, R. \& {Eichler}, D. 1987, \physrep, 154, 1

\bibitem[{{Blasi} \& {Colafrancesco}(1999)}]{1999APh....12..169B}
{Blasi}, P. \& {Colafrancesco}, S. 1999, Astroparticle Physics, 12, 169

\bibitem[{{Bonafede} {et~al.}(2012){Bonafede}, {Br{\"u}ggen}, {van Weeren},
  {Vazza}, {Giovannini}, {Ebeling}, {Edge}, {Hoeft}, \& {Klein}}]{bonafede2012}
{Bonafede}, A., {Br{\"u}ggen}, M., {van Weeren}, R., {et~al.} 2012, \mnras,
  426, 40

\bibitem[{{Bonafede} {et~al.}(2009){Bonafede}, {Feretti}, {Giovannini},
  {Govoni}, {Murgia}, {Taylor}, {Ebeling}, {Allen}, {Gentile}, \&
  {Pihlstr{\"o}m}}]{2009A&A...503..707B}
{Bonafede}, A., {Feretti}, L., {Giovannini}, G., {et~al.} 2009, \aap, 503, 707

\bibitem[{{Brada{\v c}} {et~al.}(2008){Brada{\v c}}, {Allen}, {Treu},
  {Ebeling}, {Massey}, {Morris}, {von der Linden}, \& {Applegate}}]{bradac2008}
{Brada{\v c}}, M., {Allen}, S.~W., {Treu}, T., {et~al.} 2008, \apj, 687, 959

\bibitem[{{Brada{\v c}} {et~al.}(2006){Brada{\v c}}, {Clowe}, {Gonzalez},
  {Marshall}, {Forman}, {Jones}, {Markevitch}, {Randall}, {Schrabback}, \&
  {Zaritsky}}]{2006ApJ...652..937B}
{Brada{\v c}}, M., {Clowe}, D., {Gonzalez}, A.~H., {et~al.} 2006, \apj, 652,
  937

\bibitem[{{Brunetti} {et~al.}(2009){Brunetti}, {Cassano}, {Dolag}, \&
  {Setti}}]{brunetti2009}
{Brunetti}, G., {Cassano}, R., {Dolag}, K., \& {Setti}, G. 2009, \aap, 507, 661

\bibitem[{{Brunetti} \& {Jones}(2014)}]{bj2014}
{Brunetti}, G. \& {Jones}, T.~W. 2014, International Journal of Modern Physics
  D, 23, 30007

\bibitem[{{Brunetti} {et~al.}(2001){Brunetti}, {Setti}, {Feretti}, \&
  {Giovannini}}]{2001MNRAS.320..365B}
{Brunetti}, G., {Setti}, G., {Feretti}, L., \& {Giovannini}, G. 2001, \mnras,
  320, 365

\bibitem[{{Cantwell} {et~al.}(2016){Cantwell}, {Scaife}, {Oozeer}, {Wen}, \&
  {Han}}]{tmc16}
{Cantwell}, T.~M., {Scaife}, A.~M.~M., {Oozeer}, N., {Wen}, Z.~L., \& {Han},
  J.~L. 2016, \mnras, 458, 1803

\bibitem[{{Cavaliere} \& {Fusco-Femiano}(1976)}]{Cavaliere76}
{Cavaliere}, A. \& {Fusco-Femiano}, R. 1976, \aap, 49, 137

\bibitem[{{Clowe} {et~al.}(2006){Clowe}, {Brada{\v c}}, {Gonzalez},
  {Markevitch}, {Randall}, {Jones}, \& {Zaritsky}}]{2006ApJ...648L.109C}
{Clowe}, D., {Brada{\v c}}, M., {Gonzalez}, A.~H., {et~al.} 2006, \apjl, 648,
  L109

\bibitem[{{Cohen} {et~al.}(2007){Cohen}, {Lane}, {Cotton}, {Kassim}, {Lazio},
  {Perley}, {Condon}, \& {Erickson}}]{2007AJ....134.1245C}
{Cohen}, A.~S., {Lane}, W.~M., {Cotton}, W.~D., {et~al.} 2007, \aj, 134, 1245

\bibitem[{{Condon} {et~al.}(1998){Condon}, {Cotton}, {Greisen}, {Yin},
  {Perley}, {Taylor}, \& {Broderick}}]{1998AJ....115.1693C}
{Condon}, J.~J., {Cotton}, W.~D., {Greisen}, E.~W., {et~al.} 1998, \aj, 115,
  1693

\bibitem[{{Conway} {et~al.}(1963){Conway}, {Kellermann}, \&
  {Long}}]{1963MNRAS.125..261C}
{Conway}, R.~G., {Kellermann}, K.~I., \& {Long}, R.~J. 1963, \mnras, 125, 261

\bibitem[{{Cuciti} {et~al.}(2015){Cuciti}, {Cassano}, {Brunetti}, {Dallacasa},
  {Kale}, {Ettori}, \& {Venturi}}]{2015A&A...580A..97C}
{Cuciti}, V., {Cassano}, R., {Brunetti}, G., {et~al.} 2015, \aap, 580, A97

\bibitem[{{Cutri} {et~al.}(2014){Cutri}, {Wright}, {Conrow}, {Fowler},
  {Eisenhardt}, {Grillmair}, {Kirkpatrick}, {Masci}, {McCallon}, {Wheelock},
  {Fajardo-Acosta}, {Yan}, {Benford}, {Harbut}, {Jarrett}, {Lake}, {Leisawitz},
  {Ressler}, {Stanford}, {Tsai}, {Liu}, {Helou}, {Mainzer}, {Gettings},
  {Gonzalez}, {Hoffman}, {Marsh}, {Padgett}, {Skrutskie}, {Beck}, {Papin}, \&
  {Wittman}}]{2014yCat.2328....0C}
{Cutri}, R.~M., {Wright}, E.~L., {Conrow}, T., {et~al.} 2014, VizieR Online
  Data Catalog, 2328

\bibitem[{{Czakon} {et~al.}(2015){Czakon}, {Sayers}, {Mantz}, {Golwala},
  {Downes}, {Koch}, {Lin}, {Molnar}, {Moustakas}, {Mroczkowski}, {Pierpaoli},
  {Shitanishi}, {Siegel}, \& {Umetsu}}]{2015ApJ...806...18C}
{Czakon}, N.~G., {Sayers}, J., {Mantz}, A., {et~al.} 2015, \apj, 806, 18

\bibitem[{{Dalya} {et~al.}(2016){Dalya}, {Frei}, {Galgoczi}, {Raffai}, \& {de
  Souza}}]{2016yCat.7275....0D}
{Dalya}, G., {Frei}, Z., {Galgoczi}, G., {Raffai}, P., \& {de Souza}, R.~S.
  2016, VizieR Online Data Catalog, 7275

\bibitem[{{Dawson} {et~al.}(2015){Dawson}, {Jee}, {Stroe}, {Ng}, {Golovich},
  {Wittman}, {Sobral}, {Br{\"u}ggen}, {R{\"o}ttgering}, \& {van
  Weeren}}]{dawson2015}
{Dawson}, W.~A., {Jee}, M.~J., {Stroe}, A., {et~al.} 2015, \apj, 805, 143

\bibitem[{{De Breuck} {et~al.}(2002){De Breuck}, {Tang}, {de Bruyn},
  {R{\"o}ttgering}, \& {van Breugel}}]{2002A&A...394...59D}
{De Breuck}, C., {Tang}, Y., {de Bruyn}, A.~G., {R{\"o}ttgering}, H., \& {van
  Breugel}, W. 2002, \aap, 394, 59

\bibitem[{{de Gasperin} {et~al.}(2015){de Gasperin}, {Intema}, {van Weeren},
  {Dawson}, {Golovich}, {Wittman}, {Bonafede}, \&
  {Br{\"u}ggen}}]{2015MNRAS.453.3483D}
{de Gasperin}, F., {Intema}, H.~T., {van Weeren}, R.~J., {et~al.} 2015, \mnras,
  453, 3483

\bibitem[{{de Gasperin} {et~al.}(2014){de Gasperin}, {van Weeren},
  {Br{\"u}ggen}, {Vazza}, {Bonafede}, \& {Intema}}]{gasperin2014}
{de Gasperin}, F., {van Weeren}, R.~J., {Br{\"u}ggen}, M., {et~al.} 2014,
  \mnras, 444, 3130

\bibitem[{{Dennison}(1980)}]{1980ApJ...239L..93D}
{Dennison}, B. 1980, \apjl, 239, L93

\bibitem[{{Douglas} {et~al.}(1996){Douglas}, {Bash}, {Bozyan}, {Torrence}, \&
  {Wolfe}}]{1996AJ....111.1945D}
{Douglas}, J.~N., {Bash}, F.~N., {Bozyan}, F.~A., {Torrence}, G.~W., \&
  {Wolfe}, C. 1996, \aj, 111, 1945

\bibitem[{{Ebeling} {et~al.}(2007){Ebeling}, {Barrett}, {Donovan}, {Ma},
  {Edge}, \& {van Speybroeck}}]{2007ApJ...661L..33E}
{Ebeling}, H., {Barrett}, E., {Donovan}, D., {et~al.} 2007, \apjl, 661, L33

\bibitem[{{Ebeling} {et~al.}(2001){Ebeling}, {Edge}, \&
  {Henry}}]{2001ApJ...553..668E}
{Ebeling}, H., {Edge}, A.~C., \& {Henry}, J.~P. 2001, \apj, 553, 668

\bibitem[{{Edge} {et~al.}(1959){Edge}, {Shakeshaft}, {McAdam}, {Baldwin}, \&
  {Archer}}]{1959MmRAS..68...37E}
{Edge}, D.~O., {Shakeshaft}, J.~R., {McAdam}, W.~B., {Baldwin}, J.~E., \&
  {Archer}, S. 1959, \memras, 68, 37

\bibitem[{{Fabricant} {et~al.}(1980){Fabricant}, {Lecar}, \&
  {Gorenstein}}]{Fabricant80}
{Fabricant}, D., {Lecar}, M., \& {Gorenstein}, P. 1980, \apj, 241, 552

\bibitem[{{Fanaroff} \& {Riley}(1974)}]{fr74}
{Fanaroff}, B.~L. \& {Riley}, J.~M. 1974, \mnras, 167, 31P

\bibitem[{{Feretti} {et~al.}(2006){Feretti}, {Bacchi}, {Slee}, {Giovannini},
  {Govoni}, {Andernach}, \& {Tsarevsky}}]{2006MNRAS.368..544F}
{Feretti}, L., {Bacchi}, M., {Slee}, O.~B., {et~al.} 2006, \mnras, 368, 544

\bibitem[{{Feretti} {et~al.}(2001){Feretti}, {Fusco-Femiano}, {Giovannini}, \&
  {Govoni}}]{feretti2001}
{Feretti}, L., {Fusco-Femiano}, R., {Giovannini}, G., \& {Govoni}, F. 2001,
  \aap, 373, 106

\bibitem[{{Feretti} {et~al.}(2012){Feretti}, {Giovannini}, {Govoni}, \&
  {Murgia}}]{feretti2012}
{Feretti}, L., {Giovannini}, G., {Govoni}, F., \& {Murgia}, M. 2012, \aapr, 20,
  54

\bibitem[{{Ferrari} {et~al.}(2008){Ferrari}, {Govoni}, {Schindler}, {Bykov}, \&
  {Rephaeli}}]{2008SSRv..134...93F}
{Ferrari}, C., {Govoni}, F., {Schindler}, S., {Bykov}, A.~M., \& {Rephaeli}, Y.
  2008, \ssr, 134, 93

\bibitem[{{Finoguenov} {et~al.}(2010){Finoguenov}, {Sarazin}, {Nakazawa},
  {Wik}, \& {Clarke}}]{2010ApJ...715.1143F}
{Finoguenov}, A., {Sarazin}, C.~L., {Nakazawa}, K., {Wik}, D.~R., \& {Clarke},
  T.~E. 2010, \apj, 715, 1143

\bibitem[{{Freeman} {et~al.}(2001){Freeman}, {Doe}, \&
  {Siemiginowska}}]{Sherpa01}
{Freeman}, P., {Doe}, S., \& {Siemiginowska}, A. 2001, in \procspie, Vol. 4477,
  Astronomical Data Analysis, ed. J.-L. {Starck} \& F.~D. {Murtagh}, 76--87

\bibitem[{{Garmire} {et~al.}(2003){Garmire}, {Bautz}, {Ford}, {Nousek}, \&
  {Ricker}}]{Garmire03}
{Garmire}, G.~P., {Bautz}, M.~W., {Ford}, P.~G., {Nousek}, J.~A., \& {Ricker},
  Jr., G.~R. 2003, in Society of Photo-Optical Instrumentation Engineers (SPIE)
  Conference Series, Vol. 4851, X-Ray and Gamma-Ray Telescopes and Instruments
  for Astronomy., ed. J.~E. {Truemper} \& H.~D. {Tananbaum}, 28--44

\bibitem[{{Gopal-Krishna} \& {Wiita}(2000)}]{2000A&A...363..507G}
{Gopal-Krishna} \& {Wiita}, P.~J. 2000, \aap, 363, 507

\bibitem[{{Govoni} {et~al.}(2001){Govoni}, {Feretti}, {Giovannini},
  {B{\"o}hringer}, {Reiprich}, \& {Murgia}}]{govoni2001}
{Govoni}, F., {Feretti}, L., {Giovannini}, G., {et~al.} 2001, \aap, 376, 803

\bibitem[{{Govoni} {et~al.}(2005){Govoni}, {Murgia}, {Feretti}, {Giovannini},
  {Dallacasa}, \& {Taylor}}]{2005A&A...430L...5G}
{Govoni}, F., {Murgia}, M., {Feretti}, L., {et~al.} 2005, \aap, 430, L5

\bibitem[{{Gower} {et~al.}(1967){Gower}, {Scott}, \&
  {Wills}}]{1967MmRAS..71...49G}
{Gower}, J.~F.~R., {Scott}, P.~F., \& {Wills}, D. 1967, \memras, 71, 49

\bibitem[{{Green}(2007)}]{2007BASI...35...77G}
{Green}, D.~A. 2007, Bulletin of the Astronomical Society of India, 35, 77

\bibitem[{{Green}(2011)}]{2011BASI...39..289G}
{Green}, D.~A. 2011, Bulletin of the Astronomical Society of India, 39, 289

\bibitem[{{Gregory} \& {Condon}(1991)}]{1991ApJS...75.1011G}
{Gregory}, P.~C. \& {Condon}, J.~J. 1991, \apjs, 75, 1011

\bibitem[{{Hales} {et~al.}(1995){Hales}, {Waldram}, {Rees}, \&
  {Warner}}]{1995MNRAS.274..447H}
{Hales}, S.~E.~G., {Waldram}, E.~M., {Rees}, N., \& {Warner}, P.~J. 1995,
  \mnras, 274, 447

\bibitem[{Henry {et~al.}(2004)Henry, Finoguenov, \& Briel}]{Henry2004}
Henry, J.~P., Finoguenov, A., \& Briel, U.~G. 2004, The Astrophysical Journal,
  615, 181

\bibitem[{{Intema} {et~al.}(2009){Intema}, {van der Tol}, {Cotton}, {Cohen},
  {van Bemmel}, \& {R{\"o}ttgering}}]{2009A&A...501.1185I}
{Intema}, H.~T., {van der Tol}, S., {Cotton}, W.~D., {et~al.} 2009, \aap, 501,
  1185

\bibitem[{{Jaffe}(1977)}]{1977ApJ...212....1J}
{Jaffe}, W.~J. 1977, \apj, 212, 1

\bibitem[{{Jee} {et~al.}(2014){Jee}, {Hughes}, {Menanteau}, {Sif{\'o}n},
  {Mandelbaum}, {Barrientos}, {Infante}, \& {Ng}}]{jee2014}
{Jee}, M.~J., {Hughes}, J.~P., {Menanteau}, F., {et~al.} 2014, \apj, 785, 20

\bibitem[{{Kalberla} {et~al.}(2005){Kalberla}, {Burton}, {Hartmann}, {Arnal},
  {Bajaja}, {Morras}, \& {P{\"o}ppel}}]{Kalberla05}
{Kalberla}, P.~M.~W., {Burton}, W.~B., {Hartmann}, D., {et~al.} 2005, \aap,
  440, 775

\bibitem[{{Kale} {et~al.}(2015){Kale}, {Venturi}, {Giacintucci}, {Dallacasa},
  {Cassano}, {Brunetti}, {Cuciti}, {Macario}, \& {Athreya}}]{kale2015}
{Kale}, R., {Venturi}, T., {Giacintucci}, S., {et~al.} 2015, \aap, 579, A92

\bibitem[{{Kale} {et~al.}(2013){Kale}, {Venturi}, {Giacintucci}, {Dallacasa},
  {Cassano}, {Brunetti}, {Macario}, \& {Athreya}}]{2013A&A...557A..99K}
{Kale}, R., {Venturi}, T., {Giacintucci}, S., {et~al.} 2013, \aap, 557, A99

\bibitem[{{Kassim} {et~al.}(2007){Kassim}, {Lazio}, {Erickson}, {Perley},
  {Cotton}, {Greisen}, {Cohen}, {Hicks}, {Schmitt}, \&
  {Katz}}]{2007ApJS..172..686K}
{Kassim}, N.~E., {Lazio}, T.~J.~W., {Erickson}, W.~C., {et~al.} 2007, \apjs,
  172, 686

\bibitem[{{Kellermann}(1964)}]{1964ApJ...140..969K}
{Kellermann}, K.~I. 1964, \apj, 140, 969

\bibitem[{{Kellermann} \& {Pauliny-Toth}(1973)}]{1973AJ.....78..828K}
{Kellermann}, K.~I. \& {Pauliny-Toth}, I.~I.~K. 1973, \aj, 78, 828

\bibitem[{{Kellermann} {et~al.}(1969){Kellermann}, {Pauliny-Toth}, \&
  {Williams}}]{1969ApJ...157....1K}
{Kellermann}, K.~I., {Pauliny-Toth}, I.~I.~K., \& {Williams}, P.~J.~S. 1969,
  \apj, 157, 1

\bibitem[{{Kerton}(2006)}]{2006MNRAS.373.1203K}
{Kerton}, C.~R. 2006, \mnras, 373, 1203

\bibitem[{{Kettenis} {et~al.}(2006){Kettenis}, {van Langevelde}, {Reynolds}, \&
  {Cotton}}]{2006ASPC..351..497K}
{Kettenis}, M., {van Langevelde}, H.~J., {Reynolds}, C., \& {Cotton}, B. 2006,
  in Astronomical Society of the Pacific Conference Series, Vol. 351,
  Astronomical Data Analysis Software and Systems XV, ed. C.~{Gabriel},
  C.~{Arviset}, D.~{Ponz}, \& S.~{Enrique}, 497

\bibitem[{{Konar} {et~al.}(2012){Konar}, {Hardcastle}, {Jamrozy}, {Croston}, \&
  {Nandi}}]{2012MNRAS.424.1061K}
{Konar}, C., {Hardcastle}, M.~J., {Jamrozy}, M., {Croston}, J.~H., \& {Nandi},
  S. 2012, \mnras, 424, 1061

\bibitem[{{Lane} {et~al.}(2014){Lane}, {Cotton}, {van Velzen}, {Clarke},
  {Kassim}, {Helmboldt}, {Lazio}, \& {Cohen}}]{2014MNRAS.440..327L}
{Lane}, W.~M., {Cotton}, W.~D., {van Velzen}, S., {et~al.} 2014, \mnras, 440,
  327

\bibitem[{{Leahy} {et~al.}(1997){Leahy}, {Black}, {Dennett-Thorpe},
  {Hardcastle}, {Komissarov}, {Perley}, {Riley}, \&
  {Scheuer}}]{1997MNRAS.291...20L}
{Leahy}, J.~P., {Black}, A.~R.~S., {Dennett-Thorpe}, J., {et~al.} 1997, \mnras,
  291, 20

\bibitem[{{Liedahl} {et~al.}(1995){Liedahl}, {Osterheld}, \&
  {Goldstein}}]{Liedahl95}
{Liedahl}, D.~A., {Osterheld}, A.~L., \& {Goldstein}, W.~H. 1995, \apjl, 438,
  L115

\bibitem[{{Lindner} {et~al.}(2014){Lindner}, {Baker}, {Hughes}, {Battaglia},
  {Gupta}, {Knowles}, {Marriage}, {Menanteau}, {Moodley}, {Reese}, \&
  {Srianand}}]{lindner2014}
{Lindner}, R.~R., {Baker}, A.~J., {Hughes}, J.~P., {et~al.} 2014, \apj, 786, 49

\bibitem[{{Ma} {et~al.}(2010){Ma}, {Ebeling}, {Marshall}, \&
  {Schrabback}}]{ma2010}
{Ma}, C.-J., {Ebeling}, H., {Marshall}, P., \& {Schrabback}, T. 2010, \mnras,
  406, 121

\bibitem[{{Macario} {et~al.}(2011){Macario}, {Markevitch}, {Giacintucci},
  {Brunetti}, {Venturi}, \& {Murray}}]{2011ApJ...728...82M}
{Macario}, G., {Markevitch}, M., {Giacintucci}, S., {et~al.} 2011, \apj, 728,
  82

\bibitem[{Mahdavi {et~al.}(2005)Mahdavi, Finoguenov, B{\"o}hringer, Geller, \&
  Henry}]{Mahdavi2005}
Mahdavi, A., Finoguenov, A., B{\"o}hringer, H., Geller, M.~J., \& Henry, J.~P.
  2005, The Astrophysical Journal, 622, 187

\bibitem[{{Markevitch} \& {Vikhlinin}(2007)}]{2007PhR...443....1M}
{Markevitch}, M. \& {Vikhlinin}, A. 2007, \physrep, 443, 1

\bibitem[{{Mewe} {et~al.}(1985){Mewe}, {Gronenschild}, \& {van den
  Oord}}]{Mewe85}
{Mewe}, R., {Gronenschild}, E.~H.~B.~M., \& {van den Oord}, G.~H.~J. 1985,
  \aaps, 62, 197

\bibitem[{{Ng} {et~al.}(2015){Ng}, {Dawson}, {Wittman}, {Jee}, {Hughes},
  {Menanteau}, \& {Sif{\'o}n}}]{ng2015}
{Ng}, K.~Y., {Dawson}, W.~A., {Wittman}, D., {et~al.} 2015, \mnras, 453, 1531

\bibitem[{{Ogrean} {et~al.}(2013){Ogrean}, {Br{\"u}ggen}, {van Weeren},
  {R{\"o}ttgering}, {Croston}, \& {Hoeft}}]{2013MNRAS.433..812O}
{Ogrean}, G.~A., {Br{\"u}ggen}, M., {van Weeren}, R.~J., {et~al.} 2013, \mnras,
  433, 812

\bibitem[{{Orr{\`u}} {et~al.}(2007){Orr{\`u}}, {Murgia}, {Feretti}, {Govoni},
  {Brunetti}, {Giovannini}, {Girardi}, \& {Setti}}]{orru2007}
{Orr{\`u}}, E., {Murgia}, M., {Feretti}, L., {et~al.} 2007, \aap, 467, 943

\bibitem[{{Orr{\`u}} {et~al.}(2015){Orr{\`u}}, {van Velzen}, {Pizzo},
  {Yatawatta}, {Paladino}, {Iacobelli}, {Murgia}, {Falcke}, {Morganti}, {de
  Bruyn}, {Ferrari}, {Anderson}, {Bonafede}, {Mulcahy}, {Asgekar}, {Avruch},
  {Beck}, {Bell}, {van Bemmel}, {Bentum}, {Bernardi}, {Best}, {Breitling},
  {Broderick}, {Br{\"u}ggen}, {Butcher}, {Ciardi}, {Conway}, {Corstanje}, {de
  Geus}, {Deller}, {Duscha}, {Eisl{\"o}ffel}, {Engels}, {Frieswijk}, {Garrett},
  {Grie{\ss}meier}, {Gunst}, {Hamaker}, {Heald}, {Hoeft}, {van der Horst},
  {Intema}, {Juette}, {Kohler}, {Kondratiev}, {Kuniyoshi}, {Kuper}, {Loose},
  {Maat}, {Mann}, {Markoff}, {McFadden}, {McKay-Bukowski}, {Miley}, {Moldon},
  {Molenaar}, {Munk}, {Nelles}, {Paas}, {Pandey-Pommier}, {Pandey}, {Pietka},
  {Polatidis}, {Reich}, {R{\"o}ttgering}, {Rowlinson}, {Scaife},
  {Schoenmakers}, {Schwarz}, {Serylak}, {Shulevski}, {Smirnov}, {Steinmetz},
  {Stewart}, {Swinbank}, {Tagger}, {Tasse}, {Thoudam}, {Toribio}, {Vermeulen},
  {Vocks}, {van Weeren}, {Wijers}, {Wise}, \& {Wucknitz}}]{2015A&A...584A.112O}
{Orr{\`u}}, E., {van Velzen}, S., {Pizzo}, R.~F., {et~al.} 2015, \aap, 584,
  A112

\bibitem[{{Paul} {et~al.}(2014){Paul}, {Datta}, \&
  {Intema}}]{2014arXiv1412.0285P}
{Paul}, S., {Datta}, A., \& {Intema}, H.~T. 2014, ArXiv e-prints
  [\eprint[arXiv]{1412.0285}]

\bibitem[{{Pauliny-Toth} {et~al.}(1966){Pauliny-Toth}, {Wade}, \&
  {Heeschen}}]{1966ApJS...13...65P}
{Pauliny-Toth}, I.~I.~K., {Wade}, C.~M., \& {Heeschen}, D.~S. 1966, \apjs, 13,
  65

\bibitem[{{Petrosian}(2001)}]{2001ApJ...557..560P}
{Petrosian}, V. 2001, \apj, 557, 560

\bibitem[{{Pizzo} {et~al.}(2011){Pizzo}, {de Bruyn}, {Bernardi}, \&
  {Brentjens}}]{2011A&A...525A.104P}
{Pizzo}, R.~F., {de Bruyn}, A.~G., {Bernardi}, G., \& {Brentjens}, M.~A. 2011,
  \aap, 525, A104

\bibitem[{{Planck Collaboration} {et~al.}(2014){Planck Collaboration}, {Ade},
  {Aghanim}, {Armitage-Caplan}, {Arnaud}, {Ashdown}, {Atrio-Barandela},
  {Aumont}, {Aussel}, {Baccigalupi}, \& et~al.}]{planck_clusters}
{Planck Collaboration}, {Ade}, P.~A.~R., {Aghanim}, N., {et~al.} 2014, \aap,
  571, A29

\bibitem[{{Pratt} {et~al.}(2009){Pratt}, {Croston}, {Arnaud}, \&
  {B{\"o}hringer}}]{pratt2009}
{Pratt}, G.~W., {Croston}, J.~H., {Arnaud}, M., \& {B{\"o}hringer}, H. 2009,
  \aap, 498, 361

\bibitem[{{Rengelink} {et~al.}(1997){Rengelink}, {Tang}, {de Bruyn}, {Miley},
  {Bremer}, {Roettgering}, \& {Bremer}}]{1997A&AS..124..259R}
{Rengelink}, R.~B., {Tang}, Y., {de Bruyn}, A.~G., {et~al.} 1997, \aaps, 124,
  259

\bibitem[{{Roger} {et~al.}(1973){Roger}, {Costain}, \&
  {Bridle}}]{1973AJ.....78.1030R}
{Roger}, R.~S., {Costain}, C.~H., \& {Bridle}, A.~H. 1973, \aj, 78, 1030

\bibitem[{{Roy} {et~al.}(2010){Roy}, {Gupta}, {Pen}, {Peterson}, {Kudale}, \&
  {Kodilkar}}]{2010ExA....28...25R}
{Roy}, J., {Gupta}, Y., {Pen}, U.-L., {et~al.} 2010, Experimental Astronomy,
  28, 25

\bibitem[{{Saikia} \& {Jamrozy}(2009)}]{2009BASI...37...63S}
{Saikia}, D.~J. \& {Jamrozy}, M. 2009, Bulletin of the Astronomical Society of
  India, 37 [\eprint[arXiv]{1002.1841}]

\bibitem[{{Sanders}(2006)}]{Contour_Sanders_2006}
{Sanders}, J.~S. 2006, \mnras, 371, 829

\bibitem[{{Sarazin} \& {Bahcall}(1977)}]{Sarazin77}
{Sarazin}, C.~L. \& {Bahcall}, J.~N. 1977, \apjs, 34, 451

\bibitem[{{Sayers} {et~al.}(2013){Sayers}, {Czakon}, {Mantz}, {Golwala},
  {Ameglio}, {Downes}, {Koch}, {Lin}, {Maughan}, {Molnar}, {Moustakas},
  {Mroczkowski}, {Pierpaoli}, {Shitanishi}, {Siegel}, {Umetsu}, \& {Van der
  Pyl}}]{2013ApJ...768..177S}
{Sayers}, J., {Czakon}, N.~G., {Mantz}, A., {et~al.} 2013, \apj, 768, 177

\bibitem[{{Scaife} \& {Heald}(2012)}]{2012MNRAS.423L..30S}
{Scaife}, A.~M.~M. \& {Heald}, G.~H. 2012, \mnras, 423, L30

\bibitem[{{Schilizzi} {et~al.}(2001){Schilizzi}, {Tian}, {Conway}, {Nan},
  {Miley}, {Barthel}, {Normandeau}, {Dallacasa}, \&
  {Gurvits}}]{2001A&A...368..398S}
{Schilizzi}, R.~T., {Tian}, W.~W., {Conway}, J.~E., {et~al.} 2001, \aap, 368,
  398

\bibitem[{{Schoenmakers} {et~al.}(2000){Schoenmakers}, {de Bruyn},
  {R{\"o}ttgering}, {van der Laan}, \& {Kaiser}}]{2000MNRAS.315..371S}
{Schoenmakers}, A.~P., {de Bruyn}, A.~G., {R{\"o}ttgering}, H.~J.~A., {van der
  Laan}, H., \& {Kaiser}, C.~R. 2000, \mnras, 315, 371

\bibitem[{{Shan} {et~al.}(2010){Shan}, {Qin}, {Fort}, {Tao}, {Wu}, \&
  {Zhao}}]{shan2010}
{Shan}, H., {Qin}, B., {Fort}, B., {et~al.} 2010, \mnras, 406, 1134

\bibitem[{{Shimwell} {et~al.}(2015){Shimwell}, {Markevitch}, {Brown},
  {Feretti}, {Gaensler}, {Johnston-Hollitt}, {Lage}, \&
  {Srinivasan}}]{shimwell2015}
{Shimwell}, T.~W., {Markevitch}, M., {Brown}, S., {et~al.} 2015, \mnras, 449,
  1486

\bibitem[{{Subrahmanyan} {et~al.}(2003){Subrahmanyan}, {Beasley}, {Goss},
  {Golap}, \& {Hunstead}}]{2003AJ....125.1095S}
{Subrahmanyan}, R., {Beasley}, A.~J., {Goss}, W.~M., {Golap}, K., \&
  {Hunstead}, R.~W. 2003, \aj, 125, 1095

\bibitem[{{van Haarlem} {et~al.}(2013){van Haarlem}, {Wise}, {Gunst}, {Heald},
  {McKean}, {Hessels}, {de Bruyn}, {Nijboer}, {Swinbank}, {Fallows},
  {Brentjens}, {Nelles}, {Beck}, {Falcke}, {Fender}, {H{\"o}randel},
  {Koopmans}, {Mann}, {Miley}, {R{\"o}ttgering}, {Stappers}, {Wijers},
  {Zaroubi}, {van den Akker}, {Alexov}, {Anderson}, {Anderson}, {van Ardenne},
  {Arts}, {Asgekar}, {Avruch}, {Batejat}, {B{\"a}hren}, {Bell}, {Bell}, {van
  Bemmel}, {Bennema}, {Bentum}, {Bernardi}, {Best}, {B{\^i}rzan}, {Bonafede},
  {Boonstra}, {Braun}, {Bregman}, {Breitling}, {van de Brink}, {Broderick},
  {Broekema}, {Brouw}, {Br{\"u}ggen}, {Butcher}, {van Cappellen}, {Ciardi},
  {Coenen}, {Conway}, {Coolen}, {Corstanje}, {Damstra}, {Davies}, {Deller},
  {Dettmar}, {van Diepen}, {Dijkstra}, {Donker}, {Doorduin}, {Dromer}, {Drost},
  {van Duin}, {Eisl{\"o}ffel}, {van Enst}, {Ferrari}, {Frieswijk}, {Gankema},
  {Garrett}, {de Gasperin}, {Gerbers}, {de Geus}, {Grie{\ss}meier}, {Grit},
  {Gruppen}, {Hamaker}, {Hassall}, {Hoeft}, {Holties}, {Horneffer}, {van der
  Horst}, {van Houwelingen}, {Huijgen}, {Iacobelli}, {Intema}, {Jackson},
  {Jelic}, {de Jong}, {Juette}, {Kant}, {Karastergiou}, {Koers}, {Kollen},
  {Kondratiev}, {Kooistra}, {Koopman}, {Koster}, {Kuniyoshi}, {Kramer},
  {Kuper}, {Lambropoulos}, {Law}, {van Leeuwen}, {Lemaitre}, {Loose}, {Maat},
  {Macario}, {Markoff}, {Masters}, {McFadden}, {McKay-Bukowski}, {Meijering},
  {Meulman}, {Mevius}, {Middelberg}, {Millenaar}, {Miller-Jones}, {Mohan},
  {Mol}, {Morawietz}, {Morganti}, {Mulcahy}, {Mulder}, {Munk}, {Nieuwenhuis},
  {van Nieuwpoort}, {Noordam}, {Norden}, {Noutsos}, {Offringa}, {Olofsson},
  {Omar}, {Orr{\`u}}, {Overeem}, {Paas}, {Pandey-Pommier}, {Pandey}, {Pizzo},
  {Polatidis}, {Rafferty}, {Rawlings}, {Reich}, {de Reijer}, {Reitsma},
  {Renting}, {Riemers}, {Rol}, {Romein}, {Roosjen}, {Ruiter}, {Scaife}, {van
  der Schaaf}, {Scheers}, {Schellart}, {Schoenmakers}, {Schoonderbeek},
  {Serylak}, {Shulevski}, {Sluman}, {Smirnov}, {Sobey}, {Spreeuw}, {Steinmetz},
  {Sterks}, {Stiepel}, {Stuurwold}, {Tagger}, {Tang}, {Tasse}, {Thomas},
  {Thoudam}, {Toribio}, {van der Tol}, {Usov}, {van Veelen}, {van der Veen},
  {ter Veen}, {Verbiest}, {Vermeulen}, {Vermaas}, {Vocks}, {Vogt}, {de Vos},
  {van der Wal}, {van Weeren}, {Weggemans}, {Weltevrede}, {White}, {Wijnholds},
  {Wilhelmsson}, {Wucknitz}, {Yatawatta}, {Zarka}, {Zensus}, \& {van
  Zwieten}}]{2013A&A...556A...2V}
{van Haarlem}, M.~P., {Wise}, M.~W., {Gunst}, A.~W., {et~al.} 2013, \aap, 556,
  A2

\bibitem[{{van Weeren} {et~al.}(2011){van Weeren}, {Br{\"u}ggen},
  {R{\"o}ttgering}, \& {Hoeft}}]{vw2011}
{van Weeren}, R.~J., {Br{\"u}ggen}, M., {R{\"o}ttgering}, H.~J.~A., \& {Hoeft},
  M. 2011, \mnras, 418, 230

\bibitem[{{van Weeren} {et~al.}(2009){van Weeren}, {R{\"o}ttgering},
  {Br{\"u}ggen}, \& {Cohen}}]{2009A&A...505..991V}
{van Weeren}, R.~J., {R{\"o}ttgering}, H.~J.~A., {Br{\"u}ggen}, M., \& {Cohen},
  A. 2009, \aap, 505, 991

\bibitem[{{van Weeren} {et~al.}(2010){van Weeren}, {R{\"o}ttgering},
  {Br{\"u}ggen}, \& {Hoeft}}]{2010Sci...330..347V}
{van Weeren}, R.~J., {R{\"o}ttgering}, H.~J.~A., {Br{\"u}ggen}, M., \& {Hoeft},
  M. 2010, Science, 330, 347

\bibitem[{{van Weeren} {et~al.}(2012){van Weeren}, {R{\"o}ttgering}, {Intema},
  {Rudnick}, {Br{\"u}ggen}, {Hoeft}, \& {Oonk}}]{2012A&A...546A.124V}
{van Weeren}, R.~J., {R{\"o}ttgering}, H.~J.~A., {Intema}, H.~T., {et~al.}
  2012, \aap, 546, A124

\bibitem[{{Venturi} {et~al.}(2007){Venturi}, {Giacintucci}, {Brunetti},
  {Cassano}, {Bardelli}, {Dallacasa}, \& {Setti}}]{venturi2007}
{Venturi}, T., {Giacintucci}, S., {Brunetti}, G., {et~al.} 2007, \aap, 463, 937

\bibitem[{{Venturi} {et~al.}(2008){Venturi}, {Giacintucci}, {Dallacasa},
  {Cassano}, {Brunetti}, {Bardelli}, \& {Setti}}]{2008A&A...484..327V}
{Venturi}, T., {Giacintucci}, S., {Dallacasa}, D., {et~al.} 2008, \aap, 484,
  327

\bibitem[{{Vikhlinin} {et~al.}(2005){Vikhlinin}, {Markevitch}, {Murray},
  {Jones}, {Forman}, \& {Van Speybroeck}}]{Vikhlinin05}
{Vikhlinin}, A., {Markevitch}, M., {Murray}, S.~S., {et~al.} 2005, \apj, 628,
  655

\bibitem[{{White} \& {Becker}(1992)}]{1992ApJS...79..331W}
{White}, R.~L. \& {Becker}, R.~H. 1992, \apjs, 79, 331

\bibitem[{{Wright} {et~al.}(1994){Wright}, {Griffith}, {Burke}, \&
  {Ekers}}]{1994ApJS...91..111W}
{Wright}, A.~E., {Griffith}, M.~R., {Burke}, B.~F., \& {Ekers}, R.~D. 1994,
  \apjs, 91, 111

\bibitem[{{Wright}(2006)}]{2006PASP..118.1711W}
{Wright}, E.~L. 2006, \pasp, 118, 1711

\bibitem[{{Zhang} {et~al.}(1997){Zhang}, {Zheng}, {Chen}, {Wang}, {Cao},
  {Peng}, \& {Nan}}]{1997A&AS..121...59Z}
{Zhang}, X., {Zheng}, Y., {Chen}, H., {et~al.} 1997, \aaps, 121, 59

\end{thebibliography}

\appendix
\section{An Extension to the Scaife \& Heald flux scale: 3C~468.1}\label{sec:3c468}
The radio source 3C~468.1 is available as a primary calibrator for GMRT observations, although perhaps it is the least preferred primary calibrator. No other calibrator source was available for the initial flux calibrator scan for our observations, however. The \texttt{SPAM} calibration routine makes use of the \citetalias{2012MNRAS.423L..30S} flux density scale, which does not include 3C~468.1. To avoid potential issues associated with using different flux density scales, therefore, we follow the method adopted by \citetalias{2012MNRAS.423L..30S} to fit the available flux density measurements for 3C~468.1 and add it to the \citetalias{2012MNRAS.423L..30S} flux density scale. 

\defcitealias{1973AJ.....78.1030R}{RCB}
\defcitealias{1969ApJ...157....1K}{KPW}
\defcitealias{1977A&A....61...99B}{B77}

Flux density measurements exist between 38 MHz and 10 GHz for 3C~468.1, with observations using a range of different flux density scales. Following \citetalias{2012MNRAS.423L..30S} we bring these measurements onto the flux scale of \cite{1973AJ.....78.1030R} (hereafter \citetalias{1973AJ.....78.1030R}). We present these measurements, along with the correction factor required to bring them onto the RCB scale, in Table \ref{tab:primcal}. All flux density measurements on the Baars scale we adjusted by interpolation of the values listed in Table 7 of \cite{1977A&A....61...99B} -- hereafter \citetalias{1977A&A....61...99B}. 

For the Texas Radio Survey at 365 MHz, \cite{1996AJ....111.1945D} find their flux density measurements to be consistent with a factor 96 per cent of the \citetalias{1977A&A....61...99B} flux scale. Below 325 MHz, the conversion process is a little more complex. For measurements on the \cite{1969ApJ...157....1K} (hereafter \citetalias{1969ApJ...157....1K}) scale at 178 MHz, \citetalias{1973AJ.....78.1030R} give a conversion factor 1.09. For the measurement from \cite{1967MmRAS..71...49G} on the CKL scale \citep{1963MNRAS.125..261C} Table 7 of \citetalias{1977A&A....61...99B} was used to convert to the \citetalias{1973AJ.....78.1030R} scale and then onto the \citetalias{2012MNRAS.423L..30S} scale. For the measurements at 74 MHz using the VLA (\citealt{2007AJ....134.1245C} and \citealt{2007ApJS..172..686K}) the \citetalias{1977A&A....61...99B} scale was used. \cite{2014MNRAS.440..327L} find that a mean conversion factor 1.1 is required to bring flux density measurements into line with the \citetalias{2012MNRAS.423L..30S} flux scale.

\begin{table}
\begin{center}
\caption[Flux density measurements for 3C~468.1 from the literature]{Integrated flux density measurements for 3C~468.1 from the literature, along with the flux density scale used in the original reference, and the `correction' factor required to bring the measurement onto the \citetalias{2012MNRAS.423L..30S} scale. \label{tab:primcal}}
\scalebox{0.8}{
\begin{tabular}{ccccc}
\hline
Frequency & $S_{\rm{int}}$ & Flux scale & Factor & Catalogue reference \\
$[$MHz$]$ & $[$Jy$]$& & & \\ 
\hline 
10700 & $\,0.31\pm0.02$ & KPW & - & KPT \\
5000 & $\,0.87\pm0.04$\tnote{a} & KPW & - & KPW \\
4850 & $\,0.96\pm0.09$ & B77 & 1.007 & GC91 \\
2695 & $\,1.93\pm0.10$\tnote{a} & KPW & - & KPW \\ 
1420 & $\,5.78\pm0.17$ & B77 & 0.972 & K06 \\
1400 & $\,5.00\pm0.25$\tnote{a} & KPW & - & KPW \\
1400 & $\,4.95\pm0.15$ & B77 & 0.972 & C98 \\
1400 & $\,4.89\pm0.24$ & B77 & 0.972 & WB91 \\
1400 & $\,5.16\pm0.13$ & K64 & 1.009 & PWH \\
750 & $\,\,9.70\pm0.49$\tnote{a} & KPW & - & KPW \\
750 & $10.06\pm0.15\,$ & K64 & 1.012 & PWH \\
408 & $21.50\pm0.64\,$ & B77/RCB & - & K06 \\
365 & $22.13\pm0.51\,$ & TXS\tnote{c} & 0.977 & D96 \\
325 & $26.80\pm1.34\,$ & WENSS\tnote{d} & 0.90 & R97 \\
232 & 43.98 & RCB & - & Z97 \\
178 & $30.00\pm3.00$\tnote{b}\, & KPW & 1.09 & KPW \\
178 & $30.50\pm2.40$\tnote{e}\, & CKL & 1.15\tnote{f} & G67  \\
159 & $34.00\pm3.40$\tnote{g}\, & RCB & - & B62 \\
74 & $40.70\pm0.66$ & B77 & 1.1 & K07 \\
74 & $42.09\pm4.41$ & B77 & 1.1 & C07 \\
38 & $24.60\pm2.46$\tnote{b} & RCB & - & H95 \\
\hline
\multicolumn{5}{l}{Notation:} \\
\multicolumn{5}{p{10cm}}{B62: \protect{\cite{1962MmRAS..68..163B}}, B77: \protect{\cite{1977A&A....61...99B}}, C07: \protect{\cite{2007AJ....134.1245C}}, C98: \protect\cite{1998AJ....115.1693C}, CKL: \protect\cite{1963MNRAS.125..261C}, D96: \protect{\cite{1996AJ....111.1945D}}, G67: \protect{\cite{1967MmRAS..71...49G}}, GC91: \protect\cite{1991ApJS...75.1011G}, H95: \protect{\cite{1995MNRAS.274..447H}}, K06: \protect\cite{2006MNRAS.373.1203K}, K07: \protect{\cite{2007ApJS..172..686K}}, K64: \protect\cite{1964ApJ...140..969K}, KPT: \protect\cite{1973AJ.....78..828K}, KPW: \protect\cite{1969ApJ...157....1K}, PWH: \protect{\cite{1966ApJS...13...65P}}, R97: \protect{\cite{1997A&AS..124..259R}}, RCB: \protect\cite{1973AJ.....78.1030R}, SH12: \protect\cite{2012MNRAS.423L..30S}, WB91: \protect{\cite{1992ApJS...79..331W}}, Z97: \protect{\cite{1997A&AS..121...59Z}}.} \\
\multicolumn{5}{l}{\hspace{0.2cm}}\\
\multicolumn{5}{l}{Table notes:} \\
\multicolumn{5}{p{10cm}}{$^{a}$ The uncertainty is quoted as 5 per cent of the integrated flux density. } \\
\multicolumn{5}{p{10cm}}{$^{b}$ The uncertainty is quoted as 10 per cent of the integrated flux density.  } \\
\multicolumn{5}{p{10cm}}{$^{c}$ Sources in the Texas radio survey were found to have flux densities that were consistent with a factor $0.9607\times$B77. See \protect\cite{1996AJ....111.1945D}.}\\
\multicolumn{5}{p{10cm}}{$^{d}$ The WENSS flux scale is complex. An overall correction factor of 0.9 was applied to bring the flux density onto the B77 scale, then to the RCB73 scale. See \protect\cite{2012MNRAS.423L..30S} for details. } \\
\multicolumn{5}{p{10cm}}{$^{e}$ The uncertainty is quoted as 8 per cent of the integrated flux density. } \\
\multicolumn{5}{p{10cm}}{$^{f}$ This conversion factor was derived using the factors in Table 7 of B77 to convert to the KPW scale and then onto SH12.  } \\
\multicolumn{5}{p{10cm}}{$^{g}$ An uncertainty of 10 per cent has been assumed. } \\
\end{tabular}
}
\end{center}
\end{table}

\begin{figure}
	\includegraphics[width=0.47\textwidth]{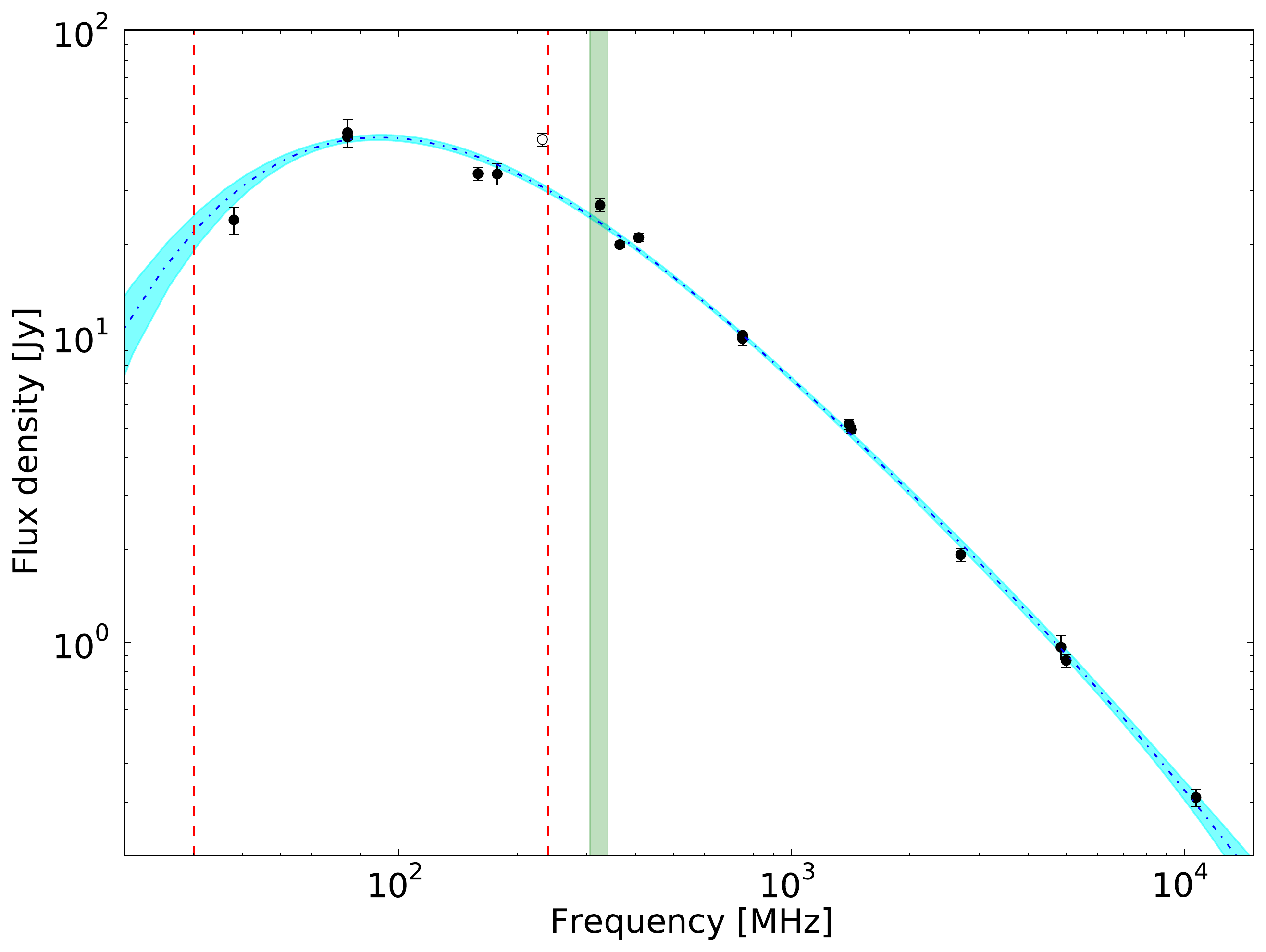}
	\caption{Integrated flux density of 3C~468.1 as a function of frequency. Dot-dashed line marks fourth-order polynomial fit to the data. Empty circle indicates measurement from the Miyun survey at 232 MHz which has been excluded from the fit. Shaded green region marks the frequency range covered by the GMRT observations in this work; shaded blue region marks $1\sigma$ uncertainty region on the model. Vertical dashed lines mark the frequency bounds of LOFAR. }
\label{fig:polyfit}
\end{figure}

\begin{figure*}
	\centering
	\includegraphics[width=0.85\textwidth]{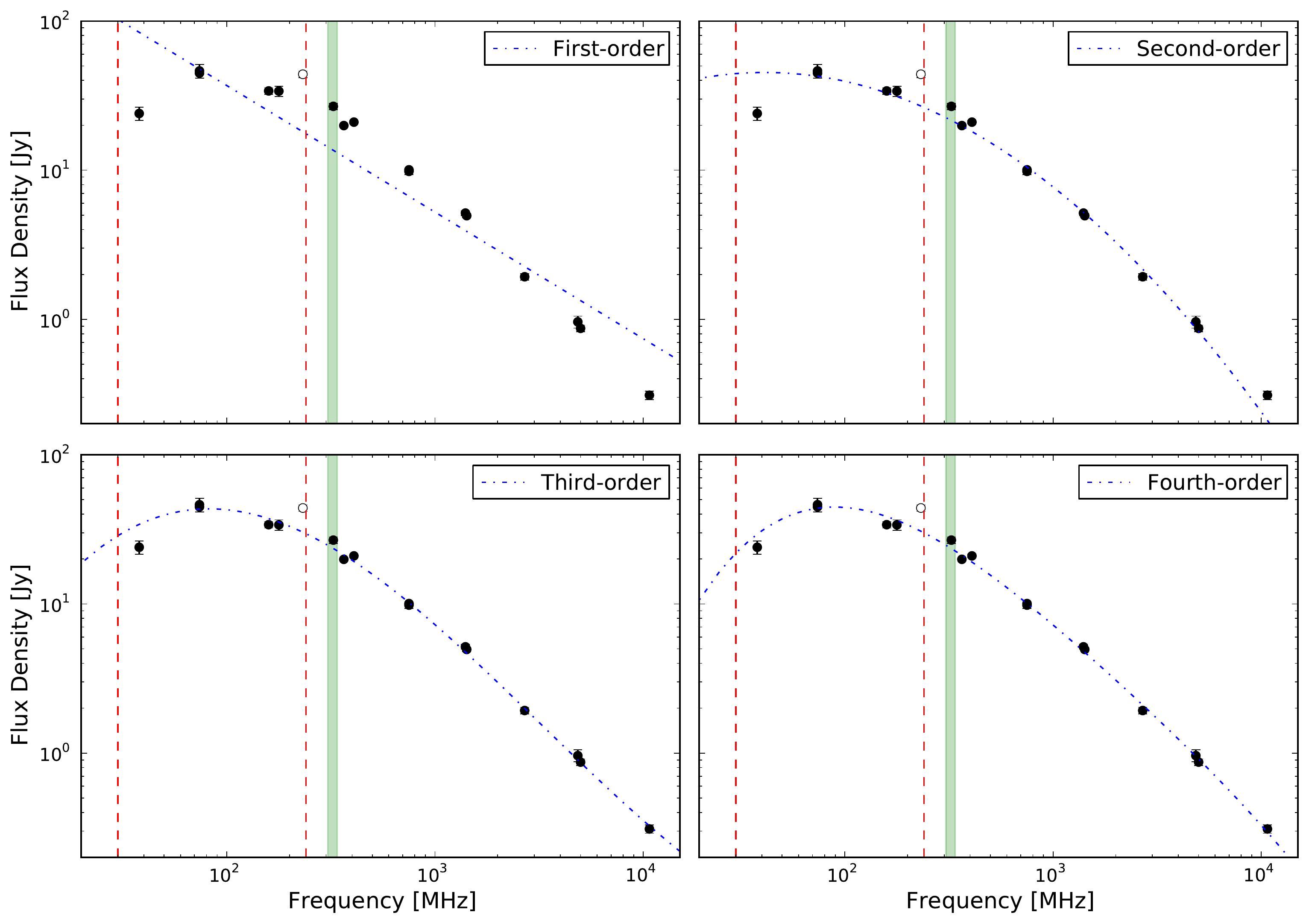}
	\caption{Polynomial spectral index models fitted to 3C~468.1, showing linear, second-, third- and fourth-order polynomial models from left to right. Empty circle indicates measurement from the Miyun survey at 232 MHz which has been excluded from the fit. Shaded green region marks the frequency range covered by the GMRT observations in this work; vertical dashed lines mark the frequency bounds of LOFAR.}
\label{fig:polyfits}
\end{figure*}

We present the flux density measurements as a function of frequency in Figure \ref{fig:polyfit}. We exclude the 232 MHz flux density from the Miyun survey from the fitting partly due to a large discrepancy between the integrated flux density (43.98 Jy) and the peak flux density (38.77 Jy) and partly due to the large offset with respect to the other flux densities from the literature (see Figure \ref{fig:polyfit}). Following \citetalias{2012MNRAS.423L..30S}, we attempt to fit a polynomial model in linear frequency space, in order to retain Gaussian noise characteristics. As such, the functional form we attempt to fit is as follows:

\begin{equation}
	S \, [ \rm{mJy} ] = A_0 \prod_{i = 1}^{N} 10^{ A_i \, \log^i [ \nu / 150 \, \rm{MHz} ]}
\end{equation}
The best-fit parameters for linear, second-, third- and fourth-order polynomial functions are listed in Table \ref{tab:polyfits} (and presented in Figure \ref{fig:polyfits}) along with the associated uncertainties and the reduced chi-squared ($\chi^2_{\rm{red}}$) which is used to evaluate the goodness of fit. From visual inspection it is clear that the linear and second-order models are a poor fit to the data; it also appears that a fourth-order polynomial describes the data marginally better at the low end of the frequency range.

\begin{table}
\begin{center}
\caption{Polynomial spectral index model parameters for 3C~468.1. \label{tab:polyfits}}
\scalebox{0.75}{
\begin{tabular}{ c | rrrr }
\hline
Order & 1 & 2 & 3 & 4 \\
\hline
$A_0$ & $26.977\pm3.138$ & $34.432\pm1.263$ & $38.751\pm1.252$ & $40.093\pm1.530$ \\
$A_1$ & $-0.840\pm0.054$ & $-0.440\pm0.035$ & $-0.392\pm0.026$ & $-0.420\pm0.030$ \\
$A_2$ & - & $-0.415\pm0.036$ & $-0.715\pm0.068$ & $-0.830\pm0.104$ \\
$A_3$ & - & - & $0.174\pm0.034$ & \,\,\,$0.389\pm0.144$ \\
$A_4$ & - & - & - & $-0.084\pm0.053$ \\
$\chi^2_{\rm{red}}$ & 5.89 & 0.96 & 0.37 & 0.24 \\
\hline
\end{tabular}
}
\end{center}
\end{table}

In the case of 3C~468.1, selection of the `best-fit' model is heavily influenced by the confidence in the single data point at 38 MHz; this flux density measurement is from the revised 3C catalogue (3CR; \citealt{1995MNRAS.274..447H}). Based on this, and the $\chi^2_{\rm{red}}$ values, we select a fourth-order polynomial as the `best-fit' model for the spectral index behaviour of 3C~468.1. In Figure \ref{fig:polyfit} we present the fourth-order fit to the data shown in Table \ref{tab:primcal}. The uncertainty in the model was derived using a Monte-Carlo simulation of the parameter space over 1000 iterations; in Figure \ref{fig:polyfit} the $1\sigma$ uncertainty is indicated by the shaded cyan region.

\section{Radio environment of MACS0025}\label{sec:app2}
In this Appendix we present postage stamp images of the extended radio sources identified in Figure \ref{fig:macs_full}. For each RG, we again measured the flux density recovered by the GMRT using \texttt{fitflux} \citep{2007BASI...35...77G}. These measurements are presented in Table \ref{tab:rg}. 

\begin{figure*}
	\centering
	\begin{subfigure}[r]{0.39\textwidth}
	\includegraphics[width=\textwidth]{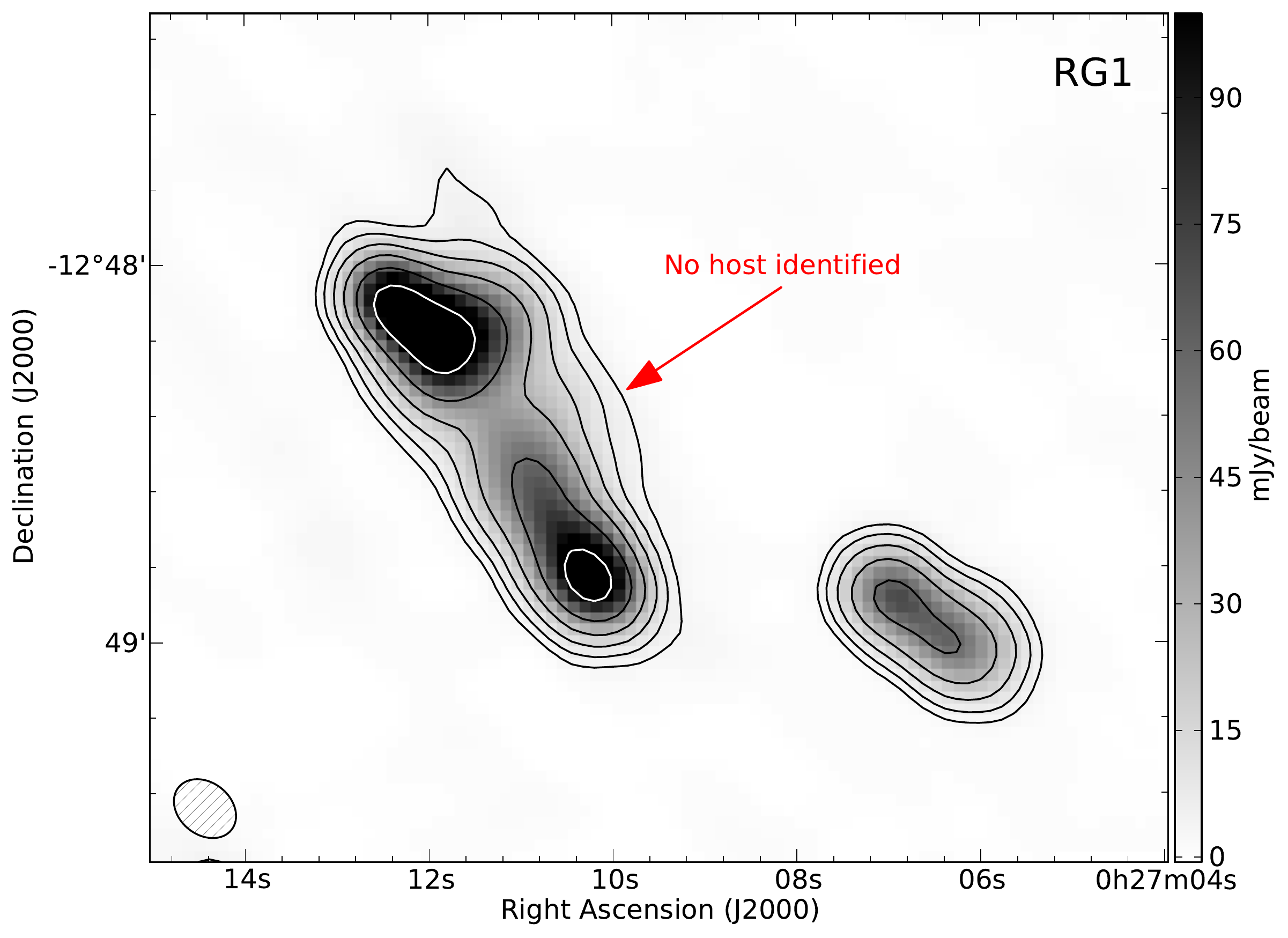}
	\end{subfigure}
	\begin{subfigure}[l]{0.39\textwidth}
	\includegraphics[width=\textwidth]{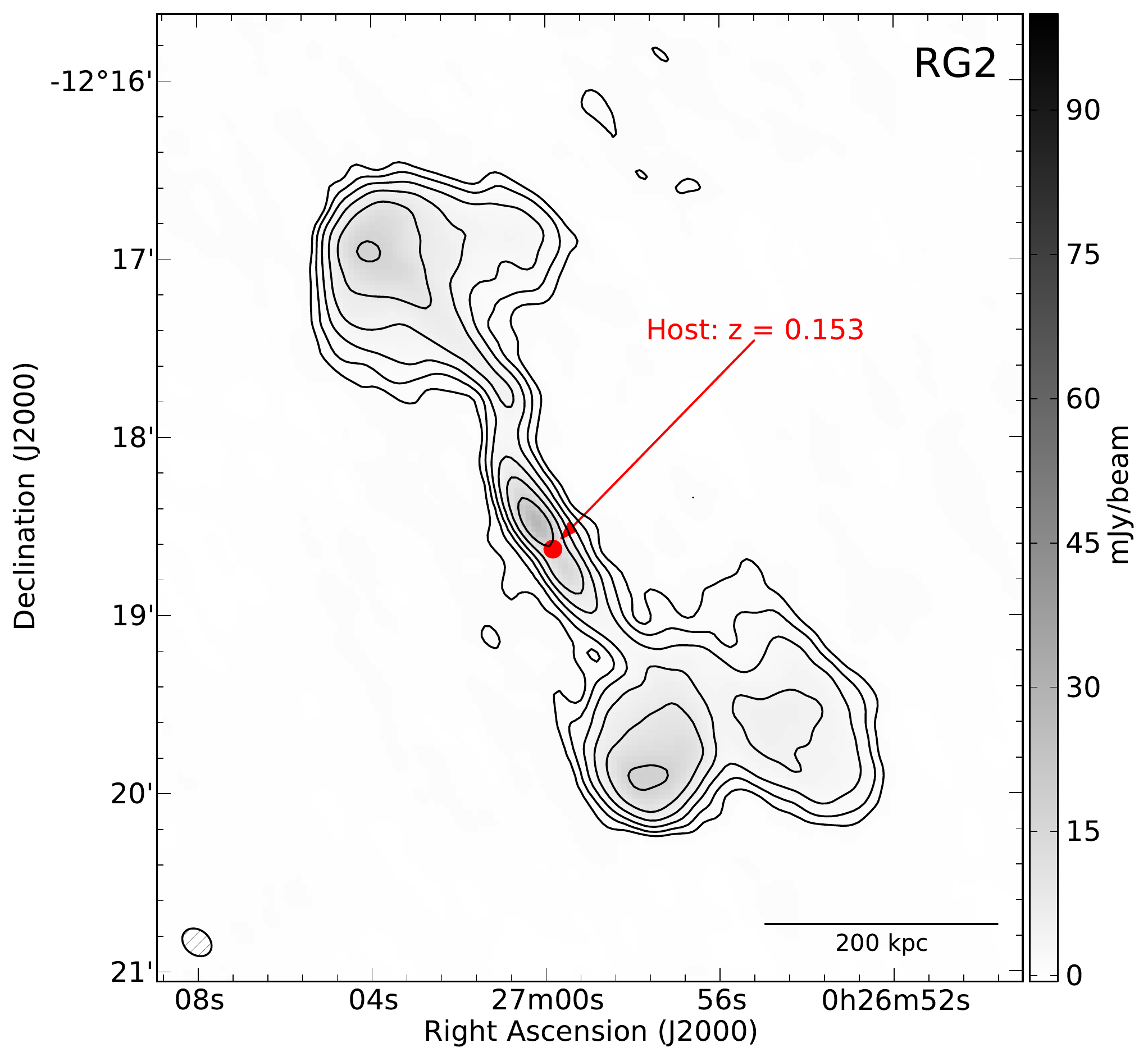}
	\end{subfigure}
	\begin{subfigure}[r]{0.39\textwidth}
	\includegraphics[width=\textwidth]{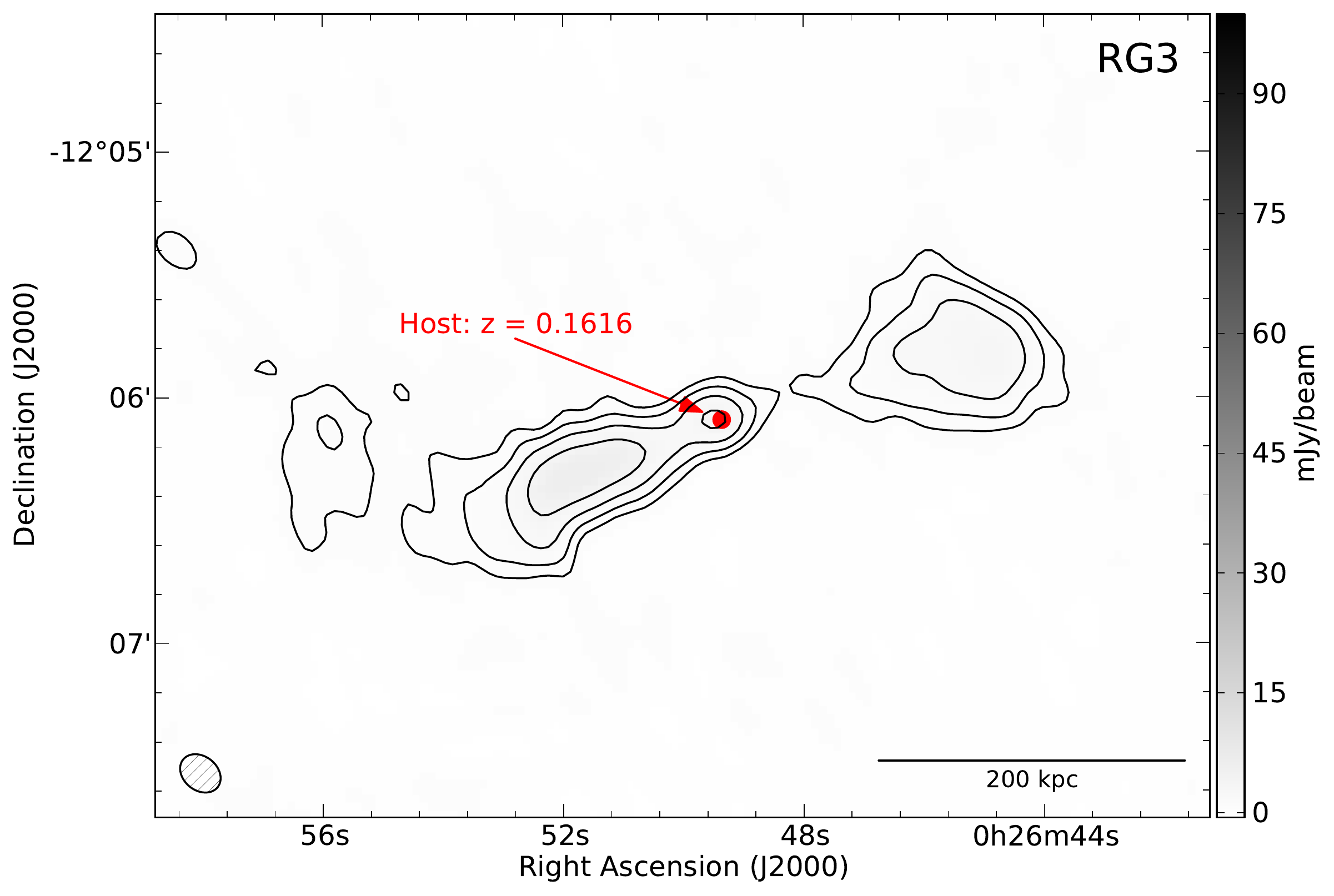}
	\end{subfigure}
	\begin{subfigure}[l]{0.39\textwidth}
	\includegraphics[width=\textwidth]{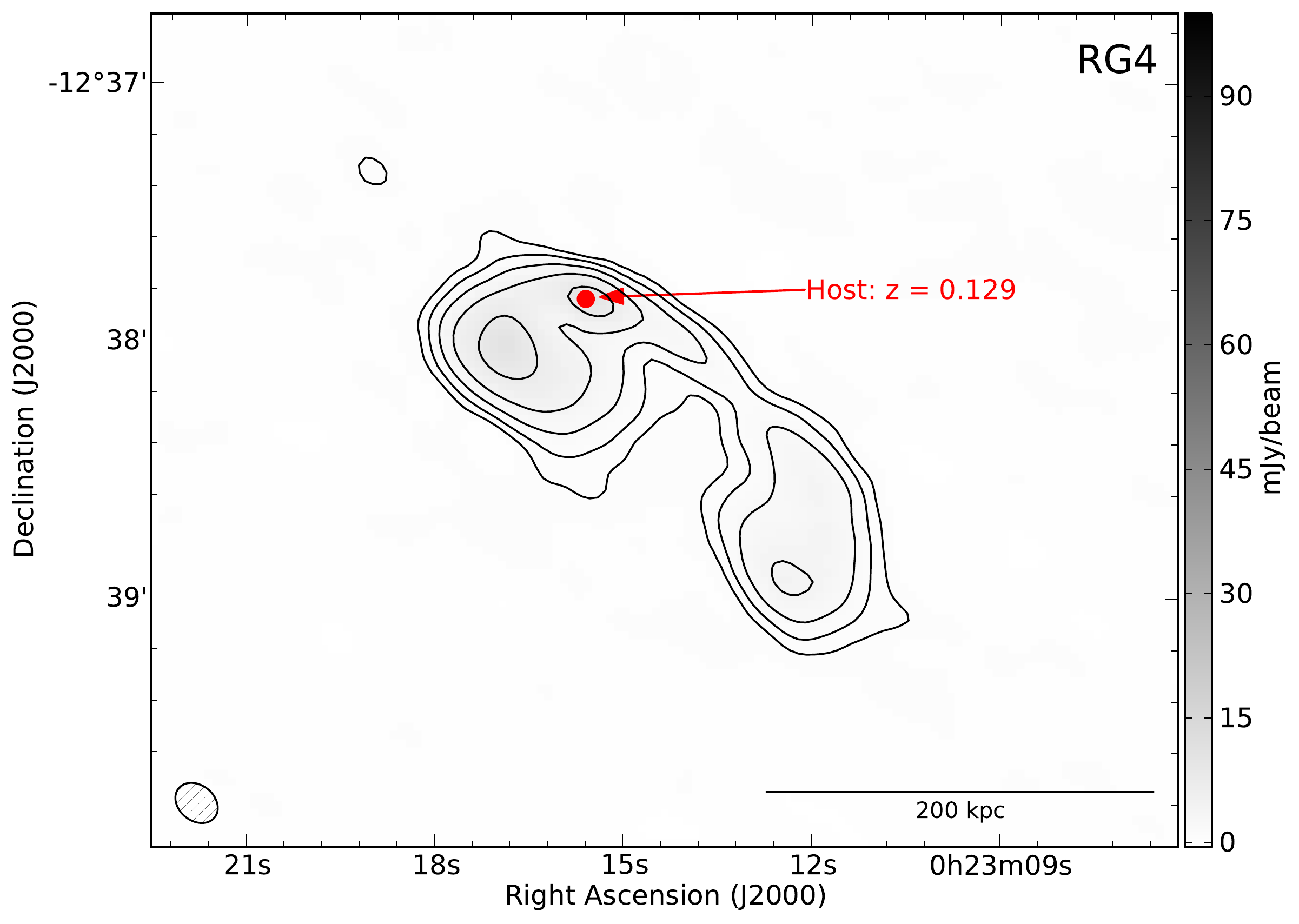}
	\end{subfigure}
	\begin{subfigure}[r]{0.39\textwidth}
	\includegraphics[width=\textwidth]{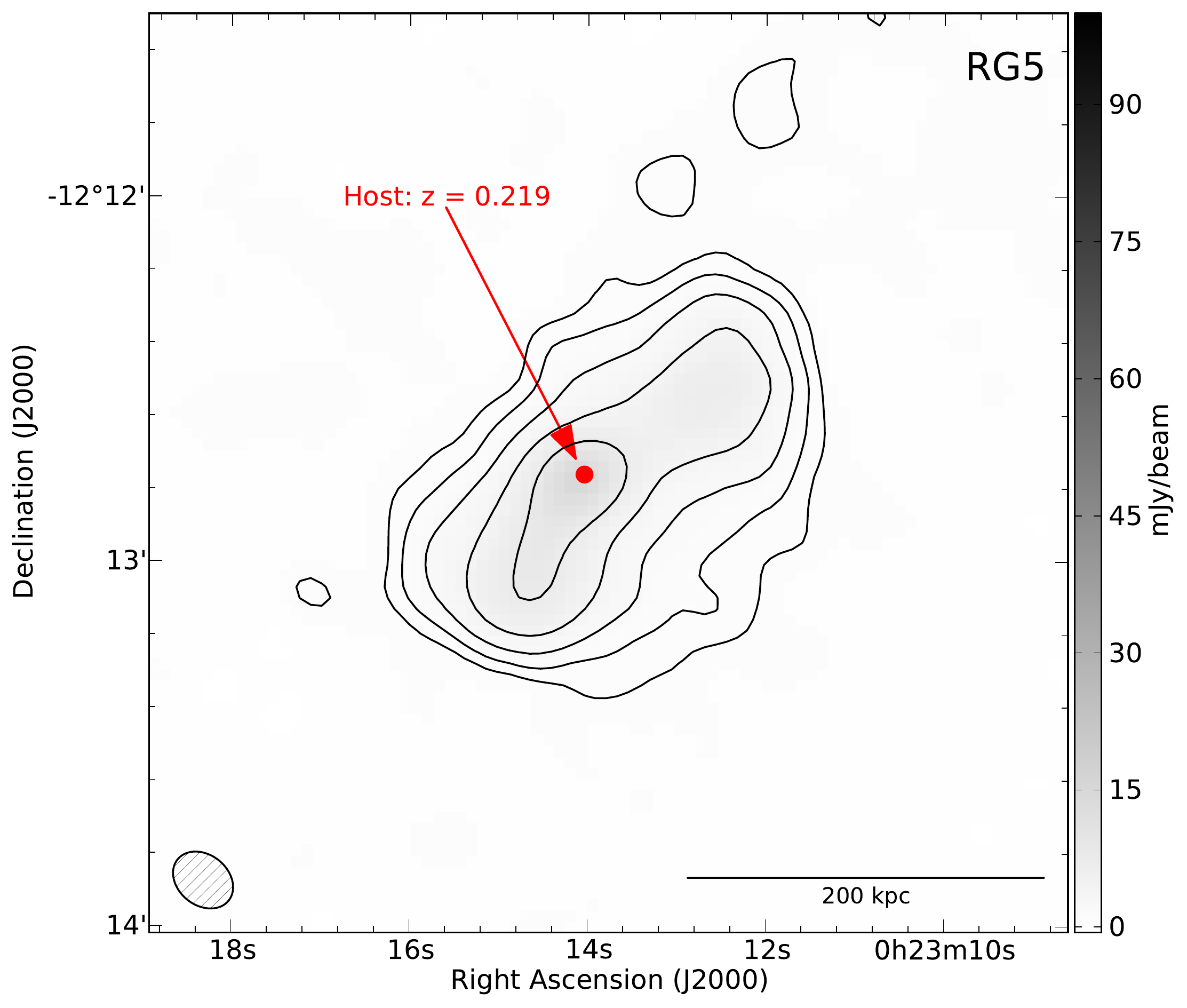}
	\end{subfigure}
	\begin{subfigure}[l]{0.39\textwidth}
	\includegraphics[width=\textwidth]{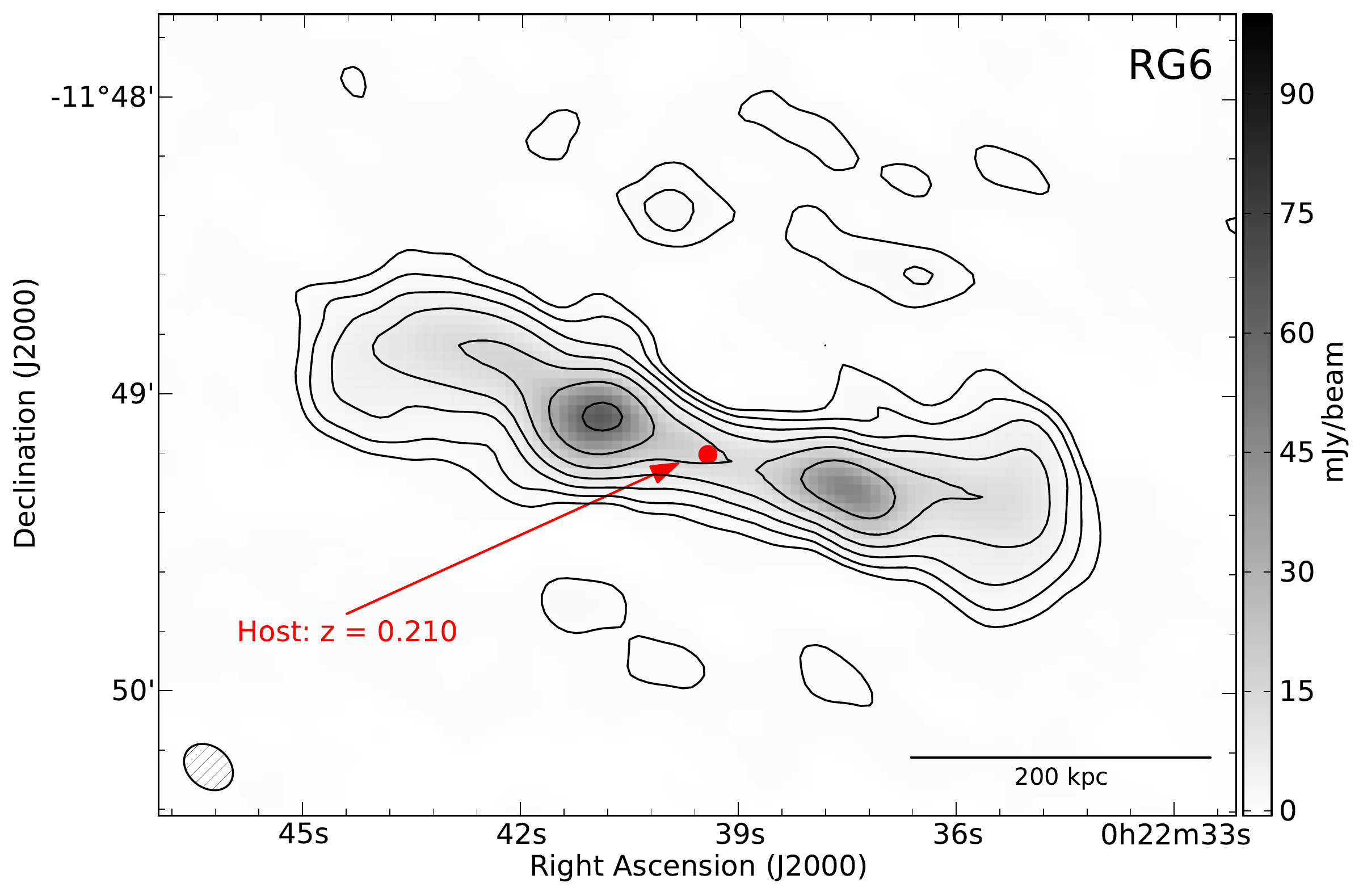}
	\end{subfigure}
	\begin{subfigure}[r]{0.39\textwidth}
	\includegraphics[width=\textwidth]{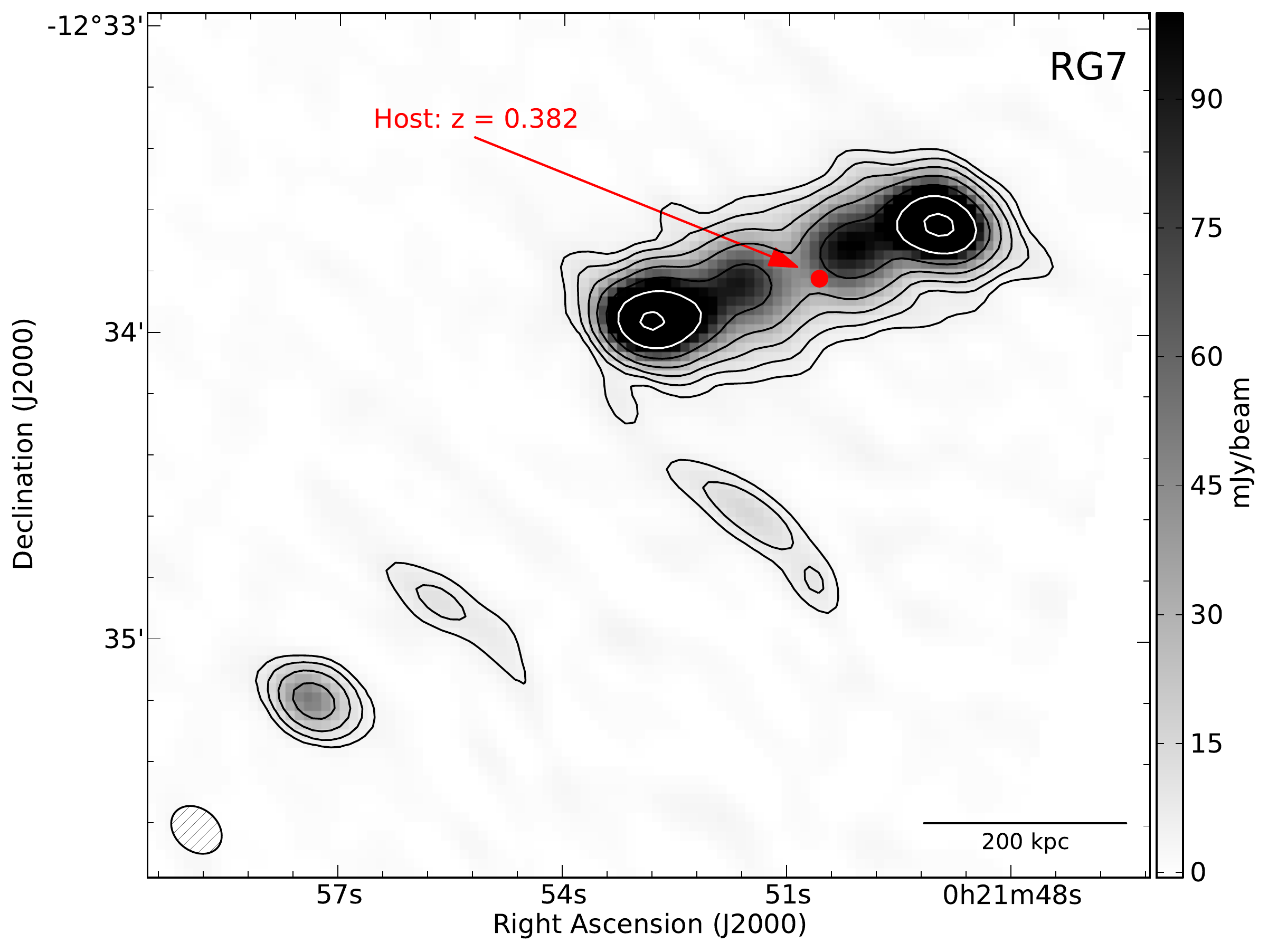}
	\end{subfigure}
\caption{Postage stamp images of resolved sources in the GMRT field of view at 325 MHz. Contours start at $5\sigma$ and scale by a factor two, where $\sigma = 697\,/\,102\,/\,89\,/\,94\,/\,93\,/\,175\,/\,875 \, \mu$Jy beam$^{-1}$ from top-left. Resolution is $10.1\times8.3$ arcsec, indicated by the hatched ellipse in the lower-left corner. Colour scales saturate at 100 mJy beam$^{-1}$. Filled circles indicate position of candidate host sources, where available. Scale bar denotes 200 kpc at the redshift of sources whose proposed hosts have measured $z$.}
\label{fig:macs_rgpostage}
\end{figure*}

Postage stamp images of these RG are presented in Figure \ref{fig:macs_rgpostage}, where contours start at $\sigma$ and scale by a factor two. The local noise for each postage stamp is noted in the caption of Figure \ref{fig:macs_rgpostage}. We have cross-referenced these sources with the literature to identify potential optical host galaxies, and we will discuss them in this appendix. 

Approximately half of this field is covered by the 12th Data Release from the Sloan Digital Sky Survey (SDSS DR12; \citealt{2015ApJS..219...12A}) from which we identify optical host galaxies with photometric redshifts for four RG. None of these objects have spectroscopic redshifts. In Figure \ref{fig:macs_rgpostage} filled circles indicate the location of any potential hosts.

\begin{table}[!htb]
\begin{center}
\caption{Measured parameters for extended radio sources in the MACS0025 field, as seen by the GMRT at 325 MHz. Flux densities were measured using \texttt{fitflux} \protect\citep{2007BASI...35...77G}. Largest angular sizes are measured directly from the postage stamp images, and converted to a largest linear size for sources with identified potential host galaxies. Redshift references are listed in the text. \label{tab:rg}}
\scalebox{0.95}{
\begin{tabular}{ccccccc}
\hline
 & & \\ 
Source & $S_{\rm{int}}$ & LAS & $z$ & LLS \\
& $[\rm{Jy}]$ & $[\rm{arcsec}]$ & & $[\rm{kpc}]$ \\
\hline
RG1 & $0.925\pm0.046$ & 83 & $-$ & $-$ \\
RG2 & $0.564\pm0.028$ & 282 & $0.153$ & $721$ \\
RG3 & $0.059\pm0.003$ & 130 & $0.162$ & $347$ \\
RG4 & $0.109\pm0.006$ & 130 & $\sim0.129$ & $\sim287$ \\
RG5 & $0.097\pm0.005$ & 78 & $0.219$ & $266$ \\
RG6 & $0.511\pm0.026$ & 170 & $0.210$ & $561$ \\
RG7 & $1.483\pm0.075$ & 86 & $0.382$ & $419$ \\
\hline
\end{tabular}
}
\end{center}
\end{table}

\subsection{RG1}
This source appears to be a double-lobed radio source, with an approximate angular extent of 83 arcsec. We do not identify any candidate host galaxy for this radio source from the literature. The nearest source catalogued is AllWISE J002711.53$-$124819.1 \citep{2014yCat.2328....0C} although this appears to be co-located with the Northern radio lobe, and is perhaps unlikely to host the observed radio emission.

\subsection{RG2: a large-angular-size radio galaxy}
The potential host for this source is identified as 2MASX~J00265983$-$1218378 (RA $00^{\rm{h}}26^{\rm{m}}59.8^{\rm{s}}$, DEC $\SI{-12}{\degree}18^{\prime}38^{\prime\prime}$) at a redshift $z=0.153$. From Figure \ref{fig:macs_rgpostage}, this host appears to be in a suitable position to host the radio emission. The angular distance between the hotspots of the radio lobes is of the order of 4.6 arcmin; with the cosmology we adopt in this work, this suggests a physical extent of order 705 kpc, slightly smaller than giant radio galaxies.

This source is well-resolved, with a marked difference in morphology between the Northern and Southern radio lobes. The North lobe appears reasonably symmetric, whereas the South lobe exhibits a highly asymmetric radio surface brightness. On the one hand, the asymmetry may indicate differences in the density of the local environment, which restricting the expansion of part of the southern lobe. On the other hand, this fainter region may be the remains of a radio lobe emitted during historic AGN activity, whereas the brighter region may be fresher plasma from a more recent phase of activity. The Northern jet also exhibits a curious `kink', which may be distorted due to the density profile of the local environment, or variation in the magnetic field orientation along the jet.
 
\subsection{RG3: a hybrid-morphology radio source}
RG3 is a complex case, exhibiting characteristics of both Fanaroff-Riley Class I (FR-I) and Class II (FR-II) radio galaxies \citep{fr74}. The Eastern lobe appears relatively bright and amorphous, with no strong evidence of a jet (FR-II). In contrast, the radio emission to the West of the core appears to be more confined for a greater distance out from the central engine. Additionally, the radio emission appears to be connected to a bright central core by a jet. Finally, further to the West there is patchy emission that if associated with this source might suggest bending of the jet.

Sources with such mixed morphology are known as hybrid-morphology radio sources (HYMORS; \citealt{2000A&A...363..507G}). A potential optical host is identified from the GLADE catalogue\footnote{Available at \url{http://aquarius.elte.hu/glade/}} \citep{2016yCat.7275....0D}. This source has a redshift $z = 0.1617$; at this redshift, the angular extent of this source (130 arcsec) corresponds to a physical extent of 347 kpc.

\subsection{RG4}
This source exhibits a curious morphology, reminiscent of the radio lobes of wide-angle-tail (WAT) RG seen in many galaxy clusters. However, no host is identified at a location consistent with this interpretation. Instead, we find a galaxy coincident with one of the `hotspots' in the NE region -- SDSS~J002315.60$-$123750.2 ($z = 0.129$) -- and no obvious host for the SW emission. 

A number of other galaxies are identified by the SDSS in this region, although they are distributed across a broad enough redshift range ($z = 0.108 - 0.414$) that we cannot suggest that they form a group. If the radio emission is associated with SDSS~J002315.60$-$123750.2, its angular extent (130 arcsec) corresponds to a physical distance of 287 kpc. The location of the host would then suggest that we are viewing this object in projection.

\subsection{RG5}
We identify a host galaxy for this object, SDSS~J002314.06$-$121245.7 ($z=0.219$). The angular extent of this source (78 arcsec) therefore corresponds to a physical extent of 266 kpc given our cosmology. At radio wavelengths, this source is also listed in the NVSS catalogue and 352 MHz WISH catalogue \citep{2002A&A...394...59D} under the identifiers NVSS 002313--121246 and WNB 0020.6--1229, respectively.

\subsection{RG6: a candidate double-double radio galaxy}
From the SDSS, we identify the host galaxy as SDSS~J002239.44$-$114912.0 ($z=0.210$). At this redshift, the angular extent of this source (170 arcsec) corresponds to a physical extent of 561 kpc. 

When viewed in concert with the host location, the source morphology suggests that RG6 may be a double-double radio galaxy (DDRG). These sources exhibit themselves as pairs of double-lobed radio sources, where the lobes are aligned along the same axis and possess a common radio core \citep{2000MNRAS.315..371S}. A number of other examples of DDRG exist in the literature (see for example \citealt{1997MNRAS.291...20L,2000MNRAS.315..371S,2001A&A...368..398S,2009BASI...37...63S,2012MNRAS.424.1061K,2015A&A...584A.112O}). DDRG are believed to arise as a result of recurrent AGN activity.

This source has also been identified in a number of radio surveys, including the 4.85\,GHz Parkes-MIT-NRAO survey (PMN; \citealt{1994ApJS...91..111W}) under the identifier PMN J0022-1148, as well as the VLSS, WISH, and NVSS.

\subsection{RG7}
The morphology of this source is reminiscent of a classic double-lobed RG, with a pair of bright hotspots diametrically opposed to each other, and extended emission trails back toward the location of a central engine. This source has been catalogued by previous surveys between 74 MHz (VLSS) and 4.85 GHz (PMN). We identify a potential host galaxy for this source -- SDSS~J002150.58$-$123349.0. This source has a photometric redshift $z = 0.382$; at this redshift, the angular size of this source (83 arcsec) corresponds to a physical extent of 419 kpc.

\end{document}